 \documentclass[11pt]{article}

\usepackage[T1]{fontenc}
\usepackage{lmodern, microtype}
\usepackage{amsmath, amsfonts, amssymb, amsthm,array,booktabs}
\usepackage{enumerate, enumitem, xcolor, soul, nicefrac, comment}
\usepackage[colorlinks=true, pdfstartview=FitV, linkcolor=blue, citecolor=black, urlcolor=blue]{hyperref}
\usepackage{subcaption}

\usepackage[small]{titlesec}
\usepackage[left=1.2in, right=1.2in, top=1.2in, bottom=1.2in]{geometry}

\newtheorem{theorem}{Theorem}
\newtheorem{lemma}{Lemma}
\newtheorem{claim}{Claim}

\newtheorem{corollary}{Corollary} 
\newtheorem{proposition}{Proposition}

\newtheorem{assumption}{Assumption}

\theoremstyle{definition}
\newtheorem{definition}{Definition}
\newtheorem{example}{Example}
\newtheorem{remark}{Remark}

\theoremstyle{remark}

\newenvironment{lemmaenum}
 {\begin{enumerate}[label=\upshape(\roman*),ref=\thelemma(\roman*)]}
 {\end{enumerate}}

\makeatletter
\renewenvironment{proof}[1][\proofname]{%
  \par\pushQED{\qed}\normalfont%
  \topsep6\p@\@plus6\p@\relax
  \trivlist\item[\hskip\labelsep\bfseries#1\@addpunct{.}]%
  \ignorespaces
}{%
  \popQED\endtrivlist\@endpefalse
}
\makeatother

\usepackage{natbib}
\setcitestyle{authoryear, open={(}, close={)}}

\newcommand{\dd}{\, \mathrm{d}}         
\newcommand{\ee}{\mathrm{e}}
         
\newcommand{\D}{\mathcal{D}}
            
\newcommand{\E}{\mathcal{E}}
            
\newcommand{\RR}{\mathbb{R}}

\DeclareMathOperator{\supp}{supp}    
 
\DeclareMathOperator{\dom}{dom} 
\DeclareMathOperator{\ri}{ri}

\DeclareMathOperator{\inte}{int}
\DeclareMathOperator*{\argmax}{arg\,max}

\newcommand{\RRcvx}{\overline{\RR}}

\renewcommand{\varepsilon}{\epsilon}

\definecolor{burgundy}{rgb}{0.5, 0.0, 0.13}
\definecolor{cadmiumgreen}{rgb}{0.0, 0.42, 0.24}
\definecolor{cottoncandy}{rgb}{1.0, 0.74, 0.85}
\definecolor{applegreen}{rgb}{0.55, 0.71, 0.0}

 \title{\bf \Large Modeling information acquisition:\\divergences, duality, and discrete choice\thanks{We thank Stefan Bucher, Andrew Caplin, Mark Dean, Francesco Fabbri, Drew Fudenberg, Duarte Gon\c{c}alves, Ben H\'{e}bert, Christian Hellwig, Philippe Jehiel, Annie Liang, Elliot Lipnowski, Jay Lu, Massimo Marinacci, Filip Mat\v{e}jka, Meg Meyer, Moritz Meyer-ter-Vehn, Stephen Morris, Jeffrey Mensch, Doron Ravid, Ilya Segal, Ludvig Sinander, Colin Stewart, Juuso Toikka, David Walker-Jones, Mirko Wiederholt, Weijie Zhong, and 
 conference and seminar audiences for helpful comments and discussions.}}

 \author{Alexander W. Bloedel\thanks{UCLA. Email: abloedel@econ.ucla.edu.}  \ \ \ \large Tommaso Denti\thanks{NYU Stern. Email: td838@stern.nyu.edu.} \ \ \
 Luciano Pomatto\thanks{Caltech. Email: luciano@caltech.edu. }
}

\begin{document}

\maketitle
 \begin{abstract}
  We introduce a new cost function over experiments, $f$-\textit{information}, based on the theory of multivariate statistical divergences, that generalizes \citeauthor{sims2003implications}'s classic model of rational inattention as well as the class of posterior-separable costs. We characterize its behavioral predictions by deriving optimality conditions that extend those of \cite{matvejka2015rational} and \cite*{caplin2019rational} beyond mutual information. Using these tools, we study the implications of $f$-information in a range of canonical decision problems. 
  The framework can be analyzed using 
  familiar methods of microeconomics---convex duality and the Arrow-Pratt approach to expected utility---and yields connections to well-known models of discrete choice.
\end{abstract}

\section{Introduction}

 Models of information acquisition describe both what a decision maker can learn before acting and the costs of that learning. While traditional models restrict the available information sources to a parametric family of experiments, more recent models adopt a non-parametric formulation, allowing the decision maker to flexibly design any experiment as their information source.  This flexibility captures the idea that the agent can fine-tune how they learn about the environment based on the decision problem at hand. Limitations on learning are then represented by an information cost function over experiments.

 Following \cite{sims2003implications}, much of the literature has assumed that the cost of information is given by Shannon's \emph{mutual information}, due in large part to its tractability. In this case, optimal behavior resembles the classic \emph{multinomial logit} model of discrete choice and the information acquisition problem can be solved via a lower dimensional optimization problem, as shown by \citet{matvejka2015rational} and \citet*{caplin2019rational}.

 This tractability comes at the price of a highly specific functional form, and a growing literature has begun to study alternative cost functions \citep*[among others]{morris2019wald, hebert2021neighborhood, caplin2022rationally, pomatto2023cost,  walker2023rational, bloedel2020cost, bordoli2025convex}. Despite much progress in this direction, extending the analysis beyond mutual information has remained challenging. Unlike utility or production functions, which are defined over familiar economic objects, information costs are defined on the abstract, infinite-dimensional space of experiments, making them inherently harder to specify. Assumptions on information costs, which are rarely observed directly, are also more difficult to test. Furthermore, no other cost function in the literature yields behavioral predictions with a structure as simple as those of mutual information. For example, the link between mutual information and logit has found no immediate generalizations.

 In this paper, we introduce a new class of information costs, \emph{$f$-information}, that encompasses mutual information and many others in the literature.  Our main result is a characterization of optimal behavior that extends those in \citet{matvejka2015rational} and \citet*{caplin2019rational} to $f$-information. Building on this result, 
 we identify a number of tractable special cases of the framework, study their implications in a range of decision problems, and relate the predicted behavior to 
 several well-known models of discrete choice:  \textit{additive perturbed utility}, \textit{nested logit}, and \emph{mixed logit}.

 \bigskip

 Formally, given a finite set $\Theta = \{\theta_1,\ldots,\theta_n\}$ of states, information is acquired by observing the outcome of an experiment $P = (S,(P_{\theta})_{\theta \in \Theta})$, where $P_\theta(s)$ is the probability of signal realization $s \in S$ in state $\theta$. The \emph{$f$-information} of experiment $P$ is defined as
 \begin{equation}\label{eq.intro_1}
    I_f(P) = \min_{\alpha \in \Delta(S)} \sum_{s \in S} \alpha(s)f\left(\frac{P_{\theta_1}(s)}{\alpha(s)},\ldots,\frac{P_{\theta_n}(s)}{\alpha(s)}\right),
 \end{equation}
 where $f$ is a non-negative convex function satisfying $f(1,\ldots,1)=0$. For a fixed distribution $\alpha$ over signal realizations, the map $f$ assigns a penalty based on the likelihood ratios between the state-contingent distributions $P_{\theta_1},\ldots,P_{\theta_n}$ and $\alpha$. In statistics, this quantity is known as the \emph{multivariate} $f$\textit{-divergence} between $P$ and $\alpha$.
 The cost of $P$ is computed by selecting the distribution $\alpha$ for which the divergence is minimal. We call the solution to the minimization problem \eqref{eq.intro_1} the \textit{$f$-mean} of $P$, which can be interpreted as the distribution over signals that best approximates the experiment as a whole. In this manner, the notion of $f$-information formalizes the idea that an experiment $P$ is more informative when its state-contingent distributions $P_{\theta_1},\ldots,P_{\theta_n}$ are more distinct from each other, and is less informative when these distributions cluster around their $f$-mean. 

 By varying the transformation $f$, we obtain a menu of different cost functions over experiments. For example, we obtain mutual information when $f$ takes the form 
 \[
    f(x) = \sum_{\theta \in \Theta} \pi(\theta)(x(\theta)\log x(\theta) - x(\theta) + 1),
 \]
 with $\pi$ the prior belief over states. Another special case of interest is the family of \emph{posterior-separable} costs \citep*{caplin2022rationally}, which includes mutual information as well as most other cost functions that have been proposed in the literature.

 \bigskip

 In the first part of the paper, we characterize optimal behavior under $f$-information. Given a decision problem with a finite action set $A$, we determine the optimal stochastic choice rule $P=(A,(P_\theta)_{\theta\in\Theta})$, where $P_\theta(a)$ is the probability of taking action $a$ in state $\theta$.
 
 It is useful to begin by recalling the case of mutual information. As shown by \citet{matvejka2015rational} and \citet*{caplin2019rational}, a stochastic choice rule is optimal 
 if and only if it satisfies two conditions. First, each conditional probability $P_\theta$ is related to the unconditional distribution $P_\pi=\sum_{\theta}\pi(\theta)P_\theta$ by the modified logit formula
 \begin{equation}\label{eq.intro_MM_1}
    P_\theta(a) = \frac{P_\pi(a)\ee^{a(\theta)}}{\sum_{b \in A} P_\pi(b) \ee^{b(\theta)}},
 \end{equation}
 where $a(\theta)$ is the payoff that action $a$ delivers in state $\theta$. Second, the unconditional probability $P_\pi$ is the solution to an auxiliary optimization problem of lower dimension.\footnote{This result has found many applications, including studies on labor economics \citep*{acharya2020rational}, insurance choice \citep{brown2024endogenous}, industrial organization \citep*{matvejka2012simple,cusumano2024competing}, optimal pricing \citep*{matvejka2015rigid,matvejka2016rationally,boyaci2018pricing}, bargaining \citep{ravid2020ultimatum,fabbri2024attention}, coordination games \citep{yang2015coordination,denti2023unrestricted}, security design \citep{yang2020optimality}, information design \citep{bloedel2021persuading}, and discrimination \citep*{echenique2025rationally}.}

 For $f$-information, we obtain a parallel two-step characterization. Central to this result is the function $f^\star$, the \emph{convex conjugate} of the transformation $f$. We show that a stochastic choice rule $P$ is optimal if and only if each conditional probability $P_\theta$ satisfies:
 \begin{equation}\label{eqn:intro_2}
     P_{\theta}(a)=\alpha(a)\nabla_{\theta}f^{\star}(a\pi-\lambda),
  \end{equation}
 where $\alpha$ is the $f$-mean of the choice rule, $\nabla_{\theta}f^{\star}$ is the partial derivative of $f^\star$ with respect to the $\theta$-th component of its argument, $\lambda \in \RR^\Theta$ is a vector of Lagrange multipliers ensuring that each of the conditional distributions $P_\theta$ sums to $1$, and the prior belief $\pi$ enters by multiplying the vector $a \in \RR^\Theta$ of state-contingent payoffs from action $a$ statewise. Furthermore, the endogenous terms $\alpha$ and $\lambda$ are the solutions to an auxiliary saddle-point problem, which is of lower dimension than the original information acquisition problem. 

 Under mutual information, condition \eqref{eqn:intro_2} reduces to \eqref{eq.intro_MM_1} because $\alpha$ equals the unconditional distribution $P_\pi$ and $\nabla_{\theta}f^{\star}(a\pi-\lambda)$ is proportional to $\ee^{a(\theta)}$, an exponential transformation of the payoff $a(\theta)$. 
 Beyond mutual information, condition \eqref{eqn:intro_2} establishes a more general relation between choice probabilities and the $f$-mean $\alpha$, with $\nabla_\theta f^\star$ replacing the exponential function as the relevant map from payoff vectors $a \in \RR^\Theta$ to probabilities.

 A notable feature of our result is that the transformation $f$ appears in \eqref{eqn:intro_2} not directly, but through its convex conjugate $f^\star$.  As in other instances of duality---Marshallian vs.~Hicksian demand, cost vs.~profit functions---these two objects provide complementary perspectives on the problem. Assumptions stated in terms of $f$ determine how the cost changes as a function of the experiment, whereas assumptions stated in terms of $f^\star$ determine how the primitives of the decision problem (i.e.\ the prior and action set)
 translate into choice probabilities. Although these two perspectives are ultimately equivalent, properties of the conjugate $f^\star$ are more directly related to optimal behavior.
 
 \bigskip

 In the second part of the paper, we focus on a tractable special case of $f$-information. We consider the additively separable specification
 \[
    f(x) = \sum_{\theta \in \Theta} \pi(\theta) \phi(x(\theta)),
 \]
 where $\phi$ is a univariate convex function. This functional form was studied in information theory by \cite{csiszar1972class}, and we accordingly refer to the resulting cost function as \emph{Csisz\'ar information}. Csiszár information preserves the separable structure of mutual information while allowing the curvature of the underlying function $\phi$ to vary, much as expected utility generalizes logarithmic utility. This 
 extra flexibility comes with much of the tractability of mutual information: $f^\star$ remains additively separable, and the Lagrange multiplier can be computed statewise. Notably, except at mutual information, this class of costs does not overlap with the posterior-separable costs studied in most prior work.
 
 We study the behavioral content of Csisz\'ar information in discrete-choice problems. We show that the optimal stochastic choice rule under Csiszár information closely resembles additive perturbed utility, a well-known model of discrete choice that generalizes logit \citep*{fudenberg2015stochastic}. For symmetric decision problems, the two models generate the same predictions. For general problems, optimal behavior under Csisz\'ar information differs by the endogenous term $\alpha$, the $f$-mean of the choice rule. We interpret $\alpha(a)$ as the \textit{salience} of action $a$, and characterize what it means for one action to be more salient than another. In special cases, $\alpha$ can be determined in closed-form. 
 
 Under Csisz\'ar information, the properties of the optimal choice rule can be analyzed through the degree of convexity of the conjugate function $\phi^\star$. To measure this convexity, we draw on tools from risk theory, in particular the Arrow-Pratt coefficient of $\phi^\star$. As we establish, the Arrow-Pratt coefficient measures the decision maker's response to a marginal increase in the stakes of the decision problem. This provides a unified way to interpret assumptions on the curvature of $\phi$, and in particular to study how behavior under Csisz\'ar information departs from the mutual information benchmark.

Using these tools, we study the canonical question of how the probability of making correct choices scales with the stakes of the decision problem. We consider a class of identification tasks in which the decision maker is rewarded for guessing the true state, and ask how the probability of a correct response varies with the reward level and the task difficulty. In perceptual experiments, subjects are often less responsive to incentives than the mutual information benchmark predicts \citep{dean2023experimental}. As we show, Csisz\'ar information allows for a much wider range of responses, and the shape of the resulting psychometric curve can be determined directly in terms of the curvature of $\phi^\star$. 
 
 \bigskip

 In the last part of the paper, we apply $f$-information to address a well-known limitation of mutual information: the fact that states enter into the analysis only through their payoff consequences. This property rules out the possibility that distinguishing between more similar states, as measured by their physical characteristics or proximity, may be costlier. In doing so, it leads to unrealistic predictions, such as sharp discontinuities in behavior where smoother adjustments would be expected \citep*{hebert2021neighborhood,morris2022coordination,dean2023experimental,pomatto2023cost}. 

 We introduce a tractable family of posterior-separable costs, \textit{Nested Shannon entropy}, that generalizes mutual information by allowing one to specify which subsets, or \emph{nests}, of states share similar attributes. These costs describe the decision maker as following an optimal two-step learning process in which they first learn about which nest contains the true state, and then learn about the states within that nest. We relate the resulting behavior to the well-known nested logit model and, in a special case, to the mixed logit model. We also connect these cost functions to \citeauthor{hebert2021neighborhood}'s (\citeyear{hebert2021neighborhood}) neighborhood-based costs and to \citeauthor{walker2023rational}'s (\citeyear{walker2023rational}) multi-attribute Shannon entropy. 

 To illustrate the model, we study a canonical one-dimensional discrimination task and derive simple sufficient conditions under which the predicted psychometric curve is S-shaped, as observed in laboratory experiments. We also apply the model to a multi-dimensional discrimination task and show that it can capture the idea that learning about a multi-dimensional state may be harder than learning about a uni-dimensional one, an intuitive prediction that does not arise under other cost functions in the literature.

 \subsection{Related literature}\label{sec.related_literature}

 Following \citeauthor{sims2003implications} (\citeyear{sims2003implications}), and building on the optimality conditions for mutual information derived by \cite*{matvejka2015rational} and \cite*{caplin2019rational}, a burgeoning literature has examined the properties and behavioral implications of information costs \citep*[see][for surveys]{mackowiak2018survey,strzalecki2025stochastic}. 
 
 \paragraph{Posterior separability.}  Most research on rational inattention has centered on the class of posterior-separable costs introduced by \cite*{caplin2022rationally}. For an experiment $P=(S,(P_\theta)_{\theta\in \Theta})$, these cost functions take the form
 \[
    C(P) = \sum_{s\in \supp P_\pi} P_\pi(s) H(p_s),
 \]
 where $P_\pi\in \Delta(S)$ is the unconditional signal distribution, $p_s\in \Delta(\Theta)$ is the posterior belief following signal realization $s$,  and $H$ is a convex entropy function assigning a cost to each posterior. By allowing for arbitrary entropy functions, posterior-separable costs generalize mutual information, which arises when $H$ is proportional to Shannon entropy.

 We show that $f$-information, despite generalizing mutual information in a seemingly distinct way, encompasses the class of posterior-separable costs. In particular, $f$ and its conjugate $f^\star$ can be expressed in terms of the entropy $H$ and its conjugate $H^\star$. 
 Hence, our analysis yields optimality conditions for posterior-separable costs as a special case. 

 While we are not the first to derive such conditions, our analysis offers a new perspective by shifting the focus from the entropy $H$ to its conjugate $H^\star$. It is well known that optimal behavior under posterior-separable costs can be characterized via concavification and related Lagrangian methods, which yield a ``primal'' first-order condition involving $\partial H$, the subgradient of the entropy $H$.\footnote{See \citet*[Lemma 1]{caplin2022rationally}, \citet[Lemma 10]{denti2022posterior}, \citet[Proposition 3]{lipnowski2022predicting}, and \citet[Corollary 2]{bloedel2025scoring}, among others.}  We instead characterize optimal behavior via a dual first-order condition involving $\nabla H^\star$, the gradient of the conjugate $H^\star$. As we explain further in Section \ref{ssec:posterior-separable} (see Remark \ref{remark:PS-primal-dual}), our dual condition is more directly related to optimal behavior: although the primal first-order condition provides an implicit characterization, in order to explicitly solve for optimal behavior one must \emph{invert} the subgradient $\partial H$, which yields precisely our dual first-order condition.\footnote{The primal condition is most useful in revealed preference and mechanism design settings, where the goal is to construct an entropy function or a decision problem to rationalize a given distribution of posteriors. In revealed preference exercises, the distribution can be inferred by the analyst from the agent's choice behavior \citep*[among others]{caplin2015revealed,caplin2022rationally,denti2022posterior}. In design problems, the designer chooses the distribution to be implemented, subject to incentive compatibility \citep*[among others]{mensch2022screening,mensch2026monopoly,bloedel2025scoring}.} 
 
 Thus, for the purpose of analyzing optimal behavior, what matters is the tractability of $H^\star$, not that of $H$. 
 Notably, there is no guarantee that both $H$ and $H^\star$ have simple closed forms. For instance, the posterior-separable costs in \cite*{hebert2021neighborhood}, \cite*{pomatto2023cost}, and \cite*{bloedel2020cost} have simple functional forms but, to the best of our knowledge, their conjugates do not. The family of Nested Shannon entropies that we introduce here provides an example of posterior-separable costs where $H$ admits a natural interpretation and, at the same time, $H^\star$ remains tractable.

 \paragraph{Beyond posterior separability.} A smaller but growing strand of the literature studies non-posterior-separable cost functions. Our paper relates to three lines of this work.

 First, several papers develop revealed-preference analyses of costly information acquisition with general cost functions (e.g., \citealp{caplin2015revealed}; \citealp*{de2017rationally}).\footnote{See also \cite*{ellis2018foundations}, \cite*{chambers2020costly}, and \cite*{lin2022stochastic}.} Although our main objectives are different, we take inspiration from this approach by using our optimality conditions to study the behavioral implications of $f$-information in a number of canonical decision problems. We leave a full revealed-preference characterization of $f$-information to future work. 

 Second, a number of papers propose non-posterior-separable costs by imposing structural restrictions directly on the cost function. Most closely related are \citet*[Theorem 2]{mu2021blackwell} and \cite*{bordoli2025convex}, which relax the linearity axioms of \cite*{pomatto2023cost} to derive costs based on R\'{e}nyi divergences between state-contingent signal distributions. Although these costs and $f$-information both build on notions of statistical divergence, we are not aware of a simple connection between them. Also related are the \emph{sequential learning-proof} costs of \cite*{bloedel2020cost}, which are defined via their robustness to dynamic optimization of the information acquisition process. While the classes of $f$-information and sequential learning-proof costs are known to overlap (both include the family of uniformly posterior-separable costs), a full characterization of their relationship remains an open question. 

 Finally, two recent papers share our interest in deriving optimality conditions for non-posterior-separable costs, albeit from complementary angles. \cite{lipnowski2022predicting} show that a version of the standard ``primal'' first-order condition for posterior-separable costs extends to the larger class of \emph{iteratively differentiable} costs, which are locally---but not globally---posterior-separable. \cite*{muller2021rational} introduce the concept of an \emph{ignorance equivalent}: a vector of state-contingent payoffs that serves as a summary statistic in information acquisition problems and, in some contexts, obviates the need to fully solve for optimal strategies. For posterior-separable costs, the ignorance equivalent collapses to (a normalized version of) the Lagrange multiplier in \eqref{eqn:intro_2}.

 \paragraph{Convex duality in discrete choice.}
 Our paper also connects to the  discrete choice literature, where dual optimality conditions related to \eqref{eqn:intro_2} date back to the Williams-Daly-Zachary Lemma for additive random utility models (see \cite*{strzalecki2025stochastic} for a survey). Most similar is the family of \emph{perturbed utility} models, in which stochastic choice arises from control costs of selecting the correct action. \cite*{hofbauer2002global} derive an analog of \eqref{eqn:intro_2} for such models. \cite*{fudenberg2015stochastic} characterize a special case, the \emph{additive perturbed utility} model, which we relate to Csisz\`ar information.

 A distinction between our paper and most of the discrete choice literature is that, in the latter, stochastic choice arises for reasons unrelated to information acquisition. \cite*{fosgerau2020discrete} study an intermediate case in which the decision maker faces a \emph{Bregman information} cost: a family of cost functions over stochastic choice rules based on Bregman divergences. They also use convex duality to extend the optimality conditions in \cite{matvejka2015rational} and connect the predicted behavior to classic models of discrete choice. Unlike our paper, however, their framework cannot generally be interpreted as a model of  information acquisition, as Bregman information costs need not be Blackwell monotone \citep*{cheng2025monotonicity}.
 
 \section{Setup}\label{sec:info-acquisiton-problems}

 We consider an agent who faces a decision problem under uncertainty and may acquire costly information before committing to an action. 
 
 Let $\Theta$ be a finite set of \emph{states}. A \emph{decision problem} is a pair $\mathcal{D}=(\pi,A)$, where $\pi\in \Delta(\Theta)$ is a full-support \emph{prior belief} over states, and $A$ is a finite set of \emph{actions}. The agent's preferences over actions are represented by a state-dependent Bernoulli utility function. For simplicity, we identify each action with its induced profile of state-dependent utilities. That is, we view $A$ as a finite subset of $\mathbb{R}^\Theta$ and interpret $a(\theta)\in \mathbb{R}$ as the utility delivered by action $a$ in state $\theta$. We write $\Delta^\circ(\Theta)$ for the set of full-support prior beliefs on $\Theta$.
 
 We model the acquisition of information as the choice of an experiment. An \emph{experiment} $P=(S,(P_\theta)_{\theta\in \Theta})$ consists of a finite \emph{signal space} $S$
 and a profile $(P_\theta)_{\theta\in \Theta}$ of state-contingent probability distributions $P_\theta\in\Delta(S)$.\footnote{Note that we allow the agent to choose the signal space $S$, which may be any finite set. Given our focus on decision problems with finite actions, restricting attention to experiments with finite signal spaces is without loss 
 from a revealed-preference perspective.} The interpretation is that, conditional on state $\theta$, the experiment yields  signal $s \in S$ with probability $P_\theta(s)$. For any prior $\pi$ and experiment $P$, we denote by $P_\pi\in \Delta(S)$ the unconditional signal distribution: 
 $
    P_\pi=\sum_{\theta\in \Theta} \pi(\theta) P_\theta.
 $
 
 The cost of information is described by a function $C$, where $C(P,\pi)\in \overline{\RR}_+$ is the cost of experiment $P$ under prior belief $\pi$.\footnote{We let $\overline{\RR}_+ = [0,+\infty]$ denote the set of non-negative extended real numbers, where $\overline{\RR} = [-\infty,+\infty]$.} Experiments with infinite cost are 
 infeasible.
 Information costs are measured in the same units as utility and enter additively. Thus, in state $\theta$, acquiring experiment $P$ and taking action $a$ yields the net payoff
$
a(\theta) - C(P,\pi).
$

 The value of information arises from the agent's ability to tailor their actions to the signals generated by the chosen experiment, and thereby indirectly to the realized state. For any finite signal space $S$, an \emph{action strategy} $\sigma=(A,(\sigma_s)_{s \in S})$ specifies a probability distribution $\sigma_s \in\Delta(A)$ over actions to be taken contingent on each signal $s \in S$. 

 Given a decision problem $\mathcal{D}=(\pi,A)$, the agent selects an experiment $P=(S,(P_\theta)_{\theta\in \Theta})$ and an action strategy $\sigma=(A,(\sigma_s)_{s \in S})$ to maximize their expected net payoff: 
 \begin{equation}\label{eq:rat_in_problem}
     \max_{P, \, \sigma} \  \sum_{\theta\in \Theta}\pi(\theta)\sum_{s\in S}P_\theta(s) \sum_{a\in A} \sigma_s(a) a(\theta)-C(P,\pi).
 \end{equation}
 We refer to \eqref{eq:rat_in_problem} as an \emph{information acquisition problem}. 

\begin{remark}
    We allow, but do not require, the cost $C(P,\pi)$ of an experiment $P$ to depend on the prior $\pi$. We do so primarily to ensure that popular cost functions from the literature, such as mutual information, are included in our framework. 
    This prior-dependence is relevant only for making predictions across decision problems with different priors; in particular, it plays no 
    role in our main theorem, where the prior is held fixed.
    \footnote{For further discussion of the interpretation and implications of prior dependence, see \cite*{denti2022experimental} and \cite{bloedel2020cost}.}
\end{remark}

\section{$f$-information cost functions}\label{sec:f-informativity}

  We study information acquisition problems under a new class of \emph{$f$-information} cost functions. These costs are based on a notion of statistical distance between probability measures known as \emph{multivariate $f$-divergence} (\citealp*{gyorfi1978f}; \citealp*{garcia2012divergences}; \citealp*{duchi2018multiclass}). 
  
 \subsection{Multivariate $f$-divergence}

 A multivariate $f$-divergence is defined by a function $f \colon \mathbb{R}_{+}^n \to \overline{\RR}_+$ that is convex, lower semicontinuous, and normalized so that $f(\mathbf{1}) = 0$, where $\mathbf{1} = (1,\dots, 1)$. 
 \begin{definition}\label{def.multivariate.divergence}
 For any $n \geq 1$, finite set $S$, profile of probability measures $(\beta_1, \ldots, \beta_n) \in \Delta(S)^n$, and reference probability measure $\alpha \in \Delta(S)$, the \emph{multivariate $f$-divergence} of $(\beta_1, \ldots, \beta_n)$ from $\alpha$ is\footnote{To evaluate the sum in this expression when $\alpha$ does not have full support, we adopt the standard continuity convention that $0\cdot f\!\left(x_1/0,\ldots,x_n/0\right) = \lim_{t\to+\infty}f(\boldsymbol{1}+tx) / t$ for all $x=(x_1,\ldots,x_n)\in\mathbb R_+^n$.}
  \[
    D_{f}(\beta_1,\ldots,\beta_{n}\Vert \alpha)=\sum_{s\in S}\alpha(s)f\left(\frac{\beta_1(s)}{\alpha(s)},\ldots,\frac{\beta_{n}(s)}{\alpha(s)}\right).
  \]
 \end{definition}

 The divergence $D_f$ quantifies the discrepancy between the profile of measures $(\beta_1, \ldots, \beta_n)$ and the reference measure $\alpha$ as the $\alpha$-weighted average of the likelihood-ratio vector $(\beta_1/\alpha, \ldots, \beta_n/\alpha)$, transformed by the function $f$. Our assumptions on $f$ imply $D_f$ is non-negative, convex, and lower semicontinuous, with $D_f(\alpha, \ldots, \alpha \Vert \alpha) = 0$ for all $\alpha$.

 When $n=1$, this definition reduces to the classical notion of \emph{$\phi$-divergence} between pairs of probability measures (\citealp{csiszar1963informationstheoretische, csiszar1967information}; \citealp*{ali1966general}). Specifically, given a function $\phi \colon \RR_+ \to \overline{\RR}_+$ that is convex, lower semicontinuous, and normalized so that $\phi(1)=0$, the $\phi$-divergence of $\beta\in \Delta(S)$ from $\alpha$ is given by
 \begin{equation}\label{eqn:phi-divergence}
    D_{\phi}(\beta \Vert \alpha)=\sum_{s\in S}\alpha(s)\phi\left(\frac{\beta(s)}{\alpha(s)}\right).
 \end{equation}
 The most familiar example in economics and statistics is the \emph{Kullback–Leibler (KL) }divergence, which is the special case of \eqref{eqn:phi-divergence} generated by $\phi(t)=t\log t - t + 1$. 

\subsection{$f$-information}\label{ssec:f-info}

 We now introduce $f$-information cost functions, building on the notion of $f$-divergence. To accommodate prior-dependence, an $f$-information cost is generated by a family 
 \[
    f=(f_\pi)_{\pi\in \Delta^\circ(\Theta)},
 \]
 where, for each full-support prior $\pi$, the function 
 $
    f_\pi\colon \mathbb{R}^\Theta_+\rightarrow\overline{\mathbb{R}}_+
 $
 is convex, lower semicontinuous, and normalized so that $f_\pi(\boldsymbol{1})=0$. For any experiment $P = (S,(P_\theta)_{\theta \in \Theta})$ and reference measure $\alpha\in \Delta(S)$, let $D_{f_\pi}(P \Vert \alpha)$ be the $f_\pi$-divergence between $(P_{\theta})_{\theta\in \Theta}$ and $\alpha$.

\begin{definition}\label{def:f_info}
 The \emph{$f$-information} of an experiment $P = (S,(P_\theta)_{\theta \in \Theta})$ 
under prior $\pi$ is 
\[
    I_{f} (P,\pi) = \min_{\alpha \in\Delta(S)}D_{f_\pi}(P \Vert \alpha).
\]
A probability measure $\alpha \in \Delta(S)$ that attains the minimum 
is called an \emph{$f$-mean} of $P$.\footnote{An $f$-mean is ensured to exist because the map $\alpha \mapsto D_{f_\pi}(P \Vert \alpha)$ is lower semicontinuous on $\Delta(S)$.} 
\end{definition}

 $f$-information reflects the principle that an experiment $P$ is more informative when its state-contingent signal distributions $P_\theta$ are more distinct from one another.
 The informativeness of $P$ is measured by the divergence between the $P_\theta$ distributions and their $f$-mean $\alpha$, which can be interpreted as a non-linear mean of the experiment. 
The more similar the distributions $(P_\theta)_{\theta \in \Theta}$ are to their $f$-mean, the more similar they are to one another---and, hence, the less informative is the experiment. 

 A similar logic underlies the more familiar concepts of mean and variance. The arithmetic mean of $n$ real numbers $x_1,\ldots,x_n$ is the unique minimizer of the quadratic distance $\sum_{i=1}^n(x_i-y)^2$ over all $y\in \mathbb{R}$. The variance of $x_1,\ldots,x_n$ 
 is then the average quadratic distance from the arithmetic mean. By analogy, the notions of $f$-mean and $f$-information can be seen as the generalized ``mean'' and ``variance'' of an experiment.  

$f$-information satisfies several standard properties, such as monotonicity with respect to the Blackwell order. Recall that  an experiment $P=(S,(P_{\theta})_{\theta\in \Theta})$ \emph{Blackwell dominates} an experiment $Q=(S',(Q_{\theta})_{\theta\in \Theta})$ if there exists a stochastic kernel $K=(K_{s})_{s\in S}$, where $K_s \in \Delta(S')$ for every $s \in S$, such that 
$ Q_\theta(s')= \sum_{s\in S} K_s(s') P_\theta(s)$ for all $\theta \in \Theta$ and $s' \in S'$. 
 An experiment is \emph{uninformative} if its signal distribution is the same in all states. 

 \begin{lemma}\label{lem:f_info_properties} For every prior $\pi$, $f$-information has the following properties:
 \begin{enumerate}
    \item If $P$ Blackwell dominates $Q$, then $I_f(P,\pi)\geq I_f(Q,\pi)$.

    \item If $P$ is uninformative, then $I_f(P,\pi)=0$.

    \item For every signal space $S$, $I_f(\,\cdot,\pi)$ is convex and lower semicontinuous on $\Delta(S)^\Theta$.
    
\end{enumerate}
\end{lemma}

Properties (i)--(iii) imply that $f$-information falls within the class of \emph{canonical} information costs from \cite{caplin2015revealed} and \cite*{de2017rationally}. As shown there, these properties are without loss of generality from a revealed-preference perspective---they cannot be falsified by observing an agent's behavior in any decision problem---and are necessary and sufficient for costs to be identifiable from choice data.\footnote{The minimization over $\alpha$ in Definition \ref{def:f_info} is essential: for fixed $\alpha\in \Delta(S)$, the function $P\mapsto D_{f_\pi}(P\|\alpha)$ is generally not Blackwell monotone on $\Delta(S)^\Theta$. Under this function, the uninformative experiment with $P_\theta = \alpha$ for all $\theta \in\Theta$ has zero cost, while other uninformative experiments generally have positive costs.}

 \subsection{Special cases}\label{subsec:f_info_example}

 By varying the generator $f$, we obtain a wide range of information costs as special cases. Examples \ref{exa:shannon} and \ref{exa:CDL} below are familiar from the rational inattention literature; Example \ref{exa:csiszar} is drawn from the statistics literature, but is new to models of information acquisition. 

 \begin{example}[Mutual information]\label{exa:shannon} Given a prior $\pi$ and an experiment $P$, the \emph{mutual information} of the state and the experiment's signal is the quantity 
 \[
    I_S(P,\pi)=\sum_{\theta\in \Theta}\pi(\theta) D_{\mathrm{KL}}(P_{\theta} \Vert P_\pi),
 \]
where $D_{\mathrm{KL}}$ denotes the Kullback-Leibler divergence. As is well known, mutual information admits the equivalent variational representation
\[
    I_S(P,\pi)=\min_{\alpha\in \Delta(S)}\sum_{\theta\in \Theta}\pi(\theta) D_{\mathrm{KL}}(P_{\theta} \Vert \alpha).
 \]
Thus, mutual information is the special case of $f$-information corresponding to
  \begin{equation}\label{eq:f_pi_mutual}
    f_\pi(x)=\sum_{\theta\in \Theta} \pi(\theta) \left(x(\theta) \log x(\theta) -x(\theta) +1\right).
 \end{equation}
In this case, the $f$-mean of $P$ 
is simply the unconditional signal distribution $P_\pi$. 
\end{example}

 \begin{example}[Posterior separable]\label{exa:CDL} Much of the rational inattention literature focuses on the class of posterior-separable costs introduced by \cite*{caplin2022rationally}.
 
 A \emph{posterior-separable} cost $C$ is generated by a family of \emph{entropy functions}
 \[
    H=(H_\pi)_{\pi\in \Delta^\circ(\Theta)},
 \]
 where, for each full-support prior $\pi$, the function $H_\pi \colon \Delta(\Theta)\rightarrow \RRcvx_+$ is convex and lower semicontinuous, with $H_\pi(\pi) = 0$. The cost of an experiment $P$ under prior $\pi$ is then
\begin{equation}\label{eq:PS_cost_CDL}
    C(P,\pi) = \sum_{s \in\supp P_{\pi}}P_{\pi}(s)H_\pi(p_{s}),
\end{equation}
 where $\supp P_\pi\subseteq S$ is the support of the unconditional signal distribution, and $p_{s}\in\Delta(\Theta)$ is the posterior belief induced by signal realization $s \in \supp P_{\pi}$ according to Bayes' rule.  

Given an entropy $H$, the cost \eqref{eq:PS_cost_CDL} can be represented as $f$-information by defining
 \begin{equation}\label{eq:f_pi_entropy}
    f_\pi(x)=\begin{cases}
    H_\pi(x\pi) & \text{if }\sum_{\theta\in \Theta}x(\theta) \pi(\theta)=1,\text{ }\\
    +\infty & \text{otherwise},
    \end{cases}
\end{equation}
 where $x\pi=(x(\theta)\pi(\theta))_{\theta\in\Theta}$ is the componentwise product of $x , \pi \in \mathbb{R}^\Theta$. Then, for every experiment $P$ and reference measure $\alpha$, $D_{f_\pi}(P \Vert \alpha)<+\infty$ implies $\alpha=P_\pi$. Consequently,
 \[
 I_f(P,\pi) = D_{f_\pi}(P \Vert P_\pi) = \sum_{s \in\supp P_{\pi}}P_{\pi}(s)H_\pi(p_{s}).
 \]
Moreover, $P_\pi$ is an $f$-mean of $P$ under $\pi$ (and it is the unique one if $I_{f}(P,\pi)<+\infty$). 

 As is well known, mutual information can be represented as a posterior-separable cost by setting $H_\pi(p) = D_{\mathrm{KL}}(p\Vert\pi)$, the KL divergence between posterior $p$ and prior $\pi$. Comparing \eqref{eq:f_pi_mutual} and \eqref{eq:f_pi_entropy} shows that mutual information has multiple representations as $f$-information.
\end{example}

 \begin{example}[Csiszár information]\label{exa:csiszar}
 Motivated by applications in statistics, \cite{csiszar1972class} introduced the following generalization of mutual information, which retains the additive separability between states but drops the specific logarithmic form \eqref{eq:f_pi_mutual}. Let $f$ be given by
 \begin{equation}
    f_\pi(x) = \sum_{\theta\in \Theta} \pi(\theta)\phi(x(\theta)), \label{eqn:separable_f}
 \end{equation}
 where $\phi \colon \mathbb{R}_+\rightarrow\overline{\mathbb{R}}_+$ is a function that is convex, lower semicontinuous, and normalized so that $\phi(1)=0$. In this case, $f$-information can be expressed as
 \begin{equation}
     I_{f}(P,\pi)=\min_{\alpha\in\Delta(S)}\sum_{\theta\in \Theta} \pi(\theta)D_{\phi}(P_{\theta}\Vert \alpha),\label{eq:csiszar_72}
 \end{equation}
 where $D_\phi$ is the conventional $\phi$-divergence on pairs of distributions. We refer to \eqref{eq:csiszar_72} as \textit{Csiszár information}. Note that Csiszár information reduces to mutual information 
 when $\phi(t) = t \log t - t+1$. More generally, in contrast to mutual information and other posterior-separable costs, under Csiszár information the $f$-mean of $P$ need not equal $P_\pi$.
\end{example}

\section{Optimality conditions}\label{sec:optimality}

 We now characterize solutions to information acquisition problems under $f$-information costs. Since $f$-information is Blackwell monotone (Lemma \ref{lem:f_info_properties}), standard arguments imply that it is without loss of generality to restrict attention to experiments whose signal space coincides with the action set, together with the identity action strategy. Any such experiment $P = (A,(P_\theta)_{\theta\in\Theta})$ describes a \emph{state-dependent stochastic choice rule}, where $P_\theta \in \Delta(A)$ represents the distribution of actions taken in state $\theta$ (\citealp{caplin2015testable}; \citealp{caplin2015revealed}). Thus, the information acquisition problem 
 reduces to 
 \begin{equation}\label{eq:ri_prob_f_infor}
    \max_{P \in \Delta(A)^\Theta} \sum_{\theta \in \Theta} \pi(\theta) \sum_{a \in A} P_\theta(a)a(\theta) - I_f(P,\pi).
\end{equation}
 We characterize the solutions to this problem, i.e.,
 the optimal stochastic choice rules. 

\subsection{Review: the case of mutual information}\label{subsec:mutual_info}

 We first review the familiar case in which the cost is given by mutual information:
 \begin{equation}\label{eq:mutual_info_problem}
    \max_{P \in \Delta(A)^\Theta} \sum_{\theta \in \Theta} \pi(\theta) \sum_{a \in A} P_\theta(a)a(\theta) - \kappa I_{\mathrm{S}}(P,\pi),   
 \end{equation}
 where the constant $\kappa>0$ is a parameter that scales the cost.

As shown by \cite{matvejka2015rational} and \cite*{caplin2019rational}, stochastic choice under mutual information takes a tractable form that resembles multinomial logit. A key feature of their characterization is that \eqref{eq:mutual_info_problem} reduces to the simpler problem

 \begin{equation}\label{eq:redux_mutual_info_problem}
    \max_{\alpha\in \Delta(A)}\kappa\sum_{\theta\in \Theta}\pi(\theta)\log \left(\sum_{a\in A}e^{\frac{a(\theta)}{\kappa}}\alpha(a)\right).
 \end{equation}
 This auxiliary problem involves maximization with respect to probability measures over the action set $A$, which are lower-dimensional objects than state-dependent choice rules.
 
 \begin{samepage}
 \begin{theorem}[\citealp{matvejka2015rational}; \citealp*{caplin2019rational}]\label{thm:ri_shannon}
 Information acquisition under mutual information has the following properties:
 \begin{enumerate}
 
    \item A stochastic choice rule $P=(A,(P_\theta)_{\theta\in \Theta})$ is a solution to (\ref{eq:mutual_info_problem}) if and only if there exists a solution $\alpha \in \Delta(A)$ of (\ref{eq:redux_mutual_info_problem}) such that, for all $\theta\in \Theta$ and $a\in A$,
     \begin{equation}\label{eq:logit}
        P_\theta(a) = \frac{\alpha(a) e^{\frac{a(\theta)}{\kappa}}}{\sum_{b\in A}\alpha(b) e^{\frac{b(\theta)}{\kappa}}}.
     \end{equation}
    Moreover, for any such $P$ and $\alpha$, it holds that $\alpha=P_\pi$. 
    
    \item The optimization problems (\ref{eq:mutual_info_problem}) and (\ref{eq:redux_mutual_info_problem}) have the same value.
 \end{enumerate}
 \end{theorem}
 \end{samepage}

 This result yields a two-step recipe for solving information acquisition problems with mutual information costs. First, find the probability measures over actions $\alpha$ that solve the auxiliary problem \eqref{eq:redux_mutual_info_problem}. Second, substitute each solution into \eqref{eq:logit} to obtain an optimal stochastic choice rule $P$. This formula is a state-dependent multinomial logit rule, modified by the baseline action distribution $\alpha$.
 Moreover, $\alpha = P_\pi$, the unconditional distribution of actions under $P$. Thus, in each state $\theta$, 
 relative choice probabilities 
 satisfy
 \begin{equation}\label{eqn:prob-ratio-MI}
 \frac{P_\theta (a)}{P_\theta(b)} = \frac{P_\pi(a)}{P_\pi(b)} e^{\frac{a(\theta)-b(\theta)}{\kappa}}.
\end{equation}
 In other words, the relative odds of choosing $a$ over $b$ equals their unconditional relative odds multiplied by the exponentiated payoff advantage of $a$ over $b$ in state $\theta$.\footnote{See \cite*{steiner2017rational}, \cite{matveenko2021attentional}, \cite{fabbri2024rational}, \cite*{hansen2025robust}, and \cite{pennesi2025rational}, among others, for extensions of Theorem~\ref{thm:ri_shannon}.}

 The tractability of this solution---as exemplified by the low-dimensionality of the auxiliary problem \eqref{eq:redux_mutual_info_problem} and the simplicity of the logit rule \eqref{eq:logit}---is perhaps the main appeal of the mutual information cost and the reason for its ubiquity in applications.\footnote{
 Mutual information also has classic information-theoretic foundations \citep{cover1999elements} and appealing axiomatic properties \citep*{de2014axiomatic,mensch2021rational,caplin2022rationally}.} In what follows, we develop a generalization of this two-step solution method for $f$-information.

\subsection{Duality}\label{subsec:fenchel_con}

 To study behavior under $f$-information, we associate with each function $f_\pi$ a dual object.
 
 \begin{definition}\label{def:conjugate}
    The \emph{Fenchel conjugate} of $f_\pi$ is the function $f^\star_\pi \colon\mathbb{R}^\Theta \to \overline{\RR}$ defined by 
    \[
        f^\star_\pi(x) = \sup_{y\in \mathbb{R}^\Theta_+} \sum_{\theta\in \Theta}x(\theta)y(\theta)-f_\pi(y). 
    \]
 \end{definition}

 Conjugation is a fundamental operation in convex analysis. In economics, it is familiar from the theory of the competitive firm, where the profit function is conjugate to the cost function. We denote by 
$
f^\star=(f_\pi^\star)_{\pi\in\Delta^\circ(\Theta)}
$ the family of conjugates associated with $f$.

 The next lemma characterizes the families of functions that can arise as conjugates of $f$-information generators. Let $g=(g_\pi)_{\pi\in\Delta^\circ(\Theta)}$ be any family of functions $g_\pi:\mathbb{R}^\Theta\to\overline{\mathbb{R}}$. For each prior $\pi$, let $\partial g_\pi(\mathbf{0})\subseteq\mathbb{R}^\Theta$ denote the subdifferential of $g_\pi$ at the zero vector $\mathbf{0}=(0,\ldots,0)$.

\begin{lemma}\label{lem:g.dual.of.f}
The following are equivalent:
\begin{lemmaenum}
\item There exists an \(f\)-information generator \(f\) such that $g=f^\star$.
\item For every prior \(\pi\), the function $g_\pi$ is convex, lower semicontinuous, and increasing, with \(g_\pi(\mathbf{0})=0\) and \(\mathbf{1}\in \partial g_\pi(\mathbf{0})\).
\end{lemmaenum}
\end{lemma}

 Given any family of functions $g$ that satisfies the conditions in Lemma \ref{lem:g.dual.of.f}(ii), the underlying generator $f$ can be recovered via conjugation.\footnote{The convexity and lower semicontinuity of $g_\pi$ are dual to the convexity and lower semicontinuity of $f_\pi$, the monotonicity of $g_\pi$ is dual to $f_\pi$ being defined on the non-negative orthant $\mathbb{R}^\Theta_+$, and the fact that $g_\pi$ satisfies $g_\pi(\boldsymbol{0}) = 0$ and $\boldsymbol{1}\in \partial g_\pi (\boldsymbol{0})$ is dual to $f_\pi$ being non-negative and satisfying $f_\pi(\boldsymbol{1}) = 0$. All these conjugacy relationships are standard \citep{rockafellar1970convex}. We thus omit the proof of Lemma \ref{lem:g.dual.of.f}.} Specifically, for each prior $\pi$, 
 \[
    f_\pi(x) = g_\pi^\star(x) = \sup_{y\in \mathbb{R}^\Theta} \sum_{\theta\in \Theta}x(\theta)y(\theta)-g_\pi(y).
 \]
Hence, the functions $f_\pi$ and $f^\star_\pi$ are linked by the duality relation $f_\pi = (f^\star_\pi)^\star$. This duality offers two equivalent perspectives on information acquisition problems with $f$-information costs, depending on whether one takes $f$ or $f^\star$ as the primitive object. 

For the remainder of the paper, we restrict attention to $f$-information generators satisfying the following regularity assumptions, which ensure that the associated conjugates are technically well-behaved. For each prior $\pi$, let
$
    \dom f_\pi
    =
    \{x\in\mathbb{R}^\Theta_+ : f_\pi(x)<+\infty\}
$
denote the effective domain of $f_\pi$, and $\ri(\dom f_\pi)$ its relative interior.
 
 \begin{assumption}\label{ass:co-finitee_etc}
     For every prior $\pi$, the function $f_\pi$ satisfies the following properties:
     \begin{enumerate}
         \item \emph{Superlinearity at infinity:} for every $y\in\dom f_\pi$ and all non-zero $x\in \mathbb{R}^\Theta_+$, 
         \[
         \lim_{t\rightarrow+\infty}\frac{f_\pi(y+tx)}{t}=+\infty.
         \]
         \item \emph{Essential strict convexity:} $f_\pi$ is strictly convex on every convex subset of 
         \[\left\{x\in\mathbb{R}_{+}^{\Theta}:\partial f_\pi(x)\neq\varnothing\right\}.
         \]
         \item  \emph{Nondegeneracy of domain:} $\boldsymbol{1}\in \ri(\dom f_\pi)$.
     \end{enumerate}
 \end{assumption}
  
 Under the first two assumptions, the conjugate $f^{\star}_\pi$ is everywhere finite (i.e., $\dom f^{\star}_\pi=\mathbb{R}^\Theta$) and  differentiable. Moreover, since $f^\star_\pi$ is convex and differentiable, its gradient 
 \[
 \nabla f^\star_\pi = \left(\nabla_\theta f_\pi^\star\right)_{\theta\in \Theta} \colon \RR^\Theta \to \RR^\Theta_+, \quad \text{where} \quad \nabla_\theta f_\pi^\star(x) = \frac{\partial f_\pi^\star}{\partial x(\theta)}(x),
 \]
 is continuous and monotone (i.e.\ $\langle \nabla f^\star_\pi(x) - \nabla f^\star_\pi(y), x - y\rangle \geq 0$ for all $x,y \in \RR^\Theta$). Conversely, if $f^{\star}_\pi$ is everywhere finite and differentiable, then $f_\pi$ is superlinear at infinity and essentially strictly convex.\footnote{See \citet*[Corollary 13.3.1 and Theorem 26.3]{rockafellar1970convex}.} The final assumption,  $\boldsymbol{1}\in \ri(\dom f_\pi)$, will serve as a constraint qualification in our main theorem.
 

\subsection{Characterization theorem}\label{subsect:charac_thm}

 We now characterize  optimal behavior under $f$-information. As in the analysis of mutual information,  
 our central observation is that the optimization problem of interest \eqref{eq:ri_prob_f_infor} can be reduced to a lower-dimensional auxiliary problem. For general $f$-information costs, this auxiliary problem now takes the form of a maxmin  problem:  
 \begin{equation}\label{eq:maxmin_problem}
    \max_{\alpha\in \Delta(A)}\min_{\lambda\in \mathbb{R}^\Theta}\sum_{a\in A}\alpha(a) f_\pi^\star(a\pi -\lambda)+\sum_{\theta\in\Theta}\lambda(\theta),
 \end{equation}
 where $a\pi=(a(\theta)\pi(\theta))_{\theta\in\Theta}$ is the componentwise product of $a,\pi \in \RR^\Theta$. Note that the objective function in this maxmin problem depends on $f_\pi$ only through its conjugate $f^\star_\pi$.
 
\begin{samepage}
\begin{theorem}\label{thm:characterization}
Information acquisition under $f$-information has the following properties:
\begin{enumerate}
 \item A stochastic choice rule $P=(A,(P_\theta)_{\theta\in \Theta})$ is a solution to (\ref{eq:ri_prob_f_infor}) if and only if there exists a saddle point $(\alpha,\lambda)$ of (\ref{eq:maxmin_problem}) such that, for all $\theta\in \Theta$ and $a\in A$, 
\begin{equation}\label{eqn:P-optimal-main}
    P_{\theta}(a)=\alpha(a) \nabla_\theta f^{\star}_\pi(a\pi-\lambda).
\end{equation}
Moreover, for any such $P$ and $(\alpha, \lambda)$, $\alpha$ is an $f$-mean of $P$ under $\pi$.
\item The optimization problem (\ref{eq:ri_prob_f_infor}) and the maxmin problem (\ref{eq:maxmin_problem}) have the same value.

\end{enumerate}

\end{theorem}
\end{samepage}

Theorem \ref{thm:characterization} provides a two-step recipe for solving information acquisition problems with $f$-information costs, paralleling Theorem \ref{thm:ri_shannon} for the special case of mutual information. 

First, find all pairs $(\alpha,\lambda)$
that are saddle points of the auxiliary problem \eqref{eq:maxmin_problem}.
As shown in the proof, this maxmin problem comes from the Lagrangian of the original problem \eqref{eq:ri_prob_f_infor}. The vector $\lambda$ collects the multipliers on the 
constraints $\sum_{a \in A} P_\theta (a) = 1$ for each $\theta \in \Theta$. Thus, $\lambda(\theta)$ can be interpreted as the shadow value of acting in state $\theta$.

Second, substitute each saddle point $(\alpha,\lambda)$ into \eqref{eqn:P-optimal-main} to obtain an optimal stochastic choice rule $P$. In particular, for any actions $a,b\in \supp P_\pi$, 
\[
\frac{P_\theta(a)}{P_\theta(b)}=\frac{\alpha(a)}{\alpha(b)}\frac{\nabla_\theta f^\star_\pi(a\pi-\lambda)}{\nabla_\theta f^\star_\pi(b\pi-\lambda)}.
\]
As under mutual information in \eqref{eqn:prob-ratio-MI}, relative choice probabilities are the product of two ratios. The first, $\alpha(a)/\alpha(b)$, captures the actions’ baseline weights under the $f$-mean $\alpha$. The second adjusts these weights according to their utility profiles through $\nabla_\theta f^\star_\pi$. Fixing $\lambda$, this adjustment increases with the utility profile of $a$.

The general $f$-information case differs from mutual information in two important respects. First, the $f$-mean $\alpha$ need not equal the unconditional action distribution $P_\pi$. Second, the relative probability of choosing $a$ over $b$ in state $\theta$ may depend on their payoffs in every state. It may also depend indirectly on all available actions, through the multiplier $\lambda$. These features allow $f$-information to generate richer choice patterns than those that can arise under mutual information. 

Because \eqref{eq:maxmin_problem} is affine in $\alpha$ and convex in $\lambda$, its saddle points are characterized by
\begin{align}
 & f_\pi^\star(a\pi -\lambda) = \max_{b\in A} f_\pi^\star(b\pi -\lambda) \quad\quad \forall a\in \supp \alpha,\label{eq:alpha}\\
 & \sum_{a\in A} \alpha(a) \nabla_\theta f_\pi^\star (a\pi -\lambda)  =1 \quad \quad \quad\, \forall \theta\in \Theta.\label{eq:lambda}
 \end{align}
Condition \eqref{eq:alpha} disciplines the support of $\alpha$, and therefore the \emph{consideration set}
\[
\supp P_\pi = \{a \in A : P_\pi(a)>0\}, 
\]
since \eqref{eqn:P-optimal-main} implies $\supp P_\pi \subseteq \supp \alpha$.\footnote{The supports coincide if the cost is posterior-separable (so $\alpha = P_\pi$) or $f^\star_\pi$ is strictly convex (so $\nabla f^\star_\pi\gg\mathbf{0}$), where the latter is dual to $f_\pi$ being \emph{essentially smooth} \citep[Theorem 26.3]{rockafellar1970convex}.} 
Condition \eqref{eq:lambda} ensures that the choice rule in \eqref{eqn:P-optimal-main} satisfies $\sum_{a \in A} P_\theta (a) = 1$ for all $\theta \in \Theta$, and hence that $\lambda$ is a valid Lagrange multiplier.\footnote{When closed-form solutions are not available, the maxmin problem can be tackled using numerical methods such as the Saddle-Point Mirror Prox algorithm \citep[pp. 315--316]{bubeck2015convex}.}

\section{Special cases}\label{sec:special-cases-characterization}

We now illustrate the optimality conditions from Theorem \ref{thm:characterization} in the context of our three leading examples: mutual information, posterior-separable costs, and Csiszár information.
 
 \subsection{Mutual information}\label{sssec:mutual_info}

 Under mutual information, the optimality conditions in Theorem \ref{thm:characterization} reduce to those of \citet{matvejka2015rational} and \citet*{caplin2019rational}, recorded in Theorem \ref{thm:ri_shannon}. This simplification follows from the additive and exponential structure of the conjugate.

 For simplicity, let $\kappa =1$. Recall from Example \ref{exa:shannon} that mutual information is the special case of $f$-information generated by \eqref{eq:f_pi_mutual}. A direct calculation gives the conjugate:
 \[
    f_\pi^\star(x) = \sum_{\theta\in \Theta} \pi(\theta) e^{\frac{x(\theta)}{\pi(\theta)}}-1.
 \]
 For this conjugate, the Lagrange multiplier $\lambda\in \RR^\Theta$ can be computed explicitly; for each candidate $\alpha\in \Delta(A)$, the optimality condition \eqref{eq:lambda} implies
\begin{equation}\label{eq:MI-special-case-lambda}
    \lambda(\theta)=\pi(\theta)\log \sum_{a\in A}\alpha(a)e^{a(\theta)} \quad \quad \forall\, \theta \in \Theta.
\end{equation}
 Thus, in each state $\theta$, the multiplier is an aggregate of the utilities delivered by the available actions, with weights given by $\alpha$. Substituting this expression for $\lambda$ into the maxmin problem \eqref{eq:maxmin_problem} reduces the search for a saddle point to the maximization over $\alpha$ in \eqref{eq:redux_mutual_info_problem}. Moreover, the optimal stochastic choice rule \eqref{eqn:P-optimal-main} reduces to the logit formula \eqref{eq:logit}:
\begin{equation}\label{eq:MI-special-case-logit}
    P_\theta(a) = \alpha(a)\nabla_\theta f^\star_\pi(a\pi - \lambda) = \alpha(a) e^{a(\theta)-\frac{\lambda(\theta)}{\pi(\theta)}} = \frac{\alpha(a) e^{a(\theta)}}{\sum_{b\in A}\alpha(b)e^{b(\theta)}}.
\end{equation}

\subsection{Posterior-separable costs}\label{ssec:posterior-separable}

For posterior-separable costs, the optimality conditions in Theorem \ref{thm:characterization} provide a dual representation of the ``primal'' optimality conditions that have been derived in the literature via concavification and related Lagrangian methods.

Recall from Example \ref{exa:CDL} that, for any family of entropy functions $H = (H_\pi)_{\pi \in \Delta^\circ(\Theta)}$, the posterior-separable cost \eqref{eq:PS_cost_CDL} is the special case of $f$-information generated by \eqref{eq:f_pi_entropy}. In this case, the conjugate of $f_\pi$ can be expressed in terms of the conjugate of $H_\pi$. In particular, the Fenchel conjugate of the entropy $H_\pi$ is the map $H^\star_\pi \colon \RR^\Theta \rightarrow \RRcvx$ given by 
 \[
    H^\star_\pi(x) = \max_{ p \in \Delta(\Theta)} \sum_{\theta\in \Theta} x(\theta)p(\theta) - H_\pi(p),
 \]
and the conjugate of the generator $f_\pi$ can be expressed as 
 \begin{equation}\label{eq:f-star-PS}
    f^\star_\pi(x) = H^\star_\pi\left(x_\pi\right), \qquad\text{where}\qquad
    x_\pi=\left(\frac{x(\theta)}{\pi(\theta)}\right)_{\theta\in \Theta}.
\end{equation}
 Since $H_\pi$ is defined on the simplex, $f^\star_\pi$ is \emph{translation invariant} with respect to the prior: 
 \[
 f^\star_\pi (x + c \pi) = f^\star_\pi(x) + c 
\]
 for every constant $c \in \mathbb{R}$. Conversely, given any $f$-information cost and any prior $\pi$, $f^\star_\pi$ satisfies this translation invariance property only if $f_\pi$ takes the form \eqref{eq:f_pi_entropy} for some entropy $H_\pi$. Thus, translation invariance is the dual counterpart of posterior separability.

In this case, Assumption \ref{ass:co-finitee_etc} holds under mild conditions on the entropy functions $H_\pi$: 

 \begin{assumption}\label{ass:strict_con_H}
For every prior $\pi$, $H_\pi$ is strictly convex on $\dom H_\pi$ and $\pi \in \ri(\dom H_\pi)$.
\end{assumption}

Under Assumption \ref{ass:strict_con_H}, which we impose for the remainder of the paper, Theorem \ref{thm:characterization} implies that an optimal stochastic choice rule $P$ satisfies 
\[
    P_\theta(a)=\frac{P_\pi(a)\nabla_\theta H^\star_\pi(a-\lambda_\pi)}{\pi(\theta)}, \qquad\text{where}\qquad
    \lambda_\pi=\left(\frac{\lambda(\theta)}{\pi(\theta)}\right)_{\theta\in \Theta}
\]
is a prior-adjusted version of the Lagrange multiplier $\lambda$.\footnote{Recall from Example \ref{exa:CDL} that $P_\pi$ is the unique $f$-mean of $P$, which must have finite cost to be optimal.} Equivalently, for each action in the consideration set, $a \in \supp P_\pi$, the associated posterior belief $p_a \in \Delta(\Theta)$ is given by 
\begin{equation}\label{eq:PS-FOC-dual}
    p_a=\nabla H^\star_\pi(a-\lambda_\pi).
\end{equation}
This provides a natural interpretation of the gradient $\nabla H^\star_\pi$: it maps the utility vector of an action $a$, shifted by the prior-adjusted multiplier $\lambda_\pi$, into the posterior belief conditional on choosing $a$. Consequently, assumptions on the conjugate $H^\star_\pi$ translate directly into properties of posterior beliefs, and hence on the decision maker's behavior.

\begin{remark}\label{remark:PS-primal-dual}
The concavification and Lagrangian methods used in the literature instead characterize the optimal action-contingent posteriors via the primal first-order condition 
\begin{equation}\label{eq:PS-FOC-primal}
   a -\lambda_\pi \in \partial H_\pi(p_a),
\end{equation}
where the subdifferential $\partial H_\pi(p_a) \subseteq \RR^\Theta$ of the entropy $H_\pi$ represents the marginal cost of producing posterior $p_a$, and $\lambda_\pi \in \RR^\Theta$ represents the Lagrange multiplier on the Bayes plausibility constraint that posteriors average to the prior, $\sum_{a \in A} P_\pi(a) p_a = \pi$.\footnote{Indeed, upon substituting our dual first-order condition \eqref{eq:PS-FOC-dual} into the optimality condition~\eqref{eq:lambda} for the Lagrange multiplier, the latter reduces precisely to the Bayes plausibility constraint $\sum_{a \in A} P_\pi(a) p_a = \pi$.}
%
However, note that \eqref{eq:PS-FOC-primal} determines the optimal posteriors only implicitly. To solve for them explicitly, one must \emph{invert} the subdifferential $\partial H_\pi$, which then yields our dual condition \eqref{eq:PS-FOC-dual}.  
This inversion operation is tractable precisely when the conjugate $H^\star_\pi$ is tractable. 
\end{remark}

 \subsection{Csisz\'ar information}\label{subsec:Csiszar-optimality}

Under Csisz\'ar information---like mutual information, but unlike general posterior-separable costs---the optimality conditions in Theorem \ref{thm:characterization} have a separable structure: the conjugate is additively separable across states, and the optimal choice rules and Lagrange multipliers can be determined statewise. This makes the analysis particularly tractable. 

Recall from Example \ref{exa:csiszar} that Csisz\'ar information is generated by $f$ with the separable form \eqref{eqn:separable_f}, for some transformation  $\phi\colon \RR_+\rightarrow \RRcvx_+$ that is convex and lower semicontinuous, with $\phi(1)=0$. In this case, the conjugate of each $f_\pi$ can be expressed in terms of conjugate of $\phi$. Specifically, the Fenchel conjugate of $\phi$ is the map $\phi^\star : \RR \to \RRcvx$ defined as 
 \[
  \phi^\star(t) = \sup_{ u \in \mathbb{R}_+} t u -\phi(u),
 \]
and, for every prior $\pi$, the conjugate of the generator $f_\pi$ can be expressed as
\begin{equation}\label{eqn:conjugate-Csiszar}
    f_\pi^\star(x) = \sum_{\theta \in \Theta} \pi(\theta) \phi^\star\left( \frac{x(\theta)}{\pi(\theta)}\right).
 \end{equation}
By Lemma \ref{lem:g.dual.of.f}, the map $\phi^\star$ is convex, lower semicontinuous,  and increasing, with $\phi^\star(0)=0$ and $1 \in \partial \phi^\star(0)$. Conversely,
 any map with these properties is the conjugate of some transformation $\phi$, and any $f$-information cost for which $f^\star$ has the form \eqref{eqn:conjugate-Csiszar} is Csisz\'ar information. Hence, Csisz\'ar information is characterized by additive separability of $f^\star$.
 
In this case, Assumption \ref{ass:co-finitee_etc} holds under mild conditions on the transformation $\phi$:

\begin{assumption}\label{ass:phi_strictly_convex}
The transformation $\phi$ is superlinear at infinity (i.e., $
\phi(t)/t\rightarrow +\infty$ as $t\rightarrow+\infty$), strictly convex on $\inte(\dom \phi)$,  and satisfies $1\in \inte(\dom \phi)$.
\end{assumption}

Under Assumption \ref{ass:phi_strictly_convex}, which we impose for the remainder of the paper, Theorem \ref{thm:characterization} implies that a stochastic choice rule $P$ is optimal if and only if it satisfies
\begin{align}
    P_\theta(a)=\alpha(a)\,(\phi^\star)'\bigl(a(\theta)-\lambda_\pi(\theta)\bigr), \qquad\text{where}\qquad \lambda_\pi(\theta) = \lambda(\theta)/\pi(\theta),
    \label{eq:P_separable_case}
\end{align}
and the $f$-mean $\alpha$ and prior-adjusted multiplier $\lambda_\pi$ are given by some saddle-point of
\begin{equation}\label{eq.lagrangian.phi.information}
    \max_{\alpha\in \Delta(A)}\min_{\lambda_\pi\in \RR^\Theta} \sum_{a\in A}\alpha(a) \sum_{\theta \in \Theta} \pi(\theta)\phi^\star(a(\theta) -\lambda_\pi(\theta)) + \sum_{\theta\in\Theta} \pi(\theta)\lambda_\pi(\theta).
 \end{equation}
Moreover, the first-order condition~\eqref{eq:lambda} for the optimal multiplier decomposes statewise:
\begin{equation}\label{eq:lambda_separable_case-1}
    \sum_{a\in A}\alpha(a)(\phi^\star)'\bigl(a(\theta)-\lambda_\pi(\theta)\bigr)=1 \quad \quad \forall\, \theta \in \Theta.
\end{equation}
For each fixed $\alpha$, conditions \eqref{eq:P_separable_case} and \eqref{eq:lambda_separable_case-1} imply that the conditional probability $P_\theta$ and the multiplier $\lambda_\pi(\theta)$ can be determined state by state, without reference to the feasible payoffs $\{a(\tau) : a \in A\}$ or values of $\lambda_\pi(\tau)$ in the other states $\tau \neq \theta$. In particular, \eqref{eq:P_separable_case} generalizes the logit rule \eqref{eq:MI-special-case-logit} under mutual information, with the increasing map $(\phi^\star)'$ replacing the exponential function, while \eqref{eq:lambda_separable_case-1} generalizes the formula  for $\lambda_\pi(\theta)$ in \eqref{eq:MI-special-case-lambda}.

\begin{table}[!t]
\centering
\small
\renewcommand{\arraystretch}{1.8}
\begin{tabular}{@{}
>{\raggedright\arraybackslash}p{0.18\textwidth}
p{0.34\textwidth}
@{\hspace{2em}}
p{0.41\textwidth}
@{}}
\toprule
Specification
& Transformation
& Conjugate \\
\midrule

Kullback-Leibler
& $\phi_{\mathrm{KL}}(t)=t\log t-t+1$
& $\phi^\star_{\mathrm{KL}}(s)=e^s-1$ \\

Pearson $\chi^2$
& $\displaystyle \phi_{\chi^2}(t)=\frac{1}{2}(t-1)^2$
& $\displaystyle \phi^\star_{\chi^2}(s)=\frac{[1+s]_+^2-1}{2}$ \\

Cressie--Read ($p>1$, $p\neq 2$)
& $\displaystyle
   \phi_p(t)=\frac{t^p-pt+p-1}{p(p-1)}$
& $\displaystyle
   \phi^\star_p(s)=
   \frac{[1+(p-1)s]_+^{p/(p-1)}-1}{p}$ \\

Jeffreys
& $\phi_{\mathrm J}(t)=(t-1)\log t$
& $\displaystyle
   \phi^\star_{\mathrm J}(s)
   =\frac{1}{J(s)}-1-\log J(s)$ \\

Double exponential
& $\displaystyle
   \phi_{\mathrm{D}}(t)
   =t\log W(et)-\frac{t}{W(et)}+1$
& $\displaystyle
   \phi^\star_{\mathrm{D}}(s)
   =\exp\!\left(e^s-1\right)-1$ \\

KL-regularized Cressie--Read
& $\displaystyle
   \phi_{p,\varepsilon}
   =\phi_p~\square_{\varepsilon}~\phi_{\mathrm{KL}}$
& $\displaystyle
   \phi^\star_{p,\varepsilon}
   =(1-\varepsilon)\phi^\star_p
   +\varepsilon\phi^\star_{\mathrm{KL}}$ \\

\bottomrule
\end{tabular}
\caption{Examples of transformations and their Fenchel conjugates.}
\label{tab:phi-conjugates}
\vspace{0.4em}
\begin{minipage}{0.98\textwidth}
\footnotesize
\emph{Notes.}
Here $[z]_+=\max\{z,0\}$, $W$ denotes the principal branch of the
Lambert $W$ function, and $J(s)=W(e^{1-s})$.
The formula for $\phi_{\mathrm{D}}$ is interpreted by continuity at
$t=0$, where its value is $1-e^{-1}$.
For the final row, $p>1$, $0<\varepsilon<1$, and
$\square_{\varepsilon}$ denotes the weighted infimal convolution
\[
    (f~\square_{\varepsilon}~g)(t)
    =
    \inf_{\substack{x,y\geq 0\\(1-\varepsilon)x+\varepsilon y=t}}
    \left\{
        (1-\varepsilon)f(x)+\varepsilon g(y)
    \right\}.
\]
\end{minipage}
\end{table}

 \paragraph{Parametric Examples.} Table~\ref{tab:phi-conjugates} collects several examples of transformations $\phi$ and their Fenchel conjugates. Each transformation satisfies Assumption~\ref{ass:phi_strictly_convex}. The first two are familiar benchmarks: the Kullback-Leibler transformation has an exponential conjugate, while the Pearson $\chi^2$ transformation has a piecewise quadratic conjugate. Historically, the Cressie--Read, or power-divergence, family was introduced to provide a unified treatment of several goodness-of-fit statistics. It includes Pearson $\chi^2$ as the case $p=2$ and converges to KL divergence as $p\downarrow1$, while retaining a simple power form for its conjugates.

 The remaining examples illustrate the more general point---relevant beyond Csisz\'ar information---that tractability can look quite different depending on whether one works with the generator $f$ or its conjugate $f^\star$. The Jeffreys transformation $\phi_{\mathrm J}$ is elementary, but its conjugate involves the Lambert function. The double-exponential transformation $\phi_{\mathrm{D}}$ illustrates the converse: its conjugate has a simple form, while the transformation itself is more involved. Finally, the KL-regularized Cressie--Read transformation $\phi_{p,\varepsilon}$ shows how different divergences can be combined while preserving tractability on the dual side. The operation of weighted infimal convolution of the two transformations $\phi_{\mathrm{KL}}$ and $\phi_p$ corresponds to the convex combination of their conjugates. The Kullback-Leibler component makes $\phi_{p,\varepsilon}^\star$ strictly increasing on $\RR$, a property not satisfied by $\phi_p$ for $p>1$.

\subsection{Csiszár information vs. posterior-separable costs}

 The optimality conditions for Csiszár information and posterior-separable costs generally have different implications. In fact, under standard smoothness assumptions, the behavioral predictions of these two classes coincide only in the case of mutual information.

\begin{definition}\label{def:csiszar_ess_smooth}
The function $\phi$ is \emph{essentially smooth} if it is finite and differentiable on $(0,+\infty)$ and satisfies the Inada condition $
\lim_{t\to 0}\phi'(t)=-\infty$.
\end{definition}
The Inada condition means that the marginal cost of acquiring conclusive evidence is infinite. By standard duality arguments, $\phi$ is essentially smooth if and only if $\phi^\star$ is strictly convex, so that $(\phi^\star)'$ is strictly increasing, and the range of $(\phi^\star)'$ is $(0,+\infty)$.\footnote{See \citet[Chapter V]{rockafellar1970convex}. These properties of $\phi^\star$ simplify the optimality conditions in Section \ref{subsec:Csiszar-optimality}. For example, they ensure uniqueness of the Lagrange multiplier: if $(\alpha_1,\lambda_1)$ and $(\alpha_2,\lambda_2)$ are two saddle points of \eqref{eq.lagrangian.phi.information}, then $\lambda_1=\lambda_2$. Indeed, condition \eqref{eq:lambda_separable_case-1} then has a unique solution in $\lambda$ for each fixed $\alpha$, which implies that the multiplier is unique because the saddle points of \eqref{eq.lagrangian.phi.information} form a product set.\label{rem:unique_multi}}

\begin{samepage}
\begin{proposition}\label{prop:phi-PS-disjoint}
Suppose that $|\Theta|\geq 3$. For any Csiszár information cost $C$ such that $\phi$ is essentially smooth and is thrice continuously differentiable with $\phi''>0$ on $(0,+\infty)$, the following statements are equivalent:
\begin{enumerate}
\item 
For every decision problem $\mathcal{D} = (\pi,A)$, there exists some posterior-separable cost that generates the same set of optimal stochastic choice rules as $C$. 
\item $C$ is proportional to mutual information: there exists $\kappa>0$ such that $C=\kappa I_S$.
\end{enumerate}
\end{proposition}
\end{samepage}

Thus, under these smoothness assumptions, mutual information is the unique point of intersection between the classes of Csiszár information and posterior-separable costs. This implies that $f$-information is a strict generalization of the posterior-separable framework.

The proof of Proposition \ref{prop:phi-PS-disjoint} focuses on a specific type of behavioral prediction: the size of the consideration set, $|\supp P_\pi|$. We show by construction that, under every (smooth) Csiszár information cost aside from mutual information, there exist decision problems for which the optimal choice rule $P$ is unique and satisfies $|\supp P_\pi|>|\Theta|$. 
In contrast, it is well known that, under every posterior-separable cost, all decision problems 
admit optimal choice rules that satisfy $|\supp P_\pi| \leq |\Theta|$ (see, e.g., \citealp[Proposition 4]{denti2022posterior}). 

More generally, as 
we illustrate 
in 
Sections \ref{sec:separable_APU}--\ref{sec:nested-entropies} below, 
the Csiszár information and posterior-separable models 
have different 
strengths. Csiszár information, with its state-separable structure, preserves much of the tractability of the mutual information benchmark, while also allowing for  richer behavioral patterns due to the flexibility to specify the function $\phi$. 
At the same time, the separability and symmetry of Csiszár information limit its applicability in settings where structural features of the state space (e.g.\ perceptual distance) are important. 
For these applications, the posterior-separable framework, which allows for entropy functions that are non-separable and asymmetric across states, is a more versatile modeling tool. We identify a new family of posterior-separable costs that is suited to these applications, 
but also remains tractable.

\begin{remark}
    Although consideration sets under $f$-information can be larger than under posterior-separable costs, we show in Online Appendix \ref{section:OA:consideration-sets} that they remain highly structured: all $f$-information costs admit optimal choice rules with $|\supp P_\pi| \leq |\Theta| +1$. This weaker bound gives a general testable restriction of the $f$-information framework.
\end{remark}




\section{Csisz\'ar information: discrete choice and closed-form solutions}\label{sec:separable_APU}

In this section and the next, we study 
the behavioral predictions 
of Csisz\'ar information. 
We focus here on two 
issues 
of 
particular 
interest for applications of rational inattention. 


First, a central insight of \citet*{matvejka2015rational} is that optimal information acquisition can provide new foundations for classic models of discrete choice. 
Their paper relates optimal 
behavior under mutual information to Luce's \emph{multinomial logit} model. 
We identify a more general connection between Csisz\'ar information and the \emph{additive perturbed utility} model \citep*{fudenberg2015stochastic}
%

Second, a known practical limitation of mutual information is that, while the logit rule \eqref{eq:logit} is tractable, the optimal $f$-mean can only be found in closed form for very special decision problems. 
We show this feature is not intrinsic to Csisz\'ar information: for certain $\phi$, the full optimal choice rule can be found in closed form in general settings.

\subsection{A foundation for additive perturbed utility}

 In the \emph{additive perturbed utility} (APU) model, given any decision problem $\D=(\pi,A)$, the agent's behavior in each state $\theta$ is described by a distribution $P_\theta \in \Delta(A)$ such that
 \begin{equation}
    P_\theta \in \argmax_{p \in \Delta(A)} \,  \sum_{a \in A} 
    \left[p(a) a(\theta) - c(p(a)) \right], \label{eqn:APU}
 \end{equation}
 where $c \colon [0,1] \to \RR\cup\{+\infty\}$ is a convex \emph{perturbation function} that incentivizes randomization. We assume that $c$ is lower semicontinuous, strictly convex on $\inte(\dom c)$, and such that $1/n\in \inte(\dom c)$, where $n=|A|$ is the number of actions. This setup is the state-dependent version of the APU model in \citet*{fudenberg2015stochastic}.

 The APU model, which has found applications in discrete choice and game theory, 
 reduces to the multinomial logit model when the perturbation takes the entropic form
 \begin{equation}\label{eq:entropic_pert}
 c(t)=\kappa \, t \log t \qquad \forall t\in [0,1],
 \end{equation}
 where $\kappa>0$ is a constant.

 To connect Csisz\'ar information and APU, we first consider decision problems in which actions are \emph{a priori} homogeneous (as in \citealp[Section II.B]{matvejka2015rational}). To formalize this symmetry condition, we fix an enumeration $A=\{a_1,\ldots,a_n\}$ of the action set and let the state space $\Theta$ be a finite subset of $\RR^n$, where the $i$th coordinate of the state represents the utility of the $i$th action: $a_i(\theta)=\theta_i$ for all $\theta\in\Theta$ and $i \in \{1,\dots,n\}$. 

 \begin{definition}\label{def:homo_actions}
 In a decision problem $\mathcal{D} = (\pi,A)$ of this form, actions are \emph{a priori homogeneous} if, for every state $\theta=(\theta_1,\ldots,\theta_n)$ and permutation $\gamma \colon \{1,\ldots,n\}\rightarrow\{1,\ldots,n\}$, 
 \[
    \theta_\gamma=\left(\theta_{\gamma(1)},\ldots,\theta_{\gamma(n)}\right)\in \Theta\quad\text{and}\quad\pi(\theta)=\pi(\theta_\gamma).
 \]
 \end{definition}

 That is, actions are \emph{a priori} homogeneous if their utility vectors are exchangeable under the prior. A simple example is a ``guess the state'' problem with $n$ equally likely states and $n$ actions, where action $i$ yields payoff $1$ in state $i$ and payoff $0$ otherwise.\footnote{In this case, each state is a permutation of the utility vector $(1,0,\ldots,0)\in\RR^n$. Other examples include any setting where the prior is uniform over all the permutations of some utility vector $x\in\RR^n$.}    
The assumption of \emph{a priori} homogeneous actions is common in applications of rational inattention \citep[e.g.,][]{matejka2012simple,bucher2021inattention,vaeth2026attention}. 

 In these symmetric environments, Csisz\'ar information gives rise to optimal choice rules that are \emph{ex-ante symmetric}, in the sense that their associated $f$-means are uniform: 
 
 \begin{lemma}\label{lem:symmetric_sol}
 Under Csisz\'ar information, if actions are a priori homogeneous, there exists an optimal choice rule $P$ whose $f$-mean $\alpha$ is uniform and coincides with the unconditional action distribution $P_\pi$. 
 \end{lemma}


We then obtain an equivalence between behavior under Csiszár information and APU:  

 \begin{proposition}\label{cor:APU-exchangeable}
   For any decision problem 
   with $n = |A|$ a priori homogeneous actions, a stochastic choice rule $P$ is ex-ante symmetric and optimal under Csisz\'ar information with transformation $\phi$ if and only if $P$ is optimal under APU with the perturbation function 
    \begin{equation}\label{eqn:APU-exchangeable}
    c(t) = \frac{1}{n}\phi\left( n t \right) \qquad \forall t\in [0,1].
    \end{equation}
    Moreover, for any perturbation function $c$ that satisfies the normalization $c(1/n)=0$ and $0\in \partial c(1/n)$,\footnote{This normalization is without loss of generality, as one can always add to $c$ an affine function that leaves the argmax set of the APU problem \eqref{eqn:APU} unchanged while ensuring that the normalization holds.} there exists a Csisz\'ar information transformation $\phi$ 
    that satisfies \eqref{eqn:APU-exchangeable}. 
\end{proposition}

When $c$ is given by \eqref{eq:entropic_pert} (up to normalization), Proposition \ref{cor:APU-exchangeable} 
yields 
the equivalence between mutual information and 
logit 
from 
\citet[Proposition 1]{matvejka2015rational}. 



\subsection{The role of $\alpha$ as salience}

For general decision problems, which need not be symmetric, optimal behavior under Csisz\'ar information corresponds to a more general \emph{salience-adjusted APU} model. Formally, consider a stochastic choice rule $P$ that is optimal under Csisz\'ar information, and let $\alpha$ be the associated $f$-mean. Then, in each state $\theta$, the distribution $P_\theta \in \Delta(A)$ satisfies\footnote{This follows from rewriting the information acquisition problem \eqref{eq:ri_prob_f_infor} as the nested optimization
\[
\max_{\alpha \in \Delta(A)} \max_{P \in \Delta(A)^\Theta} \sum_{\theta \in \Theta} \pi(\theta) \sum_{a \in A} P_\theta(a)a(\theta) - \sum_{\theta \in \Theta} \pi(\theta) D_{\phi}(P_\theta \Vert \alpha),
\]
where we write Csisz\'ar information as in \eqref{eq:csiszar_72}, and considering the inner maximization over $P$ given $\alpha$.
}
  \begin{align}
    P_\theta \in \argmax_{p \in \Delta(A)} \, \sum_{a\in A}  \left[p(a)a(\theta)  
    -  \alpha(a) \phi\left( \frac{p(a)}{\alpha(a)}\right)\right].\label{eq:A3PU}
 \end{align}
 This describes an extension of the APU model \eqref{eqn:APU} in which the perturbation function may be action- and menu-dependent, through the given probability measure $\alpha \in \Delta(A)$. 
 As we explain, 
 the term $\alpha(a) \in [0,1]$ 
 can be interpreted as the \emph{salience} of action $a \in A$. 

 To formalize this interpretation, we introduce an ordering over actions:
 
 \begin{definition}
 Given a decision problem $\mathcal{D} = (\pi, A)$ and a stochastic choice rule $P \in \Delta(A)^\Theta$, action $a\in A$ is \textit{more salient than} action $b\in A$ if, for every state $\theta \in \Theta$,
     \[
        a(\theta)=b(\theta) \ \ \implies \ \ P_\theta(a) \geq P_\theta(b).
     \]
 \end{definition}
That is, action $a$ is more salient than action $b$ if, in every state where the two actions yield the same payoff, $a$ is chosen at least as often as $b$. In this sense, the salience ordering captures the idea that one action is more likely to be chosen than another, all else equal.

For this ordering to be nontrivial, the decision problem must exhibit sufficient richness. We say that actions $a,b\in A$ are \emph{comparable} if there is a state $\theta \in \Theta$ in which $a(\theta) = b(\theta)$.  

In such environments, the salience ordering characterizes the ordinal properties of $\alpha$: 

 \begin{proposition}\label{prop.ordering.alpha.lambda}
     Let $P$ be a choice rule that is optimal under Csisz\'ar information, with $\phi$ essentially smooth. 
     For any 
     comparable actions $a,b \in A$, the following are equivalent:
     \begin{lemmaenum}
         \item Action $a$ is more salient than action $b$.
         \item $\alpha(a) \geq \alpha(b)$, where $\alpha$ is the $f$-mean of $P$. 
     \end{lemmaenum}
\end{proposition}




 The discrete choice literature has considered versions of the salience-adjusted APU model in which $\alpha$ is treated as an exogenous parameter. \citet*{mattsson2002probabilistic} and \citet*{cerreia2023multinomial} study the case where $\phi$ takes the entropic form that yields the salience-adjusted logit rule \eqref{eq:logit},
 interpreting $\alpha$ as the agent's default choice rule or initial bias. \citet*{chambers2025weighted} study the case where $\phi$ is quadratic (see Section \ref{sec:chi_squared} below), interpreting $\alpha(a)$ as the inherent salience of action $a$. These approaches are suited to modeling a decision maker's involuntary and automatic (``bottom up'') allocation of attention.

In our framework, by contrast, $\alpha $ is determined endogenously via the agent's optimal and deliberate (``top down'') allocation of attention. The optimality conditions for $\alpha$ impose extra discipline on the salience weights. For example, they imply that all strictly dominated actions $a \in A$ have zero salience: $\alpha(a) = 0$. In the next section, we illustrate additional features of the optimal $\alpha$, for a special case in which it can be found explicitly.


\begin{remark}
Luce's axiom of \emph{independence of irrelevant alternatives} (IIA) is central to the theory of random choice, providing a behavioral foundation for 
multinomial logit. 
\citet{matvejka2015rational} show that, when $\alpha$ is 
an exogenous parameter, two relaxations of IIA jointly characterize the salience-adjusted logit rule \eqref{eq:logit}. In Online Appendix \ref{sec:IIA}, we provide a complementary characterization: when $\alpha$ is endogenous, 
one of their relaxed 
IIA axioms singles out mutual information within the class of Csiszár information costs.
\end{remark}

\subsection{Closed-form solutions under $\chi^2$-information}\label{sec:chi_squared}

 Under mutual information, the optimal $f$-mean $\alpha$ is characterized as the solution to the auxiliary maximization problem \eqref{eq:redux_mutual_info_problem}. While this characterization is often useful, it rarely yields a closed-form formula for $\alpha$. 
 The main exceptions are symmetric settings where $\alpha$ is uniform \citep{matvejka2015rational}, 
 and asymmetric settings where the prior $\pi$ belongs to very special families of distributions \citep{brown2024endogenous, pellegrino2023devil}. 

This limitation is not intrinsic to Csiszár information: for certain 
transformations 
 $\phi$, the saddle point $(\alpha,\lambda)$ can be found in closed form for general decision problems.

 To 
 illustrate this point, we consider a quadratic specification. Let 
 $\phi$ be defined as 
 \begin{equation}\label{eq.Chisquare}
    \phi(t)=\frac{\kappa}{2}(t-1)^2, 
 \end{equation}
 where $\kappa>0$ is a constant. The corresponding $\phi$-divergence is 
 known as the \emph{$\chi^2$-divergence}, and we accordingly refer to the associated Csiszár information cost as $\chi^2$\textit{-information}. The $\chi^2$-divergence is one of the classical discrepancy measures between probability distributions, originating in Pearson's goodness-of-fit test. 
 This quadratic specification can also be viewed as a local approximation to a general Csiszár information cost.\footnote{If $\phi$ is twice differentiable, then its second-order expansion around $1$ is $\phi(t) = \frac{\phi''(1)}{2}(t-1)^2 + o((t-1)^2)$. Therefore, the local quadratic approximation of $\phi$ is given by \eqref{eq.Chisquare} with the constant $\kappa=\phi''(1)$.}


 Towards characterizing optimal behavior, note that the conjugate $\phi^\star$ of \eqref{eq.Chisquare} is 
 \[
    \phi^\star(t) =
    \begin{cases}
    t+\dfrac{t^2}{2\kappa} & \text{if } t > -\kappa,\\[8pt]
     -\dfrac{\kappa}{2} & \text{if } t \leq -\kappa.
    \end{cases}
 \]
 For simplicity, we focus on decision problems $\mathcal{D}=(\pi,A)$ such that the constant $\kappa$ satisfies 
\begin{equation}\label{eq.kappa.ge.R}
    \kappa > \max_{\theta \in \Theta,a \in A} a(\theta) - \min_{\theta \in \Theta,a \in A} a(\theta).
 \end{equation}
Intuitively, this ensures that information is sufficiently costly that, at the optimum, no state is conclusively ruled out. Formally, \eqref{eq.kappa.ge.R} implies that the relevant arguments of $\phi^\star$ lie in the region $(-\kappa, +\infty)$ over which $(\phi^\star)'>0$, so the optimal choice rule satisfies $\supp P_\theta = \supp \alpha$ for all $\theta \in \Theta$ (Lemma~\ref{prop.kappa.ge.R} in the appendix). 
Under this mild assumption, we obtain: 

 \begin{proposition}\label{prop.saddlepoint.quadratic}
For any $\chi^2$-information cost and decision problem $\mathcal{D} = (\pi,A)$ such that \eqref{eq.kappa.ge.R} holds, a stochastic choice rule $P = (A,(P_\theta)_{\theta \in \Theta})$ is optimal if and only if it satisfies 
\begin{equation}\label{eq:WL-P}
P_\theta(a) = \alpha(a) + \alpha(a) \left(\frac{a(\theta) - \sum_{b\in A}\alpha(b)b(\theta)}{\kappa}\right) \qquad \forall a \in A,  \theta \in \Theta
\end{equation}
for some probability measure $\alpha \in \Delta(A)$ that solves the auxiliary quadratic program:
\begin{equation}\label{eq.quadratic.program.distance}
    \max_{\alpha\in\Delta(A)} \sum_{a \in A} \alpha(a)m_a + \frac{1}{4\kappa}\sum_{a,b\in A}\alpha(a)\alpha(b)D_{ab},
 \end{equation}
 where the vector $m \in\RR^A$ and the matrix $D\in\RR^{A\times A}$ are defined componentwise as
 \[
 m_a=\sum_{\theta\in\Theta}\pi(\theta)a(\theta) \qquad \text{ and } \qquad D_{ab}=\sum_{\theta\in\Theta}\pi(\theta){(a(\theta)-b(\theta))}^2. 
 \]
 \end{proposition}

Proposition~\ref{prop.saddlepoint.quadratic} 
simplifies 
the search for optimal choice rules to the 
task of solving the quadratic program \eqref{eq.quadratic.program.distance}. 
Given the optimal $\alpha$, the optimal choice rule $P$ is determined in closed-form by the formula \eqref{eq:WL-P}, which is the state-dependent version of the \emph{weighted linear} discrete choice model \citep*{chambers2025weighted}. 
 As shown in the proof, the quadratic program \eqref{eq.quadratic.program.distance} characterizes the saddle-points 
of the maxmin problem 
\eqref{eq.lagrangian.phi.information}:
the optimal 
$\alpha$ is the $f$-mean of $P$, and it pins down the 
multiplier $\lambda_\pi = \sum_{a \in A} \alpha(a) a$ as the average payoff profile under action distribution $\alpha$. 


The objective of the quadratic program \eqref{eq.quadratic.program.distance} comprises two terms. The first term is the prior expected payoff under action distribution $\alpha$. The second term rewards action distributions $\alpha$ that induce payoff dispersion: 
$D_{ab}$ is defined as 
the squared $L^2(\pi)$ distance between the payoff profiles of actions $a$ and $b$, 
so 
 $
  \sum_{a,b\in A}\alpha(a)\alpha(b)D_{ab}
 $
 is the average squared distance between payoff profiles. As $\kappa$ increases and information becomes costlier, the coefficient $1/(4\kappa)$ decreases and the dispersion term becomes less important.


When the optimal $\alpha$ has full support, it is determined by a linear first-order condition:
\[
    m + \frac{1}{2\kappa}D\alpha=\gamma\mathbf 1,
\]
 where $\gamma \in \RR$ is the multiplier on the constraint $\sum_{a\in A}\alpha(a)=1$. The same construction also applies in the general, partial-support case: one specifies a candidate support $\hat{A}\subseteq A$, replaces $D$ and $m$ in the above with their restrictions to $\hat{A}$, solves the resulting linear system, and checks the usual complementary-slackness inequalities for actions outside $\hat{A}$. The linearity of these conditions implies that $\alpha$ can be determined in closed form.

For example, take any decision problem with \emph{uncorrelated} and \emph{non-constant} actions: 
\[
\sum_{\theta \in \Theta} \pi(\theta) a(\theta) b(\theta) = m_a m_b \qquad \text{and} \qquad \sigma_a^2 =\sum_{\theta\in\Theta}\pi(\theta){(a(\theta)-m_a)}^2>0 \qquad \forall a,b\in A, a \neq b,
\]
where $\sigma_a^2$ is the prior variance of the payoff profile $a \in \RR^\Theta$. This nests settings with independent actions, which are common in applications and include the main asymmetric cases where $\alpha$, or merely its support, can be found explicitly under mutual information (\citealp*[Section 3.2]{caplin2019rational}; \citealp{pellegrino2023devil,brown2024endogenous}). 


\begin{corollary}\label{cor.independent.actions}
Suppose that actions are uncorrelated and non-constant, and define 
\[
    v_a=\kappa m_a+\frac{1}{2}(m_a^2+\sigma_a^2) \qquad \forall a\in A.
\]
Then $\alpha\in\Delta(A)$ is optimal in \eqref{eq.quadratic.program.distance} if and only if there exists $(\beta,\gamma)\in\RR^2$ such that 
\[
    \alpha(a)
    =
    \frac{\bigl[v_a-\beta m_a - \gamma\bigr]_+}{\sigma^2_a} \qquad \forall a \in A, \quad\text{and} \qquad \beta=\sum_{b\in A}\alpha(b)m_b.
\]
\end{corollary}

Corollary~\ref{cor.independent.actions} reduces the computation of the optimal $f$-mean to the determination of two scalars. Given $(\beta,\gamma)$, the weight on action $a$ is positive exactly when its adjusted value $v_a-\beta m_a$ exceeds the cutoff $\gamma$. Conditional on passing this cutoff, the weight on action $a$ is proportional to its excess value and inversely proportional to its variance $\sigma_a^2$.


\section{Csiszár information: choice accuracy and learning incentives}\label{sec:choice_accuracy}


This section takes a closer look at the properties of 
Csiszár information. We investigate a 
central 
question for models of information acquisition: how the accuracy of learning, and hence the quality of decision making, varies with the stakes of the decision problem. We 
consider a canonical task 
in which the agent is rewarded for identifying the true state, and examine how the 
predicted
probability of 
a correct choice responds to changes in the reward level and the task difficulty. The resulting comparative statics allow 
one to 
identify the 
cost function 
non-parametrically, characterize the marginal cost of information, and determine how the 
shape 
of the transformation $\phi$ governs 
the model's predictions. 

Throughout, to simplify the exposition, we restrict attention to transformations $\phi$ that are essentially smooth (Definition \ref{def:csiszar_ess_smooth}) and  denote their conjugate functions by $\psi=\phi^\star$.

\subsection{Response functions}

 We consider decision problems with $n$ equally likely states and $n$ actions, where each action represents a bet on an event comprising $m$ states ($n$ and $m$ are any integers such that $1 \le m < n$). A successful bet, for which the realized state belongs to the chosen event, yields a reward of $w>0$. An unsuccessful bet yields a payoff of zero. To ensure symmetry, we assume that, in each state, exactly $m$ actions yield the reward $w$ and the remaining $n-m$ actions yield zero payoff. 
 We call these problems \textit{Identification Tasks}.\footnote{Note that different identification tasks may have different state spaces. In this section, we adopt the convention that the same transformation $\phi$ defines a Csiszár information cost on all state spaces.}


 To illustrate the definition, consider the following three examples. Let actions be indexed by states: $A=\{a_\theta:\theta\in \Theta\}$. When $m=1$, each action $a_\theta$ is a bet \emph{on} state $\theta$, which pays $w$ if and only if the realized state is $\theta$. When $m=n-1$, each action $a_\theta$ is a bet \emph{against} state $\theta$, which pays $w$ if and only if the realized state is \emph{not} $\theta$. Between these extremes, suppose the states are points arranged on a circle, and each action $a_\theta$ pays $w$ if and only if the realized state is $\theta$ or one of its $m-1$ immediate clockwise successors.


 
 

 Although the symmetric structure of identification tasks is somewhat special, these problems are well-suited for implementation in laboratory experiments. For example, the $m=1$ case is a leading specification in existing experimental tests of rational inattention \citep*{caplin2020rational,dewan2020estimating,dean2023experimental}. 


 For identification tasks, the main behavioral prediction of interest is the probability that the agent makes a successful bet. Theorem \ref{thm:characterization} yields the following characterization:

\begin{proposition}\label{pro:unique_combinatorial}
 For any identification task with parameters $(n,m,w)$ and any optimal stochastic choice rule $P$, the probability of making a successful bet in state $\theta$ is given by 
 \[
    P_\theta(\{a:a(\theta)=w\}) = \frac{m}{n}\psi^\prime(w-\ell),
 \]
 where $\ell \in \RR$ is the unique solution of the equation
 \[
    \frac{m}{n}\psi^\prime(w-\ell)+ \left(1-\frac{m}{n}\right)\psi^\prime(-\ell)=1.
 \]
\end{proposition}

Proposition \ref{pro:unique_combinatorial} has two notable features that reflect the symmetry of the environment. First, both the success probability and the coefficient $\ell \in \RR$, which is the multiplier $\lambda_\pi(\theta)$, are independent of the state $\theta$. Second, the parameters $n$ and $m$ enter only via the ratio $m/n$, which is the fraction of actions that yield the reward in each state. We interpret this ratio as a measure of \emph{task difficulty}, where higher values correspond to simpler tasks. 

Consequently, optimal behavior in identification tasks can be expressed in terms of two parameters: the reward level $w >0$ and the task difficulty $\gamma = \frac{m}{n} \in (0,1)$.  
%
 For every $\gamma\in (0,1)$,\footnote{Allowing for all $\gamma \in (0,1)$ is merely a matter of notational convenience, as the rational values of $\gamma$ spanned by the ratios $m/n$ of integers $1\leq m < n$ provide a dense approximation of the interval $(0,1)$.} we define the decision maker's \emph{response function} $\rho_\gamma\colon(0,+\infty)\rightarrow (0,1)$ as%
 \begin{equation}\label{eqn:Csiszar-RF-1}
    \rho_\gamma(w) =  \gamma\psi^\prime(w-\ell_\gamma(w)),
  \end{equation}
 where $\ell_\gamma(w) \in \RR$ is the unique solution of the equation
  \begin{equation}\label{eqn:Csiszar-RF-2}
    \gamma \psi^\prime(w-\ell_\gamma(w))+ (1-\gamma)\psi^\prime(-\ell_\gamma(w))=1.
 \end{equation}
 The response function $\rho_\gamma$ succinctly captures how choice accuracy responds to learning incentives: per Proposition \ref{pro:unique_combinatorial}, it describes the state-independent probability of making a successful bet as the reward level $w$ varies across tasks of fixed difficulty $\gamma$. In what follows, we study its properties and how they connect to assumptions on the cost function.


\subsection{First-order properties} 

 Under mutual information, for each $\gamma$, the response function takes the logistic form:
 \begin{equation}\label{eq:response_f_mutual_info}
    \rho_\gamma(w) = \frac{\gamma e^\frac{w}{\kappa}}{\gamma e^\frac{w}{\kappa}+1-\gamma},
 \end{equation}
 where $\kappa>0$ is the constant in \eqref{eq:mutual_info_problem}. \cite{dean2023experimental} provide evidence that, in the case of two states and two actions (i.e. $\gamma = 1/2$), the response function implied by mutual information fails to adequately fit experimental data because ``subjects are less responsive to incentives than the Shannon model would predict.'' It is natural to expect that this issue may extend to other values of $\gamma$ and alternative experimental designs, given that the full family of response functions \eqref{eq:response_f_mutual_info} is determined by the single scalar $\kappa$. 

 By contrast, under fully general information costs, the only testable restrictions on the response function, for fixed $\gamma$, are that it is increasing and lies above $\gamma$.\footnote{Formally, fixing $\gamma$:  
 For any family of choice rules $(P^w)_{w >0}$ that is optimal under some cost function, the unconditional success probabilities $\sum_{\theta \in \Theta} \pi(\theta) P^w_\theta(\{a : a(\theta)  =w\})$ are increasing in $w$ and exceed $\gamma$. Conversely, for any response function---viz., schedule of conditional, but state-independent, success probabilities $w \mapsto P^w_\theta(\{a : a(\theta)  =w\})$---that is increasing in $w$ and lies above $\gamma$, there is a cost function and family of optimal choice rules that generates it. This follows from the revealed preference analyses of \citet{caplin2015revealed} and \citet*{caplin2020rational} (cf. \citealp{dewan2020estimating}). 
 } \cite*{caplin2020rational},  \cite*{dewan2020estimating}, and \cite*{dean2023experimental} all provide experimental evidence consistent with these weaker predictions.
 %
%
%
We first show that Csisz\'ar information is flexible enough to generate almost any such response function:
 
\begin{proposition}\label{pro:rep_theorem_fixed_g}
 For each $\gamma$, the response function \eqref{eqn:Csiszar-RF-1} has the following properties:
 \begin{enumerate}
    \item $\rho_\gamma(w)$ is strictly increasing in $w$.
    \item $\rho_\gamma(w)$ is continuous in $w$.
    \item $\rho_\gamma(w)\rightarrow \gamma$ as $w\rightarrow 0$.
    \item $\rho_\gamma(w)\rightarrow 1$ as $w\rightarrow +\infty$.
 \end{enumerate}
 Conversely, for every $\gamma$, a function satisfying 
 (i)--(iv) is the response function for some $\phi$.
\end{proposition}

Properties (i) and (iii) are the main predictions: the response function is strictly increasing in the reward $w$ and converges to the minimum value $\gamma$, the success probability from 
uniformly randomizing over actions, as the reward vanishes. Property (ii), continuity, cannot be directly tested with finite data. 
Property (iv), which means that the state is learnable with arbitrary precision, is due to the hypothesis that $\phi$ is essentially smooth. 


 As the proof of Proposition \ref{pro:rep_theorem_fixed_g} clarifies, for each fixed $\gamma$, any response function satisfying these properties can be rationalized by a continuum of different Csisz\'ar information costs. Thus, it is impossible to non-parametrically identify $\phi$ by observing the agent's behavior in identification tasks with a fixed difficulty level $\gamma$.\footnote{\cite*{caplin2020rational},  \cite*{dewan2020estimating}, and \cite*{dean2023experimental} use identification tasks with fixed $\gamma$ to estimate information cost functions within specific parametric classes.} However, by varying $\gamma$ we obtain:

 
 \begin{proposition}\label{pro:identification_resp_f}
    If $\phi_1$ and $\phi_2$ induce the same response function for every $\gamma$, then $\phi_1=\phi_2$.
 \end{proposition}
 
 The proof shows that identification is ensured even in the simpler case where $\phi_1$ and $\phi_2$ induce the same response function for every $\gamma$ of the form $\gamma=1/n$ or $\gamma=(n-1)/n$. Thus, it suffices to focus on tasks in which the agent bets on or against the true state. Namely, the conjugate $\psi = \phi^\star$ can be recovered from such data as follows: for all $w>0$,
 
 \[
 \psi^\prime(w) =\sup_{n>1} n \rho_{\frac{1}{n}}(w)\qquad\text{and}\qquad\psi^\prime(-w) = \inf_{n>1} n\left(1-\rho_{\frac{n-1}{n}}(w)\right).
 \]

 Motivated by these results, we now extend Proposition~\ref{pro:rep_theorem_fixed_g} to the case where both $w$ and $\gamma$ can vary. To this end, we introduce the concept of an \textit{inverse} response function. Given a family of response functions $\left(\rho_\gamma\right)_{\gamma \in (0,1)}$, for each  $x\in (1,+\infty)$ and $y\in (0,1)$, let $w(x,y)$ and $\gamma(x,y)$ denote the unique values of $w$ and $\gamma$ that solve the equations:\footnote{To see that these quantities are well defined, note that the two equations imply $\gamma(x,y)=\frac{1-y}{x-y} \in (0,1)$. For this value of $\gamma$, we have $\gamma x\in(\gamma,1)$. The properties of $\rho_{\gamma}$ stated in Proposition \ref{pro:rep_theorem_fixed_g} imply that there is a unique $w_{x,y}>0$ such that $\rho_{\gamma}(w_{x,y})=\gamma x$. Setting $w(x,y)=w_{x,y}$ defines the inverse response function.}
 \[
    \frac{\rho_\gamma(w)}{\gamma} = x\qquad\text{and}\qquad
    \frac{1-\rho_\gamma(w)}{1-\gamma}  = y.
 \]
 We call the map $(x,y)\mapsto w(x,y)$ the \emph{inverse response function} associated with $\left(\rho_\gamma\right)_{\gamma \in (0,1)}$.

 Note that $\frac{\rho_\gamma(w)}{\gamma}$ is the likelihood ratio between the agent's probability of choosing correctly and the probability $\gamma$ of choosing correctly when no information is available. Thus, while $\rho_\gamma$ maps reward levels to choice probabilities, the inverse response function maps observed behavior---expressed in likelihood ratios---to the underlying reward. Given $\gamma(x,y)$, the quantity $w(x,y)$ is the reward level that generates the likelihood ratios $(x,y)$.

 Using this object, one can jointly test and identify the Csisz\'ar information model: 

 \begin{proposition}\label{pro:rep_theorem_inv_resp}
 For any $\phi$, the inverse response function has the following properties:
 \begin{enumerate}
    \item $w(x,y)$ is strictly increasing in $x$ and strictly decreasing in $y$.
    \item $w(x,y)$ is continuous in $x$ and $y$.
    \item  $w(x,y)\rightarrow 0$ as $x\rightarrow 1$ and $y\rightarrow 1$.
    \item $w(x,y)\rightarrow+\infty$ as $x\rightarrow+\infty$ or $y\rightarrow 0$.
    \item For all $x$ and $x^\prime$, $w(x,y)-w(x^\prime,y)$ is independent of $y$.
    \item For all $y$ and $y^\prime$, $w(x,y)-w(x,y^\prime)$ is independent of $x$.
 \end{enumerate}
 Conversely, any function satisfying (i)--(vi) is the inverse response function for some $\phi$. Furthermore, if $\phi_1$ and $\phi_2$ induce the same inverse response function, then $\phi_1=\phi_2$.
 \end{proposition}
 Properties (i)--(iv) mirror the corresponding properties of the response function in Proposition \ref{pro:rep_theorem_fixed_g}, and they admit similar interpretations. Properties (v) and (vi), meanwhile, reflect the state-separable structure that defines Csisz\'ar information. Empirically testing these properties would shed light on the extent to which this separable structure constrains the model. Finally, as shown in the proof, the transformation $\phi$ can be recovered from  the inverse response function via the following conditions: for all $x\in (1,+\infty)$ and $y\in (0,1)$,
 \[
    \phi^\prime (x) = \inf_{z\in (0,1)} w(x,z)\qquad\text{and}\qquad \phi^\prime (y) = - \inf_{z\in (1,+\infty)} w(z,y).
 \]

 

 \subsection{Tools from risk theory: the Arrow-Pratt coefficient}

 The above results characterize the range of response functions that can arise under Csiszár information and show how the transformation $\phi$ can be elicited from identification tasks. We now turn to a more local question: how does the agent respond to a small increase in learning incentives, and what does this 
reveal about the marginal cost of information?

To this end, we draw on concepts from expected utility theory. In what follows, we assume the conjugate $\psi =\phi^\star$ is twice continuously differentiable and $\psi''>0$.\footnote{Given that $\phi$ is essentially smooth, this is equivalent to $\phi$ being twice continuously differentiable with $\phi''>0$ on $(0,+\infty)$. See Corollary 4.2.10 in \citet*[Chapter X.4]{urruty1993convex}.}  We study the local properties of Csiszár information  through the function $R_\psi \colon \RR \to \RR_+$ defined as
\[
    R_\psi(t)=\frac{\psi''(t)}{\psi'(t)}.
\]
The quantity $R_\psi(t)$ is a measure of the degree of convexity of $\psi$ at $t$. By analogy with expected utility theory, we refer to $R_\psi$ as the \emph{Arrow--Pratt coefficient} of $\psi$.\footnote{The Arrow--Pratt coefficient of a utility function $u$ is usually defined as $-u^{\prime\prime}/u^\prime$, where the negative sign makes it an index of concavity (i.e. risk aversion). Here, the ratio $\psi^{\prime\prime}/\psi^\prime$ instead measures convexity.}

This coefficient has a direct interpretation in terms of the response function. Recall that, in an identification task, the functions $\rho_\gamma$ and $\ell_\gamma$ are given by \eqref{eqn:Csiszar-RF-1} and \eqref{eqn:Csiszar-RF-2}. 
Hence,
\begin{equation}\label{eq.arrow-pratt-local_zero}
    \log\frac{\rho_\gamma(w)}
             {1-\rho_\gamma(w)}
    =
    \log\frac{\gamma}{1-\gamma}+
    \int_{0}^{w} R_\psi(t-\ell_\gamma(w))\dd t .
\end{equation}
That is, the change in log-odds induced by the reward $w$ is obtained by integrating the local curvature of $\psi$ over the interval $(0,w)$. This observation suggests that $R_\psi$ governs the agent's local responsiveness to learning incentives. In particular, as $w\rightarrow 0$, we have
\begin{equation}\label{eq.arrow-pratt-integrated}
    \log\frac{\rho_\gamma(w)}
             {1-\rho_\gamma(w)}
    =
    \log\frac{\gamma}{1-\gamma}+
    w R_\psi(0) + o(w) .
\end{equation}

 Thus, the Arrow--Pratt coefficient $R_\psi(0)$ determines the first-order response of choice accuracy to a small learning incentive. It also admits an equivalent, dual interpretation in terms of the marginal cost of information. In particular, by standard duality arguments,\footnote{This follows from the standard duality $\psi' = (\phi')^{-1}$, the inverse function theorem, and $\psi'(0)=1$.}  
\[
    R_\psi(0) = \frac{1}{\phi''(1)}.
\]
 The term $\phi''(1)$ characterizes the marginal cost of acquiring a low-precision experiment: it quantifies the local convexity of $\phi$ at the likelihood ratio $t=1$, and low-precision learning induces variability in the signal-contingent likelihood ratios in a neighborhood of $t=1$. Therefore, the marginal response of choice accuracy to incentives, measured by $R_\psi(0)$, is inversely proportional to the marginal cost of that response, measured by $\phi''(1)$.

 Note that \eqref{eq.arrow-pratt-integrated} only reveals the value $R_\psi(0)$. Except in the case of mutual information, where 
 $R_\psi$ is constant, this does not pin down the full schedule $R_\psi$. 
 We now show how to recover $R_\psi$ by combining low-stakes identification tasks with general decision problems.

\subsection{Combined decision problems and marginal incentives}
 
 Given an arbitrary decision problem $(\pi_1,A_1)$ with state space $\Theta_1$,  we perturb it by adding an identification task with state space $\Theta_2=\{1,\ldots,n\}$, a uniform prior, action set $A_2$, reward $w>0$, and a fraction $\gamma=m/n$ of correct actions in each state. The agent faces the two problems simultaneously. The resulting joint decision problem has state space
 \[
    \Theta=\Theta_1\times\mathrm \Theta_2.
 \]
 The two dimensions are independent, so the joint prior $\pi \in \Delta(\Theta)$ is the product measure
 \[
    \pi(\theta_1,\theta_2)=\tfrac1n \pi(\theta_1). 
 \]
 The joint action set $A$ comprises all pairs $(a_1,a_2) \in A_1\times A_2$, and payoffs are additive:\footnote{Similar combinations of decision problems have been studied in different contexts \citep*[among others]{bergemann2022counterfactuals,sandomirskiy2025monotonicity,sandomirskiy2026separability}.}
 \[
    a(\theta_1,\theta_2)=a_1(\theta_1) + a_2(\theta_2).
 \]

 Given an optimal choice rule $P \in \Delta(A)^\Theta$ in the joint decision problem, we study the log-odds of betting correctly rather than incorrectly in the identification task, conditional on the joint state $\theta = (\theta_1,\theta_2)$ and on choosing action $a_1$ in the original decision problem:
 \[
    \log\frac{P_{\theta}(\{a_2:a_2(\theta_2) = w\} \vert a_1)}{ P_{\theta}(\{a_2 : a_2(\theta_2) = 0\} \vert a_1)}.
 \]
 Higher values of this expression correspond to greater choice accuracy in the identification task. Although the two state spaces are independent and payoffs are additive, this log-odds ratio need not be independent of $a_1$ and $\theta_1$. The reason is that information is acquired via a common cost function, and hence the marginal cost of learning about the added identification task depends on the information used to solve the original problem.
 
\begin{proposition}\label{pro:arrow-p-appr}
 There is a selection $w \mapsto P^w$ of optimal choice rules for the joint decision problem such that for every $\theta\in\Theta$, every $a_1\in A_1$, and every sequence of reward levels $(w_n)$ such that $w_n \downarrow 0$ and $P_\pi^{w_n}(a_1) > 0$ for every $n$,
 \begin{equation}\label{eq:arrow-pratt}
 \log\frac{P^{w_n}_{\theta}(\{a_2:a_2(\theta_2) = w_n\} \vert a_1)}{ P^{w_n}_{\theta}(\{a_2 : a_2(\theta_2) = 0\} \vert a_1)}
 = \log\frac{\gamma}{1-\gamma} +w_n\,R_{\psi}(a_1(\theta_1)-\lambda_{\pi_1}(\theta_1)) + o(w_n),
\end{equation}
 where $\lambda_{\pi_1}$ is the Lagrange multiplier associated with the decision problem $(\pi_1,A_1)$.
\end{proposition}

 The proposition characterizes the full Arrow--Pratt schedule in terms of marginal learning incentives.
 In the degenerate case where the original decision problem has only one action, the joint problem effectively reduces to a standalone identification task, and \eqref{eq:arrow-pratt} reduces to \eqref{eq.arrow-pratt-local_zero}. More generally, the original decision problem determines the point at which the Arrow--Pratt coefficient is evaluated, namely $a_1(\theta_1)-\lambda_{\pi_1}(\theta_1)$. In this respect, it plays the same role as initial wealth in the classical Arrow--Pratt approximation from expected utility theory.\footnote{Recall that the Arrow--Pratt coefficient of a Bernoulli utility function is linked to risk premia through the \emph{Arrow--Pratt approximation}: for a small risk, the premium is approximately proportional to the variance, with proportionality factor given by half the Arrow--Pratt coefficient evaluated at initial wealth.} By varying over a sufficiently rich family of baseline decision problems, one can recover from \eqref{eq:arrow-pratt} the full schedule $R_\psi$. In particular, it suffices to use any family of baseline problems that contains all identification tasks.\footnote{In identification tasks, the term $a_1(\theta_1)-\lambda_{\pi_1}(\theta_1)$ takes the form $w - \ell_\gamma(w)$ for actions that yield the reward in state $\theta_1$, and $-\ell_\gamma(w)$ for actions that do not. 
 As shown in the proof of Proposition \ref{pro:identification_resp_f}, by varying over all $w>0$ and $\gamma \in (0,1)$, these two terms span the full real line. Thus, \eqref{eq:arrow-pratt} pins down $R_\psi$.}

 \subsection{Second-order properties}\label{sec:second_order_p}

 We now study the second-order properties of response functions, such as concavity and convexity, and relate them to the shape of the cost function. These curvature properties are of interest for two reasons. First, they are naturally connected to the marginal cost of information, as we explain below. Second, they can be used to 
 behaviorally distinguish between different forms of the transformations $\phi$. For example, \citet{dewan2020estimating} provide experimental evidence for both concave and S-shaped response functions, and use this data to test and calibrate several parametric families of information costs.

 In what follows, we assume $\psi = \phi^\star$ is thrice continuously differentiable and $\psi''>0$.\footnote{Equivalently, $\phi$ is thrice continuously differentiable with $\phi''>0$ on $(0,+\infty)$. This follows from Corollary 4.2.10 in \citet*[Chapter X.4]{urruty1993convex} and the inverse function theorem.} 
 The relevant property of $\psi$ is no longer its convexity, as measured by the Arrow--Pratt coefficient $R_\psi$, but rather 
 the rate at which its convexity changes, which we measure by
\[
    R_{\psi'}(t)=\frac{\psi'''(t)}{\psi''(t)}.
\]
 In expected utility theory, the map $R_{\psi'} \colon \RR \to \RR$ is known as the \emph{prudence index} of $\psi$, and it plays a central role in the study of precautionary savings \citep{kimball1993standard}.\footnote{The prudence index of a concave utility function $u$ is defined as $-u^{\prime\prime\prime}/u^{\prime\prime}$, where the negative sign and $u^{\prime\prime}<0$ make it a measure of the convexity of $u'$. Here, we omit the negative sign because $\psi^{\prime\prime}>0$.} In our context, the monotonicity properties of $R_{\psi'}$ govern the curvature of response functions. 

We first characterize the conditions under which response functions are concave, i.e., the agent's choice accuracy exhibits decreasing marginal sensitivity to learning incentives.


\begin{proposition}\label{pro:response_f_concave}
The following statements are equivalent:
\begin{enumerate}
    \item For all $\gamma$, $\rho_\gamma$ is concave.
    \item $R_{\psi^\prime}(t)\geq 0$ for $t<0$, and $R_{\psi^\prime}(t)\leq 0$ for $t>0$.
    \item 
    $\phi'''(t) \leq 0$ for $t\in (0,1)$, and $\phi'''(t) \geq 0$ for $t >1$. 
\end{enumerate}
\end{proposition}

 Thus, the response function $\rho_\gamma$ is concave for every task difficulty $\gamma$ if and only if the prudence index $R_{\psi^{\prime}}$ single-crosses zero from above at $t=0$. Equivalently, concavity arises whenever $\phi''$ is decreasing on $(0,1)$ and increasing on $(1,+\infty)$, which means that $\phi$ is locally more convex at likelihood ratios that are further away from $t=1$. We interpret condition (iii) as formalizing one notion of \emph{increasing marginal cost} of information: since acquiring more information corresponds to inducing more variability of the likelihood ratio around $t=1$, it means that each marginal piece of information is more expensive.\footnote{For a concrete example, let $\phi_{\mathrm B}(t)=t-1-\log t$ denote the Burg transformation and let $\phi_3(t)=(t^3-3t+2)/6$ denote the $p=3$ Cressie--Read transformation. Then
$
    \phi=\phi_{\mathrm B}+2\phi_3
$
satisfies Proposition \ref{pro:response_f_concave}-(iii), and therefore generates a concave response function.}


 Next, we investigate the conditions under which response functions are S-shaped, i.e., first convex and later concave. This shape arises in the mutual information benchmark \eqref{eq:response_f_mutual_info}, and is commonly observed in a wide range of perceptual experiments.\footnote{In various related settings, experiments in psychophysics often find that choice probabilities form an S-shaped ``psychometric function'' of the stimulus \citep*{woodford2020modeling,khaw2021cognitive}.} Intuitively, the pattern of initially increasing marginal sensitivity to incentives reflects a marginal cost of information that is decreasing, at least initially, starting from zero information.
 
 Formally, we say that the response function $\rho_\gamma$ is \emph{S-shaped} if it satisfies the condition:\footnote{Given our assumptions on $\psi$, $\rho_\gamma$ is twice continuously differentiable and $\rho'_\gamma>0$ (see Appendix \ref{app:proofs-second_order_p}).} 
 \[
    w_1 \geq w_2\quad\text{and}\quad\rho_\gamma^{\prime\prime}(w_1) \geq 0 \quad\Longrightarrow\quad \rho_\gamma^{\prime\prime}(w_2)\geq 0.
 \]
We call $\rho_\gamma$ \emph{strongly S-shaped} if its Arrow--Pratt coefficient, $R_{\rho_\gamma} = \rho''_\gamma/\rho'_\gamma$, is decreasing. By construction, every response function that is strongly S-shaped is also S-shaped. For example, the response function \eqref{eq:response_f_mutual_info} under mutual information is strongly S-shaped.




\begin{proposition}\label{pro:S_shape_resp} The following statements are equivalent:
\begin{enumerate}
    \item For all $\gamma$, $\rho_\gamma$ is strongly S-shaped.
    \item $R_{\psi^\prime}$ is decreasing.
\end{enumerate}
Moreover, under these conditions, $R_{\phi^\prime} = \phi''' / \phi''$ is increasing.
\end{proposition}

Thus, the response function $\rho_\gamma$ is strongly S-shaped for every task difficulty if and only if the prudence index $R_{\psi'}$ is decreasing. It follows that a decreasing prudence index $R_{\psi'}$ is a sufficient condition for S-shaped response functions.\footnote{A simple example is obtained by smoothing the conjugate of the Pearson $\chi^2$ transformation. Define 
\[
    \psi(s)
    =
    \frac{1}{\log(1+e)}
    \int_0^s \log\bigl(1+e^{1+u}\bigr)\,du.
\]
Simple calculations show $R_{\psi'}$ is strictly decreasing.}

\section{Perceptual distance and nested entropies}\label{sec:nested-entropies}

 A well-known limitation of mutual information is that states enter into the analysis only via their prior probabilities and payoff consequences, while structural features of the state space---such as the physical or perceptual distance between different states---are irrelevant for learning costs and behavior. This invariance property, which is also shared by Csiszár information costs, implies that optimal behavior is identical across payoff-equivalent states \citep*{caplin2022rationally}. While this is natural in some applications, such as the symmetric settings of Sections \ref{sec:separable_APU}--\ref{sec:choice_accuracy}, it is understood that invariance can yield unrealistic predictions in settings where states that are more similar are are inherently harder to distinguish \citep*[e.g.,][]{hebert2021neighborhood,morris2022coordination,dean2023experimental,pomatto2023cost}.

 To address this limitation, we introduce a new family of posterior-separable costs. These costs, which generalize mutual information, provide a framework for modeling structural features of the state space  and retain tractability by yielding closed-form conjugates. We connect the predicted behavior to the classic \emph{nested logit} and \emph{mixed logit} models of discrete choice, study its implications in two canonical discrimination tasks, and relate the model to \citeauthor{hebert2021neighborhood}'s (\citeyear{hebert2021neighborhood}) neighborhood-based costs and \citeauthor{walker2023rational}'s (\citeyear{walker2023rational}) multi-attribute Shannon entropy.

 \subsection{Nested Shannon entropy}


 Our approach is based on the idea that, in complex settings, decision makers often do not learn directly about the full state space. Instead, they simplify the learning problem by organizing states into a smaller collection of nests that capture the features regarded as focal, and then acquire information at that coarser level. Learning therefore has two dimensions: it may shift beliefs across nests, and it may refine beliefs within nests.

\begin{definition}\label{def.personal_state_space}
 Given a prior belief $\pi \in \Delta^\circ(\Theta)$, a \textit{nest structure} $(N,\nu)$ consists of a finite set $N$ of \emph{nests} and a joint distribution $\nu\in \Delta(N\times \Theta)$ over nests and states such that: (i) the marginal of $\nu$  on $N$ has full support, and (ii) the marginal of $\nu$ on $\Theta$ is $\pi$.
\end{definition}

 The nest set $N$ collects the categories through which the agent views the state space. If states are one-dimensional, the nests might correspond to intervals such as ``low,'' ``medium,'' and ``high.'' If states describe the characteristics of a health plan, the nests might be coarse labels such as  ``high deductible,'' ``narrow network,'' and ``open formulary.'' 

 The joint distribution $\nu$ models the agent's \emph{ex-ante} perception of how states relate to nests.\footnote{Conditions (i) and (ii) rule out redundant nests and ensure consistency with the prior, respectively.} Let $\overline{\nu} \in \Delta(N)$ be its marginal on $N$, $\nu_i \in \Delta(\Theta)$ be the conditional distribution over states given nest $i \in N$, and $B_i = \supp \nu_i $ be the set of states related to nest $i \in N$. 

In settings where the agent deterministically categorizes each state into a unique nest, 
 $(B_i)_{i \in N}$ forms a partition of $\Theta$ and $\nu$ is pinned down by the prior: $\overline{\nu}(i) = \pi(B_i)$ and $\nu_i = \pi(\cdot \mid B_i)$. In many settings, however, nests may overlap. For example, if states are one-dimensional, the agent may only have a vague notion of the threshold between ``medium'' and ``high.'' If states are multi-dimensional, as in the context of health plans, the agent may categorize separately along each dimension. 
 In these cases, $(B_i)_{i \in N}$ forms a non-disjoint covering of $\Theta$ and, given any state $\theta$, the conditional distribution over nests $\nu_\theta \in \Delta(N)$ specifies the strength of association between $\theta$ and each nest that contains it.

 We study the posterior-separable costs generated by the following entropy functions: 


\begin{definition}\label{def:nested_s_entropy}
 Given a prior belief $\pi \in \Delta^\circ(\Theta)$, nest structure $(N,\nu)$, and strictly positive weights $\zeta$ and $\eta$, the \emph{nested Shannon entropy} $H_{\pi}\colon\Delta(\Theta)\to\overline{\mathbb{R}}_+$ is defined as
\begin{equation}\label{eqn:nested-shannon-entropy}
 H_{\pi}(p) =
 \min\left\{ \zeta\, D_{\mathrm{KL}}(\bar{\mu}\Vert \bar{\nu})
 +
 \eta\,\sum_{i\in N} \bar{\mu}(i) D_{\mathrm{KL}}(\mu_i\Vert \nu_i)\right\},
\end{equation}
 where the minimization is over all $\mu\in\Delta(N\times\Theta)$ for which the marginal on $\Theta$ is $p$.\footnote{\mbox{As above, $\overline{\mu} \in \Delta(N)$ is the marginal on $N$ and $\mu_i \in \Delta(\Theta)$ is the conditional distribution on $\Theta$ given $i$.}}
\end{definition}

 Nested Shannon entropy can be viewed as describing an optimal two-step learning process. Given a target posterior $p \in \Delta(\Theta)$ over states, the agent chooses the least costly way to produce a joint belief $\mu \in \Delta(N \times \Theta)$ over nests and states that marginalizes to $p$.  
 The objective 
 in \eqref{eqn:nested-shannon-entropy} evaluates the cost of $\mu$ in two stages. The first term measures the change in beliefs \emph{across} nests: $D_{\mathrm{KL}}(\bar{\mu}\Vert \bar{\nu})$ is the KL divergence between the posterior and prior marginal beliefs over $N$. 
 The second term measures the expected change in beliefs \emph{within} nests: it is the expectation, under the posterior marginal $\overline{\mu}$, of the KL divergence $D_{\mathrm{KL}}(\mu_i\Vert \nu_i)$ between the nest-contingent posterior and prior beliefs over $\Theta$. Hence, 
 it is as if the agent first learns indirectly about nests and then learns directly about the states within each nest, with the weights $\zeta$ and $\eta$ scaling the costs of these two respective steps.


\begin{remark}
Definition \ref{def:nested_s_entropy} readily extends to the case where learning within some nests may be costlier than learning within others: simply replace $\eta$ with a vector $(\eta_i)_{i \in N}$ of nest-specific weights.
More generally, as we explain in Online Appendix \ref{sec:nested_f}, the construction underlying nested Shannon entropy can be extended to arbitrary $f$-information costs by, in effect, replacing KL divergence with any generator $f$. This provides a general method for incorporating structural features of the state space into the $f$-information framework and, in particular, into the Csiszár information model. 
\end{remark}

\subsection{Special cases}

 Before proceeding with the analysis, we highlight several special cases of nested Shannon entropy that help clarify the role of its parameters and the minimization step in \eqref{eqn:nested-shannon-entropy}.

 First, consider the weights $\zeta$ and $\eta$. Assuming that $\zeta \leq \eta$ captures the idea that learning indirectly about nests is cheaper than learning directly about states. In this case, the nested Shannon entropy of a posterior $p$ is bounded above by the KL divergence between $p$ and the prior $\pi$, scaled by the within-nest weight $\eta$:
\[
    H_{\pi}(p) \leq \eta D_{\mathrm{KL}}(p\Vert \pi).
\]
 Since KL divergence is the entropy function underlying mutual information (Example \ref{exa:CDL}), it follows that the nested Shannon cost function is bounded above by mutual information scaled by $\eta$. Intuitively, relative to the mutual information benchmark, categorizing states into nests allows the agent to simplify the learning problem and thus reduce costs. In the special case where $\zeta = \eta$, so that such categorization has no cost-saving benefit, these bounds become equalities and the nested Shannon cost reduces to mutual information.\footnote{These bounds and equalities follow from the chain rule for KL divergence \citep{cover1999elements}.}

 Next, consider the joint distribution $\nu$ over nests and states. We give three examples. First, in the extreme case of \emph{perfect categorization}, each nest contains a single state: $|B_i| = 1$ for all $i \in N$. Since learning the nest fully reveals the state, learning is only costly in the first step regarding nests, and the cost function reduces to mutual information scaled by the across-nest weight $\zeta$: $H_{\pi}(p) = \zeta D_{\mathrm{KL}}(p\Vert \pi)$. Second, in the opposite extreme of \emph{null categorization}, nests and states are independent: $\nu_i=\nu_j$ for all $i,j\in N$. Since learning the nest reveals nothing about the state, the agent optimally defers all learning to the second step regarding states, and the cost function reduces to mutual information scaled by the within-nest weight $\eta$: $H_{\pi}(p) = \eta D_{\mathrm{KL}}(p\Vert \pi)$. Finally, in the intermediate case of \emph{deterministic categorization}, each state is contained in a unique nest: the collection of events $(B_i)_{i \in N}$ is a partition of $\Theta$. The minimizer in \eqref{eqn:nested-shannon-entropy} then has a closed-form solution: 

\begin{proposition}\label{cor:MASE}
    Under deterministic categorization, for any target posterior $p \in \Delta(\Theta)$, the minimum in \eqref{eqn:nested-shannon-entropy} is attained if and only if, for all $i\in N$ with $p(B_i)>0$ and $\theta\in B_i$,
    \[
    \bar{\mu}(i)=p(B_i)\quad\text{and}\quad \mu_i(\theta)=p(\theta\vert B_i).
    \]
\end{proposition}

\subsection{Closed-form conjugate and nested logit posteriors}

An important feature of the nested Shannon model is that it yields closed-form conjugates:


 \begin{proposition}\label{prop:nested_logit}
 The conjugate of the entropy $H_\pi$ in \eqref{eqn:nested-shannon-entropy} is given by, for every $x\in \RR^\Theta$, 
 \[
    H^\star_{\pi}(x)= \zeta \log \left( \sum_{i\in N } \bar{\nu}(i) \left(\sum_{\theta \in \Theta} \nu_i(\theta) e^{x(\theta)/\eta} \right)^{\eta/\zeta} \right).
 \]
 \end{proposition}
 
 In discrete choice theory, this functional form is known as the surplus function of the \emph{generalized nested logit} model \citep{wen2001generalized}. 
 While in discrete choice, nests represent sets of actions that the agent regards as similar, in our setting, nests represent sets of states that share attributes regarded as similar in the learning process.\footnote{The discrete choice literature typically also assumes $\zeta \geq \eta$, which ensures that the model admits a random utility representation. Here, by contrast, it is often natural to assume $\zeta \leq \eta$, as discussed above.} 
 %
%
%
%
 For the special case of deterministic categorization, where nests are disjoint, the conjugate reduces to the surplus function of the classic \emph{nested logit} model \citep[Chapter 4]{train2009discrete}:
 \[
    H^\star_{\pi}(x)= \zeta \log \left( \sum_{i\in N } \pi(B_i) \left(\sum_{\theta \in B_i} \pi(\theta\vert B_i) e^{x(\theta)/\eta} \right)^{\eta/\zeta} \right).
 \]


 Leveraging Proposition \ref{prop:nested_logit}, the optimality conditions from Theorem \ref{thm:characterization} can be expressed in closed form as a function of the $f$-mean $\alpha=P_\pi$ and the prior-adjusted multiplier $\lambda_\pi$. In particular, 
 for every action $a \in \supp P_\pi$ in the consideration set, the posterior belief $p_a \in \Delta(\Theta)$ at which action $a$ is chosen is given by \eqref{eq:PS-FOC-dual} as the gradient of the conjugate:
 \[
    p_a(\theta) = \nabla_\theta H_{\pi}^\star(a-\lambda_\pi)= \frac{ \sum_{i\in N }\nu(i,\theta)e^{\frac{a(\theta)-\lambda_\pi(\theta)}{\eta}}\left(\sum_{\tau \in \Theta} \nu_i(\tau) e^{\frac{a(\tau)-\lambda_\pi(\tau)}{\eta}} \right)^{\frac{\eta-\zeta}{\zeta}}}{\sum_{i\in N } \bar{\nu}(i) \left(\sum_{\tau \in \Theta} \nu_i(\tau) e^{\frac{a(\tau)-\lambda_\pi(\tau)}{\eta}} \right)^{\frac{\eta}{\zeta}}}.
 \]

 These optimal posteriors admit a decomposition reminiscent of nested logit models: 

 \begin{proposition}\label{pro:minimizers_nested_s}
 For each posterior $p_a \in \Delta(\Theta)$, the minimum in \eqref{eqn:nested-shannon-entropy} is achieved by the joint belief $\mu_a \in \Delta(N \times \Theta)$ whose marginal distribution $\overline{\mu}_a \in \Delta(N)$ over nests is given by
 \[
 \bar{\mu}_a(i) = \frac{ \bar{\nu}(i) \left(\sum_{\tau \in \Theta} \nu_i(\tau) e^{\frac{a(\tau)-\lambda_\pi(\tau)}{\eta}} \right)^{\frac{\eta}{\zeta}}}{\sum_{j\in N } \bar{\nu}(j) \left(\sum_{\tau \in \Theta} \nu_j(\tau) e^{\frac{a(\tau)-\lambda_\pi(\tau)}{\eta}} \right)^{\frac{\eta}{\zeta}}} \qquad \forall i \in N,
 \]
 and whose nest-contingent conditional distributions $\mu_{(a,i)} \in \Delta(\Theta)$ over states are given by 
 \[
 \mu_{(a,i)}(\theta) = \frac{\nu_i(\theta)e^{\frac{a(\theta)-\lambda_\pi(\theta)}{\eta}}}{\sum_{\tau \in \Theta} \nu_i(\tau) e^{\frac{a(\tau)-\lambda_\pi(\tau)}{\eta}}} \qquad  \forall  i  \in N, \, \theta \in \Theta.
 \]
 \end{proposition}

Hence, under the optimal choice rule implied by Theorem \ref{thm:characterization} and the optimal two-step learning strategy implied by the entropy function \eqref{eqn:nested-shannon-entropy}, the posterior $p_a$ decomposes as
\[
p_a = \sum_{i \in N} \bar{\mu}_a(i) \mu_{(a,i)}.
\]
This is the generalized nested logit rule of \citet{wen2001generalized}, except that the roles of actions and states are interchanged (and payoffs are shifted by the multiplier $\lambda_\pi$). In this manner, assumptions on the nest structure $(N,\nu)$ translate directly into properties of joint posterior beliefs over nests and states, and hence on the agent's optimal behavior. 

\subsection{Hard learning constraints and mixed logit choice}

 The case where learning within nests is infinitely costly, that is, $\eta\to+\infty$, is especially tractable. In this limit, the agent cannot refine beliefs within a nest and can only learn across nests, so an extension $\mu$ of a posterior $p$ to nest-state pairs has finite cost only if $\mu_i = \nu_i$ for all $i$ with $\bar{\mu}(i)>0$. Consequently, the limiting nested Shannon entropy is



 \begin{equation}\label{eq:nested_entropy_limit_case}
    H_{\pi}(p) = \inf \left\{ \zeta 
    D_\text{KL}(\bar{\mu} \Vert \bar{\nu}) : \bar{\mu} \in \Delta(N), \ \sum_{i \in N} \bar{\mu}(i)\nu_i = p \right\}.
 \end{equation}
 If, for simplicity, the set $\{\nu_i : i \in N\}$ is assumed to be affinely independent, the feasible set in \eqref{eq:nested_entropy_limit_case} reduces to a singleton whenever $H_\pi(p)<+\infty$ (equivalently, $p$ is in the convex hull of $\{\nu_i  : i \in N\}$).\footnote{Affine independence also implies $|N| \leq |\Theta|$, in line with the idea that nests are coarser than states.} Thus, the agent is restricted to learning only about nests, and each feasible target posterior $p$ uniquely determines the associated joint posterior $\mu$.

In this case, as under mutual information, the Lagrange multiplier admits a closed-form expression, and the associated $f$-mean solves an auxiliary maximization problem:
\begin{equation}\label{eq:redux_mutual_info_problem_2}
    \max_{\alpha\in \Delta(A)}
    \zeta\sum_{i\in N}\bar{\nu}(i)\log\left(\sum_{a\in A}
    e^{\frac{\bar{a}(i)}{\zeta}}\alpha(a)\right),
\end{equation}
 where
\[ 
    \bar{a}(i) = \sum_{\theta \in \Theta} \nu_i(\theta) a(\theta)
\]
 is the prior expected payoff of action $a$ conditional on nest $i \in N$. Furthermore, generalizing the logit formula \eqref{eq:logit}, the optimal choice rule takes the form of a state-dependent \emph{mixed logit} rule \citep[Chapter 6]{train2009discrete}. The mixing distribution is the conditional prior distribution $\nu_\theta\in\Delta(N)$ over nests given the state $\theta$.

\begin{samepage}
\begin{proposition}\label{prop:perceptual_mm}
 Let costs be posterior-separable, with entropy function \eqref{eq:nested_entropy_limit_case}. Suppose the set $\{\nu_i:i\in N\}$ is affinely independent. Then,
    a stochastic choice rule $P=(A,(P_\theta)_{\theta\in\Theta})$ is optimal if and only if there is a solution $\alpha$ of \eqref{eq:redux_mutual_info_problem_2} such that, for all $\theta\in\Theta$ and $a\in A$,
    \begin{equation}\label{eq:logit_2}
        P_\theta(a)
        =
        \sum_{i\in N}\nu_\theta(i)\, Q_i(a),  \qquad \text{where} \qquad  Q_i(a) = \frac{\alpha(a)e^{\frac{\bar{a}(i)}{\zeta}}}
{\sum_{b\in A}\alpha(b)e^{\frac{\bar{b}(i)}{\zeta}}}.
    \end{equation}
    Moreover, for any such $P$ and $\alpha$, it holds that $\alpha=P_\pi$.

\end{proposition}
\end{samepage}

Intuitively, since the decision maker is constrained to learn only about nests, the information acquisition problem can be solved in two steps. First, consider the reduced-form problem in which the nest set $N$ serves as the state space, the marginal distribution $\overline{\nu} \in \Delta(N)$ serves as the prior, the multiset $\{(\bar{a}(i))_{i \in N} : a \in A\} \subseteq \RR^N$ of nest-contingent expected payoff profiles serves as the action set,\footnote{It is a multiset because there may exist actions $a,b \in A$ such that $a \neq b$, yet $\overline{a}(i) = \overline{b}(i)$ for all $i \in N$.} and the cost function is mutual information scaled by $\zeta$. The solutions of this reduced-form problem are characterized by the auxiliary problem \eqref{eq:redux_mutual_info_problem_2} and the nest-dependent stochastic choice rule $Q = (A, (Q_i)_{i \in N})$ in \eqref{eq:logit_2}, mirroring \citet*{matvejka2015rational} and \citet*{caplin2019rational}. Second, in each state $\theta$, the mixing measure $\nu_\theta \in \Delta(N)$ determines the probability of $\theta$ being perceived as a member of each nest, as mechanically specified by the nest structure.


This representation implies that perceptually similar states induce similar behavior. Formally, if two states $\theta$ and $\tau$ have similar conditional distributions over nests  $\nu_\theta, \nu_\tau \in \Delta(N)$ under the given nest structure, then their choice probabilities must also be similar: 
\[
|P_\theta(a)-P_\tau(a)|
\leq
\sum_{i\in N}|\nu_\theta(i)-\nu_\tau(i)| \qquad \forall a\in A.
\]
Hence, in the nested Shannon model with hard learning constraints, the shape of optimal behavior is directly governed by the shape of the conditional mixing distributions $(\nu_\theta)_{\theta \in \Theta}$. We now use this model to study perceptual distance in a canonical discrimination task.

\subsection{Perceptual distance in one-dimensional problems}\label{subsec:perc_distance}

 We study a one-dimensional problem where the state space is a finite, equally spaced subset of the real line. For clarity, we index the states in increasing order and write $\Theta = \{\theta_1, \ldots, \theta_n\}$, with $\theta_{i+1} - \theta_i = \Delta > 0$ for all $i = 1, \ldots, n-1$. Since discrimination tasks are typically formulated in continuous settings, this construction can be viewed as a uniform discretization.

 The agent chooses between a risky action $r$ and a safe action $s$. The payoff of the risky action varies monotonically with the state: $\theta \geq \tau$ implies $r(\theta) \geq r(\tau)$. A simple example is a binary bet where action $r$ pays 1 if the state is positive, and $-1$ if the state is negative.  The safe action's payoff is normalized to zero.

 We consider a decision maker whose perceptual acuity diminishes with proximity between states. To encode this structure, we set $N = \{1, \ldots, n\}$ and interpret $\nu_{\theta_i}(j)=\nu(j,\theta_i)/\pi(\theta_i)$ as the probability of perceiving state $\theta_i$ as $\theta_j$.

 A central object of interest in discrimination tasks is the relationship between stimulus intensity and choice frequency. In our framework, this is captured by the function $$\theta \mapsto P_\theta(r),$$ commonly referred to as \emph{psychometric function}. Psychometric functions observed in experiments are typically S-shaped \citep*{woodford2020modeling,khaw2021cognitive}. This means that $P_\theta(r)$ increases with $\theta$, consistent with action $r$ being more appealing in high states, and that this function is convex at low stimulus levels and concave at high ones. In our discrete setting, we say that the psychometric function is \emph{convex at} $\theta_i$ if
 \[
    P_{\theta_{i}}(r)-P_{\theta_{i-1}}(r) \leq P_{\theta_{i+1}}(r)-P_{\theta_{i}}(r),
 \]
 and we say that it is \emph{concave at} $\theta_i$ if the inequality is reversed. The next proposition relates the shape of the psychometric function to properties of the nest structure.

 \begin{proposition}\label{pro:1dim_task}
 Let costs be posterior-separable, with the entropy function \eqref{eq:nested_entropy_limit_case} and the nest structure described above. Suppose the set $\{\nu_i:i\in N\}$ is affinely independent. The psychometric function has the following properties:
 \begin{enumerate}
     \item  The psychometric function is increasing if the nest structure satisfies the monotone likelihood ratio property (MLRP): 
    \[
        \theta \geq \tau\text{ and }i\geq j\quad\text{implies}\quad \nu_\theta(i)\nu_{\tau}(j)\geq \nu_{\tau}(i)\nu_\theta(j).
    \]

    \item Assume the MLRP. The psychometric function is convex at $\theta_i$ if $\frac{1}{2}\nu_{\theta_{i-1}}+\frac{1}{2}\nu_{\theta_{i+1}}$ first-order stochastically dominates $\nu_{\theta_i}$.

    \item Assume the MLRP. The psychometric function is concave at $\theta_i$ if $\nu_{\theta_i}$ first-order stochastically dominates $\frac{1}{2}\nu_{\theta_{i-1}}+\frac{1}{2}\nu_{\theta_{i+1}}$.
     \end{enumerate}
\end{proposition}

 The MLRP captures the idea that higher states are more likely to be perceived as higher nests, reflecting perceptual consistency with the ordering of states. Next we provide an example satisfying the MLRP.

 \begin{example} 
 For all $\theta\in \Theta$ and $i\in N$, define 
 \[
    \nu_\theta(i) = \frac{\gamma(\vert \theta-\theta_i\vert)}{\sum_{j\in N}\gamma(\vert\theta-\theta_j\vert)},
 \]
 where $\gamma\colon \RR_+\rightarrow (0,+\infty)$ is a decreasing function. This specification assigns higher encoding probability to nearby states, with the decay governed by $\gamma$. The nest structure satisfies the MLRP if $\gamma$ is log-concave. 
\end{example}

\subsection{Relation to neighborhood-based costs}\label{ssec:HW}

 Nested Shannon entropy closely resembles two classes of cost functions in the literature: \citeauthor{hebert2021neighborhood}'s (\citeyear{hebert2021neighborhood}) \emph{neighborhood-based costs} and \citeauthor{walker2023rational}'s (\citeyear{walker2023rational}) \emph{multi-attribute Shannon entropy} (MASE).  
 To facilitate comparison, we focus here on the leading specification of neighborhood-based cost, which includes MASE as a special case.\footnote{The full family of neighborhood-based costs can be analogously related to the general family of nested entropies that we define in Online Appendix \ref{sec:nested_f}.}

 Given a collection $\{B_i\}_{i=0}^n$ of events $B_i \subseteq \Theta$ that cover $\Theta$ 
 and an associated collection of positive weights $\{\kappa_i\}_{i=0}^n$, \citet{hebert2021neighborhood} define the entropy function
 \begin{equation}\label{eqn:neighborhood}
    HW_\pi(p) = \sum_{i=0}^n   \kappa_i\,\bar{p}(i)  D_\text{KL}(p_i \Vert \pi_i),
 \end{equation}
 where $\overline{p}(i) = p(B_i)$, $p_i = p(\cdot \mid B_i)\in\Delta(B_i)$ is the conditional distribution of $p$ on $B_i$, and $\pi_i = \pi(\cdot \mid B_i)\in\Delta(B_i)$ is the analogous conditional distribution of the prior $\pi$. \citet{hebert2021neighborhood} interpret each event $B_i$ as a \emph{neighborhood} of states that are costly to distinguish. The neighborhood structure $\{B_i\}_{i=0}^n$ is analogous to a nest structure in the nested Shannon model, and \eqref{eqn:neighborhood} resembles the $\zeta \to 0$ limit of \eqref{eqn:nested-shannon-entropy}, but without the minimization step.

The most direct connection between the nested Shannon, neighborhood-based, and MASE models arises when the neighborhood structure is hierarchical. Let $B_0=\Theta$ and let $\{B_i\}_{i=1}^n$ form a partition of $\Theta$. In this case, the chain rule for KL divergence gives
 \[
    HW_\pi(p) = \kappa_0 \, D_\text{KL}(\bar{p} \Vert \bar{\pi}) + \sum_{i=1}^n (\kappa_0 + \kappa_i) \, \bar{p}(i) D_\text{KL}(p_i \Vert \pi_i). 
 \]
 This expression is exactly the MASE entropy function of \cite{walker2023rational}. It also coincides with the nested Shannon entropy \eqref{eqn:nested-shannon-entropy} under deterministic categorization, with nests $\{B_i\}_{i=1}^n$ and scaling parameters $\zeta = \kappa_0$ and $\eta_i = \kappa_0 + \kappa_i$ (Proposition \ref{cor:MASE}). Thus, this special case of the nested Shannon model aligns with the subclass of neighborhood-based models that exhibit tree-like neighborhood structures, and it is equivalent to MASE. 

 Beyond this special case, however, the nested Shannon and neighborhood-based models diverge in subtle but important ways. When neighborhoods overlap, there are multiple ways to extend a posterior belief $p$ to state-neighborhood pairs.  If state $\theta$ lies in two distinct neighborhoods $B_i \neq B_j$, nested Shannon entropy splits the probability mass $p(\theta)$ across the two events and, when several such splits are possible, it selects one with minimal cost. By contrast, the neighborhood-based entropy accounts for the probability $p(\theta)$ twice, since $\theta$ is  included in both events.\footnote{Thus, the induced measure $\bar{p}$ on $\{0,\dots,n\}$ in \eqref{eqn:neighborhood} typically has total mass strictly greater than one.} In the next section, we study a simple class of decision problems where the two models yield qualitatively different predictions.

\subsection{The challenge of multi-dimensional learning}

We study a setting where the state has two dimensions, and the agent finds \emph{multi-dimensional learning} difficult: learning about either dimension separately is easy, but learning about both at once is harder. 

This idea arises in several contexts. In the perceptual experiments of \citet{tversky1969substitutability}, subjects can easily tell which of two rectangles has the larger area when the rectangles differ only in width or only in height, but they make more mistakes when the rectangles differ along both dimensions. In a market setting, a consumer may find it easy to compare products that differ only with respect to a single attribute, such as quality or price, but harder to compare products that differ along multiple attributes.

To illustrate this idea, consider a simple environment in which the state space is
\[
\Theta=\{u,d\}\times\{l,r\},
\]
with uniform prior \(\pi(\theta)=1/4\) for every \(\theta\in\Theta\). One can think of the state as the location of a visual stimulus: the first coordinate indicates whether it is \emph{up} or \emph{down}, and the second whether it is \emph{left} or \emph{right}. Define the events
\[
    U=\{u\} \times \{l,r\}, \qquad D=\{d\} \times \{l,r\},
\]
\[
    L=\{u,d\} \times \{l\}, \qquad R=\{u,d\} \times \{r\}.
\]
Thus, \(\{U,D\}\) partitions states according to the vertical dimension, while \(\{L,R\}\) partitions them according to the horizontal dimension.

For each event \(i\in\{U,D,L,R\}\), let \(a_i\in \mathbb{R}^{\Theta}\) be the action that pays \(1\) if \(\theta\in i\) and \(0\) otherwise. We also define two additional actions, \(a_{\mathrm{diag}}\) and \(a_{\mathrm{off}}\), by
\[
a_{\mathrm{diag}}(\theta)=
\begin{cases}
1 & \text{if } \theta\in\{(u,l),(d,r)\},\\
0 & \text{otherwise,}
\end{cases}
\qquad
a_{\mathrm{off}}(\theta)=
\begin{cases}
1 & \text{if } \theta\in\{(u,r),(d,l)\},\\
0 & \text{otherwise.}
\end{cases}
\]
So \(a_{\mathrm{diag}}\) pays \(1\) when the state lies on the diagonal, and \(a_{\mathrm{off}}\) pays \(1\) when it lies on the off-diagonal. Using these actions, we construct three binary decision problems:
\[
A_1=\{a_U,a_D\}, \qquad A_2=\{a_L,a_R\}, \qquad A_3=\{a_{\mathrm{diag}},a_{\mathrm{off}}\}.
\]
In problem 1, the decision maker bets on the vertical dimension of the state; in problem 2, on the horizontal dimension; and in problem 3, on whether the state lies on the diagonal. 

These three problems are simply permutations of one another. Thus, any cost function that is symmetric under all permutations of the state space, such as mutual information, predicts the same choice accuracy in all three problems. This seems restrictive. If the decision maker finds multi-dimensional learning difficult, one would instead expect higher accuracy in problems 1 and 2, which require learning only one dimension of the state, than in problem 3, which requires combining information about both dimensions.

We now show that this pattern arises under a nested Shannon specification, in the limiting case where learning about a single dimension is essentially costless.

Consider a nested Shannon cost with nest set
\[
N=\{U,D,L,R\},
\]
where each nest corresponds to one of the four events above. Let \(\nu\in\Delta(N\times \Theta)\) extend the prior \(\pi\) as follows: each nest has ex ante probability \(1/4\), so \(\bar{\nu}(i)=1/4\) for all \(i\in N\), and conditional on nest \(i\), the two states in that nest are equally likely, so \(\nu_i(\theta)=1/2\) for all \(\theta\in B_i\). We fix the within-nest weight \(\eta>0\), and consider the limit as the across-nest weight \(\zeta\to 0\). This specification captures the idea that the decision maker represents the environment in terms of the two underlying dimensions of uncertainty, where the limit \(\zeta\to 0\) isolates the cost of learning across dimensions rather than within a single one.

\begin{proposition}\label{cor:multi-tasking}
For each decision problem \(j\in\{1,2,3\}\), let \(P^j\in\Delta(A_j)^\Theta\) be an optimal stochastic choice rule under the nested Shannon cost described above. As \(\zeta\to 0\), for all $a_1 \in A_1$ and $a_2 \in A_2$,
\[
    P^1_\theta(a_1)\to a_1(\theta) \quad\text{and}\quad
    P^2_\theta(a_2)\to a_2(\theta).
\]
Moreover, for every $\zeta>0$, and $a_3 \in A_3$,
\[
P^3_\theta(a_3)=
\begin{cases}
\dfrac{e^{1/\eta}}{e^{1/\eta}+1} & \text{if } a_3(\theta)=1,\\[6pt]
\dfrac{1}{e^{1/\eta}+1} & \text{otherwise.}
\end{cases}
\]
\end{proposition}

The intuition is straightforward. As \(\zeta\to 0\), it becomes essentially costless to distinguish perfectly between \(U\) and \(D\), and likewise between \(L\) and \(R\). Therefore, in problems 1 and 2, choice becomes nearly perfect. By contrast, in problem 3 the decision maker must combine information from both dimensions, and the resulting accuracy is determined by the parameter $\eta$, which measures the difficulty of joint learning.

Although this behavioral pattern is intuitive, it cannot be generated by any specification of the neighborhood-based model \eqref{eqn:neighborhood}. In that class of models, if choice is nearly perfect in both problems 1 and 2, then it must also be nearly perfect in problem 3, and indeed the cost function must be close to identically zero. The reason is that the two models aggregate costs differently: the Nested Shannon model separates the cost of learning \emph{about} nests from the cost of learning \emph{within} nests, while neighborhood-based models generally do not. See Online Appendix~\ref{sec:online.appendix.nests} for the formal details of this comparison.

\bibliography{refs}

\newpage
\appendix
\begin{center}
    {\bf \Large Appendix}
\end{center}

\section{Proof of Lemma \ref{lem:f_info_properties}}

 (i). Let $K= (K_{s})_{s\in S}$ be a stochastic kernel that satisfies $Q_\theta(s')= \sum_{s\in S} K_s(s') P_\theta(s)$ for every $s' \in S'$. For each $\alpha\in \Delta(S)$, define $K\alpha\in \Delta(S^\prime)$ by 
\[
    (K\alpha)(s^\prime) = \sum_{s\in S} K_s(s^\prime)\alpha(s).
\]
 Note that $Q_\theta=KP_\theta$ for all states $\theta$. Then,
\begin{align*}
    I_{f}(Q,\pi) &= \inf_{\alpha \in\Delta(S^\prime)}D_{f_\pi}(Q \Vert \alpha)\\
    & \leq\inf_{\alpha \in\Delta(S)}D_{f_\pi}(Q\Vert K \alpha)\leq\inf_{\alpha\in\Delta(S)}D_{f_\pi}(P\Vert \alpha )=I_{f}(P,\pi),
\end{align*}
 where the second inequality follows from the data-processing inequality for $f$-divergences (\citealp*{duchi2018multiclass}, Proposition 2).

 (ii). Since $f$ is non-negative, $I_f(P,\pi)\geq 0$ for every experiment $P$. If $P$ is uninformative, then we can pick $\alpha\in \Delta(S)$ such that $\alpha=P_\theta$ for all $\theta\in \Theta$, and obtain 
 \[
    0 = f(\boldsymbol{1}) = D_{f_\pi}(P\Vert \alpha)\geq I_f(P,\pi). 
 \]
 Thus, when $P$ is uninformative, $I_f(P,\pi)= 0$.

 (iii). To verify convexity, take $P,Q\in \Delta(S)^\Theta$ and $t\in[0,1]$. Let $\alpha,\beta\in\Delta(S)$ such that $I_{f}(P,\pi)=D_{f_\pi}(P\Vert \alpha)$ and $I_{f}(Q,\pi)=D_{f_\pi}(Q\Vert \beta)$. Then, since $D_{f_\pi}$ is convex on $\Delta(S)^\Theta\times\Delta(S)$,
\begin{align*}
    tI_{f}(P,\pi)+(1-t)I_{f}(Q,\pi) & = t D_{f_\pi}(P\Vert \alpha)+(1-t)D_{f_\pi}(Q \Vert \beta)\\
    & \geq D_{f_\pi}(tP+(1-t) Q \Vert t\alpha +(1-t) \beta)\\
    & \geq I_{f}(t P+(1-t) Q).
\end{align*}

 To verify lower semicontinuity, let $(P^n)$ be a sequence in $\Delta(S)^\Theta$ with limit $P$. For every $n$, take $\alpha^n\in\Delta(S)$ such that $I_{f}(P^n,\pi)=D_{f_\pi}(P^{n}\Vert \alpha^{n})$. Since $\Delta(S)$ is compact, we can assume that the sequence $(\alpha^n)$ is convergent without loss of generality. Setting $\alpha =\lim_{n\rightarrow\infty}\alpha^n$, we obtain 
\[
    \liminf_{n\rightarrow\infty}I_{f}(P^n,\pi)=\liminf_{n\rightarrow\infty}D_{f_\pi}(P^{n}\Vert \alpha^{n})\geq D_{f_\pi}(P\Vert \alpha)\geq I_{f}(P,\pi),
\]
 where we use the lower semicontinuity of $D_{f}$ on $\Delta(S)^\Theta\times\Delta(S)$. 

\section{Proof of Theorem \ref{thm:characterization}}

We begin by recasting (\ref{eq:ri_prob_f_infor}) as a constrained optimization problem. With a slight abuse of notation, we write $P=(A,(P_{\theta})_{\theta\in\Theta})$ to denote the \emph{improper choice rule} that specifies, for every $\theta\in \Theta$, a non-negative measure over actions $P_{\theta}\in\mathbb{R}_{+}^{A}$. For every $x\in \mathbb{R}_+^\Theta$, let $\mathcal{P}_x$ be the set of improper choice rules $P\in \mathbb{R}_+^{A\times\Theta}$ such that $\sum_{a\in A}P_\theta(a)=x(\theta)$ for all $\theta\in \Theta$. The set $\mathcal{P}_{\boldsymbol{1}}$ consists of the proper choice rules considered in the main text.

 The $f_\pi$-divergence $D_{f_\pi}(P\Vert\alpha)$  between an improper choice rule $P$ and a probability distribution $\alpha\in\Delta(A)$ is defined in the obvious way, extending Definition \ref{def.multivariate.divergence} to non-negative measures. The function 
 $
     (P,\alpha) \mapsto D_{f_\pi}(P\Vert \alpha)
 $
 is lower semicontinuous and convex on $\mathbb{R}_+^{\Theta\times A}\times \Delta(A)$. In addition,  
 $
    D_{f_\pi}(P\Vert \alpha)\geq f_\pi(x)
 $
 for all $x\in \mathbb{R}^\Theta_+$ and $P\in \mathcal{P}_x$.

Let $I_{f}(P,\pi)$ be the $f$-information of an improper choice rule $P$:
\[
I_{f}(P,\pi) = \min_{\alpha\in \Delta(A)} D_{f_\pi}(P\Vert \alpha).
\]
The function 
$
 P \mapsto I_{f}(P,\pi)
$
is lower semicontinuous and convex on $\mathbb{R}_+^{\Theta\times A}$.

For every $x\in\mathbb{R}_{+}^{\Theta}$, we consider the constrained optimization problem 
\begin{align}\label{eq:proof_opt_prob}
\max_{P\in\mathcal{P}_x}\sum_{\theta\in \Theta}\pi(\theta)\sum_{a\in A}P_{\theta}(a)a(\theta)-I_{f}(P,\pi).
\end{align}
We go back to (\ref{eq:ri_prob_f_infor}) when $x=\boldsymbol{1}$. We denote by $V_\pi(x)$ the value of (\ref{eq:proof_opt_prob}). We say that $\lambda\in\mathbb{R}^{\Theta}$ is a \emph{Lagrange multiplier} for (\ref{eq:ri_prob_f_infor}) if
\[
    V_\pi(\boldsymbol{1})=\sup_{P\in\mathbb{R}_{+}^{A\times\Theta}}\sum_{\theta\in \Theta}\pi(\theta)\sum_{a\in A}P_{\theta}(a)a(\theta)-I_{f}(P,\pi)-\sum_{\theta\in\Theta}\lambda(\theta)\sum_{a\in A}(P_{\theta}(a)-1).
\]

\begin{lemma}
The value function $V_\pi\colon \mathbb{R}_{+}^{\Theta}\rightarrow [-\infty,+\infty)$
satisfies the following properties:
\begin{enumerate}
\item $\dom V_\pi=\dom f_\pi$.
\item For every $x\in\mathbb{R}_{+}^{\Theta}$, there exists $P\in\mathcal{P}_{x}$
such that
\[
V_\pi(x)=\sum_{\theta\in \Theta}\pi(\theta)\sum_{a\in A}P_{\theta}(a)a(\theta)-I_{f}(P,\pi).
\]
\item $V_\pi$ is concave.
\end{enumerate}
\end{lemma}
\begin{proof}
(i). Fix $x\in\mathbb{R}_{+}^{\Theta}$. If $V_\pi(x)>-\infty$, then
there exist $P\in\mathcal{P}_{x}$ and $\alpha\in\Delta(A)$ such
that $D_{f_\pi}(P\Vert\alpha)<+\infty$. Since 
$f_\pi(x)\leq D_{f_\pi}(P\Vert\alpha)$, we obtain $x\in\dom f_\pi$.

Conversely, suppose that $f_\pi(x)<+\infty$. Given a distribution $\alpha\in\Delta(A)$,
we define $P\in\mathbb{R}_{+}^{A\times\Theta}$ by $P_{\theta}(a)=\alpha(a)x(\theta)$.
Note that $\sum_{a\in A}P_{\theta}(a)=x(\theta)$ for all $\theta\in\Theta$.
Moreover, $D_{f_\pi}(P\Vert\alpha)=f_\pi(x)$. Thus, $P\in\mathcal{P}_{x}$
and $I_{f}(P,\pi)<+\infty$. We deduce that $x\in\dom V_\pi$.

(ii). If $V_\pi(x)=-\infty$, then $f_\pi(x)=+\infty$ by (i). It follows that $I_f(P,\pi)\geq f_\pi(x)=+\infty$ for all $P\in\mathcal{P}_{x}$. We obtain that
\[
V_\pi(x)=-\infty=\sum_{\theta\in \Theta}\pi(\theta)\sum_{a\in A}P_{\theta}(a)a(\theta)-I_{f}(P,\pi).
\]
for all $P\in\mathcal{P}_{x}$.

Suppose instead that $V_\pi(x)\in \mathbb{R}$. Since the function
\[
 P\mapsto \sum_{\theta\in \Theta}\pi(\theta)\sum_{a\in A}P_{\theta}(a)a(\theta)-I_{f}(P,\pi)
\]
is upper semicontinuous, the desired result follows from the compactness
of $\mathcal{P}_{x}$.

(iii). Take $x,y\in\mathbb{R}_{+}^{\Theta}$ and $t\in(0,1)$. By (ii),
there are $P\in\mathcal{P}_{x}$ and $Q\in\mathcal{P}_{y}$ such that
\begin{align*}
V_\pi(x) & =\sum_{\theta\in \Theta}\pi(\theta)\sum_{a\in A}P_{\theta}(a)a(\theta)-I_{f}(P,\pi),\\
V_\pi(y) & =\sum_{\theta\in \Theta}\pi(\theta)\sum_{a\in A}Q_{\theta}(a)a(\theta)-I_{f}(Q,\pi).
\end{align*}
Using the fact that $I_{f}(\cdot,\pi)$ is convex on $\mathcal{P}_x$,
we obtain
\begin{align*}
& (1-t)V_\pi(x)+tV_\pi(y) \\
 = & \sum_{\theta\in \Theta}\pi(\theta)\sum_{a\in A}[(1-t)P_{\theta}(a)+tQ_{\theta}(a)]a(\theta) -(1-t)I_{f}(P,\pi)-tI_{f}(Q,\pi) \\
  \leq & \sum_{\theta\in \Theta}\pi(\theta)\sum_{a\in A}[(1-t)P_{\theta}(a)+tQ_{\theta}(a)]a(\theta)-I_{f}((1-t)P+tQ,\pi)\\
 \leq &  V_\pi((1-t)x+ty).
\end{align*}
This demonstrates that $V$ is concave.
\end{proof}
 Since $\dom f_\pi=\dom V_\pi$ and $\boldsymbol{1}\in\ri(\dom f_\pi)$ by Assumption \ref{ass:co-finitee_etc}, we have $\boldsymbol{1}\in\ri(\dom V_\pi)$.  Given that $V_\pi$ is concave, the superdifferential of $V_\pi$ at $x=\boldsymbol{1}$, defined as $\partial V_\pi(\boldsymbol{1})$, is nonempty \citep[Theorem 23.4]{rockafellar1970convex}: there exists $\lambda\in \mathbb{R}^\Theta$ such that for all $x\in \mathbb{R}^\Theta_+$,
\[
    V_\pi(\boldsymbol{1}) - \sum_{\theta\in \Theta}\lambda(\theta) \geq  V_\pi(x) - \sum_{\theta\in \Theta}\lambda(\theta)x(\theta).
\]
 Note that $\lambda\in\partial V_\pi(\boldsymbol{1})$ if and only if $\lambda$ is a Lagrange multiplier for (\ref{eq:ri_prob_f_infor}), as defined above.

 We define the Lagrangian function $\mathcal{L}_\pi\colon \mathbb{R}_{+}^{A\times\Theta}\times\mathbb{R}^{\Theta}\rightarrow [-\infty,+\infty)$ by
\begin{align*}
 \mathcal{L}_\pi(P,\lambda) & =\sum_{\theta\in \Theta}\pi(\theta)\sum_{a\in A}P_{\theta}(a)a(\theta)-I_{f}(P,\pi)-\sum_{\theta\in\Theta}\lambda(\theta)\left(\sum_{a\in A}P_{\theta}(a)-1\right)\\
 & =\sum_{a\in A}\sum_{\theta\in \Theta}(a(\theta)\pi(\theta)-\lambda(\theta))P_{\theta}(a)-I_{f}(P,\pi)+\sum_{\theta\in \Theta}\lambda(\theta).
\end{align*}
The Lagrangian function is concave in $P$ and affine in $\lambda$. It defines the maxmin problem
\begin{equation}\label{eq:maxmin_lagrangian}
    \max_{P\in \mathbb{R}_+^{\Theta\times A}}\min_{\lambda\in \mathbb{R}^\Theta} \mathcal{L}_\pi(P,\lambda).
\end{equation}
 By standard arguments (see, e.g., \citealp{rockafellar1970convex}, Theorem 28.3), a pair $(P,\lambda)$ is a saddle point of (\ref{eq:maxmin_lagrangian}) if and only $P$ is a solution of (\ref{eq:ri_prob_f_infor}) and $\lambda$ is a Lagrange multiplier for (\ref{eq:ri_prob_f_infor}). We have shown that (\ref{eq:ri_prob_f_infor}) admits a solution and a Lagrange multiplier. Thus,  the maxmin problem (\ref{eq:maxmin_lagrangian}) admits a saddle point. Moreover, the saddle value of (\ref{eq:maxmin_lagrangian}) is $V_\pi(\boldsymbol{1})$.

 The next step is crucial and it allows us to connect the Lagrangian function $\mathcal{L}_\pi$ to the conjugate function $f^\star_\pi$.

\begin{lemma}\label{lem:connecting_Lagrangian}
 For all $\lambda\in\mathbb{R}^{\Theta}$ and $\alpha\in\Delta(A)$,
\[
\max_{P\in\mathbb{R}_{+}^{A\times\Theta}}\sum_{a\in A}\sum_{\theta\in \Theta}(a(\theta)\pi(\theta)-\lambda(\theta))P_{\theta}(a)-D_{f_\pi}(P\Vert\alpha)=\sum_{a\in A}\alpha(a)f^{\star}_\pi(a\pi-\lambda).
\]
The maximum is achieved by $P\in\mathbb{R}_{+}^{A\times\Theta}$
such that, for all $\theta\in\Theta$ and $a\in A$,
\[
P_{\theta}(a)=\alpha(a)\nabla_\theta f^{\star}_\pi(a\pi-\lambda).
\]
\end{lemma}
\begin{proof}
Since $f_\pi$ is superlinear at infinity (Assumption \ref{ass:co-finitee_etc}), if $D_{f_\pi}(P\Vert\alpha)<+\infty$ and $\alpha(a)=0$,
then $P_{\theta}(a)=0$ for all $\theta\in\Theta$. Thus, direct computations show that
\begin{align*}
 & \sup_{P\in\mathbb{R}_{+}^{A\times\Theta}}\sum_{a\in A}\sum_{\theta\in \Theta}(a(\theta)\pi(\theta)-\lambda(\theta))P_{\theta}(a)-D_{f_\pi}(P\Vert\alpha)\\
= &\sum_{a\in\supp(\alpha)}\alpha(a)\sup_{x\in\mathbb{R}_{+}^{\Theta}}\sum_{\theta\in\Theta}(a(\theta)\pi(\theta)-\lambda(\theta))\frac{x(\theta)}{\alpha(a)}-f_\pi\left(\frac{x}{\alpha(a)}\right)\\
= & \sum_{a\in\supp(\alpha)}\alpha(a)\sup_{y\in\mathbb{R}_{+}^{\Theta}}\sum_{\theta\in \Theta}(a(\theta)\pi(\theta)-\lambda(\theta))y(\theta)-f_\pi(y)\\
= & \sum_{a\in\supp(\alpha)}\alpha(a)f_\pi^{\star}(a\pi-\lambda)=\sum_{a\in A}\alpha(a)f^{\star}_\pi(a\pi-\lambda).
\end{align*}
The second part of the statement follows from the fact that
\[
f^{\star}_\pi(a\pi-\lambda)=\sum_{\theta\in \Theta}(a(\theta)\pi(\theta)-\lambda(\theta))y(\theta)-f_\pi(y)\quad\Longleftrightarrow\quad y \in\partial f^{\star}_\pi(a\pi-\lambda).
\]
See \citet*[Theorem 23.5]{rockafellar1970convex}. Since $f^{\star}_\pi$ is differentiable (being $f_\pi$ superlinear at infinity and essentially strictly convex, see Assumption \ref{ass:co-finitee_etc}), $\partial f^{\star}_\pi(a\pi-\lambda)=\{\nabla f^{\star}_\pi(a\pi-\lambda)\}$.
\end{proof}

The next lemmas establish a relationship between the maxmin problems (\ref{eq:maxmin_problem}) and (\ref{eq:maxmin_lagrangian}). To ease the exposition, we denote by $L_\pi$ the function
\[
    (\alpha,\lambda)\mapsto \sum_{a\in A}\alpha(a)f^\star_\pi(a\pi-\lambda)+\sum_{\theta\in\Theta}\lambda(\theta). 
\]
\begin{lemma}
\label{lem:values}The maxmin problems (\ref{eq:maxmin_problem}) and (\ref{eq:maxmin_lagrangian}) have the same value, $V_\pi(\boldsymbol{1})$.
\end{lemma}
\begin{proof}
By a minimax theorem \citep[Corollary 37.3.1]{rockafellar1970convex}, the maxmin problem (\ref{eq:maxmin_problem}) has a saddle value. We have argued above that the saddle value of (\ref{eq:maxmin_lagrangian}) is $V_\pi(\boldsymbol{1})$. By Lemma \ref{lem:connecting_Lagrangian},
\begin{align*}
\inf_{\lambda\in\mathbb{R}^{\Theta}}\sup_{P\in\mathbb{R}_{+}^{A\times\Theta}}\mathcal{L}_\pi(P,\lambda) & =\inf_{\lambda\in\mathbb{R}^{\Theta}}\sup_{\alpha\in\Delta(A)}L_\pi(\alpha,\lambda).
\end{align*}
Hence, (\ref{eq:maxmin_problem}) and (\ref{eq:maxmin_lagrangian}) have the same value, $V_\pi(\boldsymbol{1})$.
\end{proof}
\begin{lemma}
\label{lem:necessity}Let $(P,\lambda)$ be saddle point of (\ref{eq:maxmin_lagrangian}).
Take any $\alpha\in\Delta(A)$ such that $D_{f_\pi}(P\Vert\alpha)=I_{f}(P,\pi)$.
Then, $(\alpha,\lambda)$ is a saddle point of $(\ref{eq:maxmin_problem})$. Moreover, $P_{\theta}(a)=\alpha(a)\nabla_{\theta}f^{\star}_\pi(a\pi-\lambda)$
for all $\theta\in\Theta$ and $a\in A$.
\end{lemma}
\begin{proof}
Since $(P,\lambda)$ is saddle point of (\ref{eq:maxmin_lagrangian}), $\mathcal{L}_\pi(P,\lambda)\geq\mathcal{L}_\pi(Q,\lambda)$
for all $Q\in\mathbb{R}_{+}^{A\times\Theta}$. In other terms,
\begin{align*}
&\sum_{a\in A}\sum_{\theta\in\Theta}(a(\theta)\pi(\theta)-\lambda(\theta))P_{\theta}(a)-D_{f_\pi}(P\Vert\alpha)\\
= &\sup_{\beta\in\Delta(A)}\sup_{Q\in\mathbb{R}_{+}^{A\times\Theta}}\sum_{a\in A}\sum_{\theta\in\Theta}(a(\theta)\pi(\theta)-\lambda(\theta))Q_{\theta}(a)-D_{f_\pi}(Q\Vert\beta).
\end{align*}
It follows from Lemma \ref{lem:connecting_Lagrangian} that $L_\pi(\alpha,\lambda)\geq L_\pi(\beta,\lambda)$
for all $\beta\in\Delta(A)$. Moreover, $P_{\theta}(a)=\alpha(a)\nabla_{\theta}f^{\star}_\pi(a\pi-\lambda)$
for all $\theta\in\Theta$ and $a\in A$.

It remains to verify that $L_\pi(\alpha,\lambda)\leq L_\pi(\alpha,l)$ for
all $l\in\mathbb{R}^{\Theta}$. Since $(P,\lambda)$ is saddle point
of (\ref{eq:maxmin_lagrangian}), $P$ is a solution of (\ref{eq:ri_prob_f_infor}). Thus, in particular, $\sum_{a\in A}P_{\theta}(a)=1$ for all $\theta\in\Theta$.
We have just argued that $P_{\theta}(a)=\alpha(a)\nabla_{\theta}f^{\star}_\pi(a\pi-\lambda)$
for all $\theta\in\Theta$ and $a\in A$. Thus, for all $\theta\in\Theta$,
\[
\sum_{a\in A}\alpha(a)\nabla_{\theta}f^{\star}_\pi(a\pi-\lambda)=1.
\]
This is the first-order condition for the problem of minimizing $L_\pi(\alpha,l)$
over $l\in\mathbb{R}^{\Theta}$. We conclude that $L_\pi(\alpha,\lambda)\leq L_\pi(\alpha,l)$
for all $l\in\mathbb{R}^{\Theta}$.
\end{proof}

\begin{lemma}
\label{lem:sufficiency} Let $(\alpha,\lambda)$ be a saddle point
of (\ref{eq:maxmin_problem}). Define $P\in\mathbb{R}_{+}^{A\times\Theta}$ by $P_{\theta}(a)=\alpha(a)\nabla_{\theta}f^{\star}(a\pi-\lambda)$.
Then, $(P,\lambda)$ is a saddle point of (\ref{eq:maxmin_lagrangian}). Moreover,
$I_{f}(P,\pi)=D_{f_\pi}(P\Vert\alpha)$.
\end{lemma}
\begin{proof}
By Lemma \ref{lem:connecting_Lagrangian}, 
\[
\sum_{a\in A}\sum_{\theta\in\Theta}(a(\theta)\pi(\theta)-\lambda(\theta))P_{\theta}(a)-D_{f_\pi}(P\Vert\alpha)+\sum_{\theta\in\Theta}\lambda(\theta)=L_\pi(\alpha,\lambda).
\]
Since $(\alpha,\lambda)$ is a saddle point of $L_\pi$, $L_\pi(\alpha,\lambda)\geq L_\pi(\beta,\lambda)$
for all $\beta\in\Delta(A)$. It follows from Lemma \ref{lem:connecting_Lagrangian}
that $L_\pi(\alpha,\lambda)$ is equal to
\[
\sup_{\beta\in\Delta(A)}\sup_{Q\in\mathbb{R}_{+}^{A\times\Theta}}\sum_{a\in A}\sum_{\theta\in\Theta}(a(\theta)\pi(\theta)-\lambda(\theta))Q_{\theta}(a)-D_{f_\pi}(Q\Vert\beta)+\sum_{\theta\in\Theta}\lambda(\theta).
\]
Overall, we deduce that $\mathcal{L}_\pi(P,\lambda)\geq\mathcal{L}_\pi(Q,\lambda)$
for all $Q\in\mathbb{R}_{+}^{A\times\Theta}$. Moreover, $I_{f}(P,\pi)=D_{f_\pi}(P\Vert\alpha)$. It remains to show that $\mathcal{L}_\pi(P,\lambda)\leq\mathcal{L}_\pi(P,l)$
for all $l\in\mathbb{R}^{\Theta}$. The first-order condition for
the problem of minimizing $L_\pi(\alpha,l)$ over $l\in\mathbb{R}^{\Theta}$
is
\[
\sum_{a\in A}\alpha(a)\nabla_{\theta}f^{\star}_\pi(a\pi-\lambda)=1.
\]
Thus, $P\in\mathcal{P}_{\boldsymbol{1}}$. As a result, $\mathcal{L}_\pi(P,\lambda)=\mathcal{L}_\pi(P,l)$
for all $l\in\mathbb{R}^{\Theta}$.
\end{proof}
Theorem \ref{thm:characterization} follows from Lemmas \ref{lem:values}--\ref{lem:sufficiency}.

\begin{remark}\label{rem:uniqueness}
The proof of Theorem \ref{thm:characterization} shows that, analogously to the case of mutual information, the solution set under $f$-information can be identified with a closed convex subset of $\Delta(A)$. The set of saddle points of \eqref{eq:maxmin_problem} is a closed convex product set in $\Delta(A)\times\mathbb{R}^{\Theta}$. Moreover, Lemmas \ref{lem:necessity} and \ref{lem:sufficiency} 
imply that any Lagrange multiplier $\lambda$ associated with a saddle point can generate any optimal choice rule $P$: for every saddle point $(\alpha,\lambda)$ and every optimal $P$, there exists $\beta\in\Delta(A)$ such that $(\beta,\lambda)$ is also a saddle point and
\[
P_\theta(a)=\beta(a)\nabla_\theta f^\star_\pi(a\pi-\lambda),
\qquad \forall \theta\in\Theta,\ a\in A.
\]
It follows that the set of optimal choice rules can be identified with a closed convex subset of $\Delta(A)$, namely, the projection of the product set of saddle points onto $\Delta(A)$.
\end{remark}




\section{Proof of Proposition \ref{prop:phi-PS-disjoint}}\label{sec:proof_ps_csiszar}

 That (ii) implies (i) is immediate: mutual information is both a Csiszár information cost and a posterior-separable cost (Examples \ref{exa:CDL}--\ref{exa:csiszar}). We prove the converse implication. 

Let $n=|\Theta| \geq 3$. Let $C$ be a Csiszár information cost generated by $\phi$, where $\phi$ is essentially smooth and thrice continuously differentiable on $(0,+\infty)$ with $\phi''>0$. Write $\psi=\phi^\star$. Then $\psi$ is strictly convex and thrice continuously differentiable, with $\psi'(\mathbb R)=(0,\infty)$ and $\psi''>0$.\footnote{As noted in the main text, essential smoothness of $\phi$ is equivalent to $\psi$ being strictly convex and $\psi'(\mathbb R)=(0,+\infty)$. By standard duality arguments,
$
\psi'=(\phi')^{-1}.
$
By the inverse function theorem, because $\phi'$ is twice continuously differentiable and $\phi''>0$, $\psi'$ is twice continuously differentiable and $\psi''>0$.}  Let $R_{\psi}=\psi^{\prime\prime}/\psi^\prime$ denote the Arrow--Pratt coefficient of $\psi$. Note that $R_\psi$ is continuously differentiable and $R_\psi>0$. We use the following lemma:

\begin{lemma}\label{lem:example_n_one}
 If $R_{\psi}$ is strictly monotone on some open interval, then there exists a decision problem whose unique optimal choice rule $P$ satisfies $\vert \supp P_\pi\vert = n+1$.
\end{lemma}

Now, suppose that, for every decision problem, there exists some posterior-separable cost that generates the same set of optimal stochastic choice rules as $C$. Under any posterior-separable cost, every decision problem admits an optimal choice rule such that $\vert \supp P_\pi\vert\leq n$ (see, e.g., \citealp[Proposition 4]{denti2022posterior}, or Proposition \ref{pro:consideration_set} in Online Appendix \ref{section:OA:consideration-sets}). Thus, the supposition and Lemma \ref{lem:example_n_one} imply that there does not exist any open interval on which $R_\psi$ is strictly monotone. Since $R_{\psi}$ is continuously differentiable, it follows that $R_{\psi}$ is constant. In particular, since $R_\psi>0$, there exists $\kappa>0$ such that
\[
R_{\psi}(t)=\frac{1}{\kappa}
\qquad\text{for all } t\in\mathbb{R}.
\]
Solving this differential equation (using the facts that $\psi(0)=0$ and $\psi^\prime(0)=1$) gives
\[
\psi(t)=\kappa\left(e^{t/\kappa}-1\right)
\qquad\text{for all } t\in\mathbb{R}.
\]
Taking convex conjugates, we obtain
\[
\phi(t)=\psi^\star(t)=\kappa\left(t\log t-t+1\right)
\qquad\text{for all } t\in\mathbb{R}_+.
\]
Thus, $C=\kappa I_S$. It remains to prove Lemma \ref{lem:example_n_one}.

\begin{proof}[Proof of Lemma \ref{lem:example_n_one}]

Fix enumerations $\Theta=\{\theta_1,\ldots,\theta_n\}$ and $A=\{a_1,\ldots,a_{m}\}$. We construct a decision problem $\mathcal{D}=(\pi,A)$ and a pair $(\alpha,\lambda)\in \Delta(A)\times \mathbb{R}^\Theta$ such that $m=n+1$, $\supp \alpha=A$, and $(\alpha,\lambda)$ is the unique saddle point of $\mathcal{D}$.

To this end, we parametrize a candidate decision problem $\mathcal D$ and a
candidate saddle point $(\alpha,\lambda)$ as follows. Given
$\bar\pi\in(0,1)$, states $\theta_1,\ldots,\theta_{n-1}$ each have prior
probability $\bar\pi/(n-1)$, while state $\theta_n$ has prior probability
$1-\bar\pi$. The payoff function is defined as follows:
\begin{itemize}
    \item for each $i=1,\ldots,n-1$, action $a_i$ pays $\rho>0$ in state
    $\theta_i$, pays $z\in\mathbb R$ in state $\theta_n$, and pays $0$ in
    all other states;
    \item action $a_n$ pays $z$ in state $\theta_n$ and
    $\sigma\in(0,\rho)$ in every other state;
    \item action $a_{n+1}$ pays $y\in\mathbb R$ in state $\theta_n$ and
    $x\in\mathbb R$ in every other state.
\end{itemize}
Given $\bar\alpha\in(0,1)$, define $\alpha\in\Delta(A)$ by
\[
\alpha(a_1)=\cdots=\alpha(a_n)=\frac{\bar\alpha}{n},
\qquad
\alpha(a_{n+1})=1-\bar\alpha.
\]
Given $\bar\lambda\in\mathbb R$, define $\lambda\in\mathbb R^\Theta$ by
\[
\lambda(\theta_1)=\cdots=\lambda(\theta_{n-1})
=\frac{\bar\pi}{n-1}\bar\lambda,
\qquad
\lambda(\theta_n)=0.
\]

Overall, the decision problem $\mathcal D$ is parametrized by
\[
\bar\pi\in(0,1),\quad \rho>0,\quad \sigma\in(0,\rho),
\quad x,y,z\in\mathbb R,
\]
while the candidate saddle point $(\alpha,\lambda)$ is parametrized by
\[
\bar\alpha\in(0,1)
\qquad\text{and}\qquad
\bar\lambda\in\mathbb R.
\]
By construction, $\vert A\vert =n+1$ and $\supp\alpha=A$. It remains to choose the parameter values
so that $(\alpha,\lambda)$ is the unique saddle point of $\mathcal D$.

For $(\alpha,\lambda)$ to be a saddle point, the necessary and sufficient conditions are as follows. For $\lambda$ to be optimal given  $\alpha$, we need:
\begin{align}
\frac{\bar{\alpha}}{n}\psi^{\prime}\left(\rho-\bar{\lambda}\right)+\frac{\bar{\alpha}(n-2)}{n}\psi^{\prime}\left(0-\bar{\lambda}\right)+\frac{\bar{\alpha}}{n}\psi^{\prime}\left(\sigma-\bar{\lambda}\right)+(1-\bar{\alpha})\psi^{\prime}\left(x-\bar{\lambda}\right) & =\psi^{\prime}(0),\label{eq:opt1x}\\
\bar{\alpha}\psi^{\prime}(z)+(1-\bar{\alpha})\psi^{\prime}(y) & =\psi^{\prime}(0).\label{eq:opt2x}
\end{align}
For $\alpha$ to be optimal given $\lambda$, we need:
\begin{align}
\frac{\bar{\pi}}{n-1}\psi\left(\rho-\bar{\lambda}\right)+\frac{\bar{\pi}(n-2)}{n-1}\psi\left(0-\bar{\lambda}\right)+(1-\bar{\pi})\psi(z) & =\bar{\pi}\psi\left(x-\bar{\lambda}\right)+(1-\bar{\pi})\psi(y),\label{eq:opt3x}\\
\bar{\pi}\psi\left(\sigma-\bar{\lambda}\right)+(1-\bar{\pi})\psi(z) & =\bar{\pi}\psi\left(x-\bar{\lambda}\right)+(1-\bar{\pi})\psi(y).\label{eq:opt4x}
\end{align}

We denote by $X$ a non-empty open interval of the real line such that $R_{\psi}$ is strictly monotone on $X$. To simplify the exposition, we assume that $X\cap(-\infty,0)\neq\varnothing$. Similar arguments apply to the case in which $X\cap(0,+\infty)\neq\varnothing$.

We choose $\bar{\lambda}>0$ such that $R_{\psi}$ is strictly monotone
on 
\[
\left(-\bar{\lambda}-\varepsilon,-\bar{\lambda}+\varepsilon\right)
\]
for all $\varepsilon$ sufficiently small. Take $\rho\in (0,\varepsilon)$ and $\sigma\in(0,\rho)$ such that 
\begin{equation}
\frac{1}{n-1}\psi\left(\rho-\bar{\lambda}\right)+\frac{n-2}{n-1}\psi\left(0-\bar{\lambda}\right)=\psi\left(\sigma-\bar{\lambda}\right).\label{eq:first_equality_psix}
\end{equation}
Note that $\sigma$ admits an explicit expression: 
\[
\sigma=\bar{\lambda}+\psi^{-1}\left(\frac{1}{n-1}\psi\left(\rho-\bar{\lambda}\right)+\frac{n-2}{n-1}\psi\left(0-\bar{\lambda}\right)\right).\label{eq:first_equality_psi}
\]
By choosing $\varepsilon$ sufficiently small, we can be sure
that
\[
\rho-\bar{\lambda}<0\quad\text{and}\quad\sigma-\bar{\lambda}<0.
\]

To satisfy \eqref{eq:opt1x}, we impose the restriction that $x>\bar{\lambda}$, and we define $\bar{\alpha}$ by:
\[
\bar{\alpha} = \frac{n\left(\psi^\prime\left(x-\bar{\lambda}\right)-\psi^\prime(0)\right)}{n\psi^\prime\left(x-\bar{\lambda}\right)-\psi^{\prime}\left(\rho-\bar{\lambda}\right)-(n-2)\psi^{\prime}\left(0-\bar{\lambda}\right)-\psi^{\prime}\left(\sigma-\bar{\lambda}\right)}.
\]
Note that, as $x\downarrow \bar{\lambda}$, we have $\bar{\alpha}\downarrow 0$. Next, to satisfy (\ref{eq:opt4x}), we impose the restrictions that $z>0$ and $y<0$, and we define $\bar{\pi}$ by:
\[
\bar{\pi}=\frac{\psi(z)-\psi(y)}{\psi\left(x-\bar{\lambda}\right)+\psi(z)-\psi\left(\sigma-\bar{\lambda}\right)-\psi(y)}.
\]
Then, thanks to (\ref{eq:first_equality_psix}), equation (\ref{eq:opt3x}) is automatically satisfied. Finally, to satisfy \eqref{eq:opt2x}, we define $y$ by: 
\[
    y= (\psi^\prime)^{-1}\left(\frac{\psi^\prime(0) - \bar{\alpha}\psi^\prime(z)}{1-\bar{\alpha}}\right).
\]
Note that $y<0$ is well defined as long as we choose $z$ sufficiently close to zero. Overall, this parameter choice ensures that $(\alpha,\lambda)$ is a saddle point of $\mathcal{D}$.

Now we prove that $(\alpha,\lambda)$ is the unique saddle point of $\mathcal{D}$. We will use the following result:
\begin{claim}
We have:
\begin{equation}\label{eq:key_disequality_psi}
\frac{1}{n-1}\psi^\prime\left(\rho-\bar{\lambda}\right)+\frac{n-2}{n-1}\psi^\prime\left(0-\bar{\lambda}\right) \neq \psi^\prime\left(\sigma-\bar{\lambda}\right).
\end{equation}
\end{claim}
\begin{proof}
Notice that 
\[
\frac{\partial}{\partial t} \psi^{\prime}\left(\psi^{-1}(t)\right) = R_\psi\left(\psi^{-1}(t)\right).
\]
Since $R_\psi$ is strictly monotone on the interval $\left(-\bar{\lambda}-\varepsilon,-\bar{\lambda}+\varepsilon\right)$, the composite function $\psi^{\prime}\circ\psi^{-1}$ is strictly convex or strictly concave on the interval $\left(\psi\left(-\bar{\lambda}-\varepsilon\right),\psi\left(-\bar{\lambda}+\varepsilon\right)\right)$. Then, the desired result follows from applying $\psi^{\prime}\circ\psi^{-1}$ to both sides of \eqref{eq:first_equality_psix}.
\end{proof}
Take any other saddle point $(\beta,l)$. Since $\phi$ is essentially smooth, the Lagrange multiplier is unique: $\lambda=l$ (Footnote \ref{rem:unique_multi}). Next, we verify that $\beta(a_i)=\beta(a_j)$ for all $i,j=1,\ldots,n-1$.
To show this, we can use the optimality conditions for the Lagrange
multiplier in states $\theta_i$ and $\theta_j$, which imply:
\[
\beta\left(a_{i}\right)\left(\psi^{\prime}\left(\rho-\bar{\lambda}\right)-\psi^{\prime}\left(0-\bar{\lambda}\right)\right)=\beta\left(a_j\right)\left(\psi^{\prime}\left(\rho-\bar{\lambda}\right)-\psi^{\prime}\left(0-\bar{\lambda}\right)\right).
\]
We conclude that $\beta(a_i)=\beta(a_j)$.

Next we argue that $\beta\left(a_{n+1}\right)=1-\bar{\alpha}$. This follows immediately
from the optimality condition for the Lagrange multiplier in
state $\theta_n$:
\[
(1-\beta(a_{n+1}))\psi^{\prime}(z)+\beta(a_{n+1})\psi^{\prime}(y)=\psi^{\prime}(0)\quad\Longrightarrow\quad\beta(a_{n+1})=1-\bar{\alpha}.
\]
Hence, we are done with proving uniqueness as soon as we show that $\beta(a_n)=\beta(a_1)$.
To do so, we use the optimality condition for the Lagrange multiplier
in the first state, which, together with \eqref{eq:opt1x}, imply:
\begin{align*}
  & \frac{(n-1)\beta(a_1)}{\bar{\alpha}}\left(\frac{1}{n-1}\psi^{\prime}\left(\rho-\bar{\lambda}\right)+\frac{n-2}{n-1}\psi^{\prime}\left(0-\bar{\lambda}\right)\right)+\frac{\beta(a_n)}{\bar{\alpha}}\psi^{\prime}\left(\sigma-\bar{\lambda}\right)\\
= & \frac{n-1}{n}\left(\frac{1}{n-1}\psi^{\prime}\left(\rho-\bar{\lambda}\right)+\frac{n-2}{n-1}\psi^{\prime}\left(0-\bar{\lambda}\right)\right)+\frac{1}{n}\psi^{\prime}\left(\sigma-\bar{\lambda}\right).
\end{align*}
By \eqref{eq:key_disequality_psi}, we must have $\beta(a_1)=\beta(a_n)$. Hence, $(\alpha,\lambda)$ is the unique saddle point of $\mathcal{D}$.
\end{proof}

\section{Proofs of the results in Section \ref{sec:separable_APU}}

\subsection{Proof of Lemma \ref{lem:symmetric_sol}}

Proposition \ref{pro:symmetry} in Online Appendix \ref{sec:symmetric_decision_problems} (which treats symmetric decision problems more generally) implies that there exists an optimal choice rule $P$ such that, for the associated saddle point $(\alpha, \lambda)$, $\alpha$ is uniform and $P_\pi = \alpha$. The lemma then follows immediately.


\subsection{Proof of Proposition \ref{cor:APU-exchangeable}}

Fix any decision problem with $n = |A|$ \emph{a priori} homogeneous actions. 

 For the ``only if'' direction, let $P$ be any ex-ante symmetric optimal choice rule under Csisz\'ar information (Lemma \ref{lem:symmetric_sol}). By symmetry and condition \eqref{eq:P_separable_case}, $\alpha(a) = 1/n$ and 
\begin{equation}\label{eq:another_opt}
    P_\theta(a) = \frac{1}{n}(\phi^\star)^\prime(a(\theta)-\lambda_\pi(\theta))=\left(\frac{1}{n}\phi^\star\right)^\prime(a(\theta)-\lambda_\pi(\theta))
\end{equation}
for every $a \in A$. Note that
\[
    \left(\frac{1}{n}\phi^\star\right)^\star(t) = \frac{1}{n}\phi(nt)=:\varphi(t)\qquad\forall t\in \RR_+.
\]
 Thus, for every $a \in A$, by standard duality arguments, \eqref{eq:another_opt} implies
 $a(\theta)-\lambda_\pi(\theta)\in \partial \varphi(P_\theta(a))$, which is equivalent to
\[
    P_\theta(a)\in \arg\max_{t\in \mathbb{R_+}} t(a(\theta)-\lambda_\pi(\theta)) - \varphi(t).
\]
 Hence, 
\[
    P_\theta \in \arg\max_{p\in \Delta(A)}\sum_{a\in A}p(a)a(\theta) -\varphi(p(a)).  
\]
 Since the function $c$ defined in \eqref{eqn:APU-exchangeable} satisfies $c(t) = \varphi(t)$ for all $t\in [0,1]$, it follows that $P$ is optimal in the APU model indexed by $c$. (Note that $c$ is a well-defined perturbation function because the Csisz\'ar information transformation $\phi$ satisfies Assumption \ref{ass:phi_strictly_convex}.) 

 For the ``if'' direction, note that the function $c$ in \eqref{eqn:APU-exchangeable} is strictly convex on $\dom c \subseteq [0,1]$, being convex on $[0,1]$ and strictly convex on $\inte(\dom c)$. Therefore, the APU model indexed by $c$ admits a unique optimal choice rule, $P$. Now, let $\mathcal{P} \subseteq \Delta(A)^\Theta$ be the set of choice rules that are ex-ante symmetric and optimal under Csisz\'ar information. Lemma \ref{lem:symmetric_sol} implies $|\mathcal{P}| \geq1$, and the ``only if'' direction proved above implies $\mathcal{P} \subseteq \{P\}$. It follows that $\mathcal{P} = \{P\}$. Hence, $P$ is ex-ante symmetric and optimal under Csisz\'ar information. 

Finally, the last part of the proposition is obtained by setting
\[
\phi(t) = \begin{cases}
n c(\frac{t}{n}) & t\in [0,n],\\
+\infty & \text{otherwise},
\end{cases}
\]
which satisfies Assumption \ref{ass:phi_strictly_convex} by the assumptions on $c$. 

\subsection{Proof of Proposition \ref{prop.ordering.alpha.lambda}}

\textit{(i) implies (ii).} Since $a$ and $b$ are comparable, there is a state $\theta$ such that $a(\theta)=b(\theta)$. Since $a$ is more salient than $b$, we have $P_\theta(a)\geq P_\theta(b)$. It follows from  \eqref{eq:P_separable_case} that 
\begin{equation}\label{eq:salience_one_d}
\alpha(a)(\phi^\star)'(a(\theta) - \lambda_\pi(\theta))\geq \alpha(b)(\phi^\star)'(b(\theta) - \lambda_\pi(\theta))= \alpha(b)(\phi^\star)'(a(\theta) - \lambda_\pi(\theta)).
\end{equation}
Since $\phi$ is essentially smooth, $(\phi^\star)'>0$. It follows from \eqref{eq:salience_one_d} that $\alpha(a)\geq \alpha(b)$.

\textit{(ii) implies (i).} Take a state $\theta$ such that $a(\theta)=b(\theta)$. Since $\alpha(a)\geq \alpha(b)$ and $(\phi^\star)^\prime\geq 0$,
\[
\alpha(a)(\phi^\star)^\prime (a(\theta)-\lambda_\pi(\theta)) \geq \alpha(b)(\phi^\star)^\prime (b(\theta)-\lambda_\pi(\theta)).
\]
It follows from \eqref{eq:P_separable_case} that $P_\theta(a)\geq P_\theta(b)$. We conclude that $a$ is more salient than $b$.

\subsection{Proofs of the results in Section \ref{sec:chi_squared}}
Set $\psi=\phi^\star$. Note that 
\begin{equation}\label{eq:derivative_a_linear}
\psi^\prime(t) = \max\left\{1+\frac{t}{\kappa},0\right\}\qquad\forall t \in \RR.
\end{equation}
For each $\alpha\in \Delta(A)$ and $\lambda_\pi\in \RR^\Theta$, define the Lagrangian
\[
L(\alpha,\lambda_\pi) = \sum_{a\in A}\alpha(a) \sum_{\theta \in \Theta} \pi(\theta)\psi(a(\theta) -\lambda_\pi(\theta)) + \sum_{\theta\in\Theta} \pi(\theta)\lambda_\pi(\theta)
\]

\begin{lemma}\label{prop.kappa.ge.R}
Assume \eqref{eq.kappa.ge.R}. Fix $\alpha\in\Delta(A)$, and let
\[
\lambda_\pi^* \in \arg\min_{\lambda_\pi\in\mathbb R^\Theta} L(\alpha,\lambda_\pi).
\]
Then
\[
a(\theta)-\lambda_\pi^*(\theta)>-\kappa
\]
for every $a\in A$ and every $\theta\in\Theta$.
 \end{lemma}
 \begin{proof}
 Fix $\theta\in \Theta$. Arguing by contradiction, suppose
 \[
    \lambda_\pi^*(\theta) \geq \min_{a\in A} a(\theta)+\kappa.
 \]
 Then, for every $a\in A$,
 \[
    a(\theta) - \lambda^*_\pi(\theta) \leq      a(\theta)-\min_{b\in A} b(\theta)-\kappa.
 \]
 It follows from \eqref{eq.kappa.ge.R} that the right-hand side is negative. Thus,
 \[
    a(\theta) - \lambda_\pi^*(\theta)<0\qquad\forall a\in A.
 \]
 This implies that
 \[
    \psi'(a(\theta) - \lambda_\pi^*(\theta))<1\qquad\forall a\in A.
 \]
We obtain a violation of the first-order condition
 \[
    \sum_{a\in A} \alpha(a)\psi'\left(a(\theta) - \lambda_\pi^*(\theta)\right) = 1.
 \]
 A contradiction.
 \end{proof}

\subsubsection{Proof of Proposition \ref{prop.saddlepoint.quadratic}}

 Fix $\alpha\in\Delta(A)$. 
 Combining Lemma \ref{prop.kappa.ge.R} with the functional form \eqref{eq:derivative_a_linear}, we obtain that
\begin{equation}\label{eqn:psi-prime-chi2}
    \psi^\prime\left(a(\theta)-\lambda_\pi^*(\theta)\right)= 1 + \frac{a(\theta)-\lambda_\pi^*(\theta)}{\kappa}\qquad\forall a\in A,\theta\in \Theta.
\end{equation}
 Hence, the optimality conditions for $\lambda^*_\pi$ in \eqref{eq:lambda_separable_case-1} reduce to
\[
    \sum_{a\in A}\alpha(a)\bigl(a(\theta)-\lambda_\pi^*(\theta)\bigr)= 0\qquad\forall \theta\in \Theta.
\]
 It follows that $\lambda^*_\pi=\sum_{a\in A}\alpha(a)a$. Substituting the expression for $\lambda_\pi^*$ into \eqref{eqn:psi-prime-chi2} and the optimality condition \eqref{eq:P_separable_case} delivers the expression for the optimal choice rule in \eqref{eq:WL-P}. For the quadratic program \eqref{eq.quadratic.program.distance}, substituting the expression for $\lambda_\pi^*$ into the Lagrangian gives
\[
\min_{\lambda_\pi\in \RR^\Theta}L(\alpha,\lambda_\pi)
= \sum_{\theta\in\Theta}\pi(\theta)\left[\sum_{a\in A}\alpha(a)\psi\left(a(\theta)-\sum_{b\in A}\alpha(b)b(\theta)\right)+\sum_{a\in A}\alpha(a)a(\theta)\right].
\]
By \eqref{eq.kappa.ge.R},
\[
\psi\left(a(\theta)-\sum_{b\in A}\alpha(b)b(\theta)\right) = a(\theta)-\sum_{b\in A}\alpha(b)b(\theta)+\frac{1}{2\kappa}\left(a(\theta)-\sum_{b\in A}\alpha(b)b(\theta)\right)^2
\]
for all $a\in A$ and $\theta\in \Theta$. Thus, 
\[
\sum_{a\in A}\alpha(a) \psi\left(a(\theta)-\sum_{b\in A}\alpha(b)b(\theta)\right)=\frac{1}{2\kappa}\sum_{a\in A}\alpha(a)\left(a(\theta)-\sum_{b\in A}\alpha(b)b(\theta)\right)^2
\]
for all $\theta\in \Theta$. Simple algebra shows that
\[
    \sum_{a\in A}\alpha(a)\left(a(\theta)-\sum_{b\in A}\alpha(b)b(\theta)\right)^2 = \frac{1}{2}\sum_{a,b\in A}\alpha(a)\alpha(b)(a(\theta)-b(\theta))^2
\]
for all $\theta\in \Theta$. It follows that 
\begin{align*}
\min_{\lambda_\pi\in \RR^\Theta}L(\alpha,\lambda_\pi)
& = \sum_{\theta\in\Theta}\pi(\theta)\left[\frac{1}{4\kappa}\sum_{a,b\in A}\alpha(a)\alpha(b)(a(\theta)-b(\theta))^2+\sum_{a\in A}\alpha(a)a(\theta)\right]\\
& = \sum_{a \in A} \alpha(a)m_a + \frac{1}{4\kappa}\sum_{a,b\in A}\alpha(a)\alpha(b)D_{ab}.
\end{align*}
The desired result then follows immediately from Theorem \ref{thm:characterization}.

\subsubsection{Proof of Corollary \ref{cor.independent.actions}}
Suppose that actions are uncorrelated. If $a\neq b$, then
\[
D_{ab}=\sum_{\theta\in\Theta}\pi(\theta){(a(\theta)-b(\theta))}^2 = \sigma^2_a+m_a^2+ \sigma^2_b+m_b^2-2m_a m_b.
\]
If instead $a=b$, then $D_{aa}=0$. Thus, 
\begin{align*}
\sum_{a,b\in A}\alpha(a)\alpha(b)D_{ab}
& =\sum_{a\in A}\sum_{b\neq a}\alpha(a)\alpha(b)\left(\sigma^2_a+m_a^2+ \sigma^2_b+m_b^2-2m_a m_b\right)\\
& =2\sum_{a\in A}\alpha(a)(\sigma^2_a+m_a^2)-2\left(\sum_{a\in A}\alpha(a)m_a\right)^2 - 2\sum_{a\in A}\alpha(a)^2\sigma^2_a.
\end{align*}
We obtain that 
\[
\min_{\lambda_\pi\in \RR^\Theta}L(\alpha,\lambda_\pi)
 = \sum_{a \in A} \alpha(a)\frac{2\kappa m_a+m_a^2+\sigma^2_a}{2\kappa} - \frac{1}{2\kappa}\left(\sum_{a\in A} \alpha(a)m_a \right)^2-\frac{1}{2\kappa}\sum_{a\in A}\alpha(a)^2\sigma^2_a.
\]
We consider the problem
\[
\max_{\alpha\in \Delta(A)}\sum_{a \in A} \alpha(a)v_a - \frac{1}{2}\left(\sum_{a\in A} \alpha(a)m_a \right)^2-\frac{1}{2}\sum_{a\in A}\alpha(a)^2\sigma^2_a.
\]
The (necessary and sufficient) first-order condition is: there exists $\gamma\in \RR$ such that
\begin{align*}
v_a - m_a\sum_{b\in A} \alpha(b)m_b-\sigma^2_a\alpha(a) & \leq \gamma\qquad\forall a\in A,\\
v_a - m_a\sum_{b\in A} \alpha(b)m_b-\sigma^2_a\alpha(a) & = \gamma\qquad\forall a\in \supp \alpha,
\end{align*}
where $\gamma$ is the Lagrange multiplier on $\sum_{a\in A}\alpha(a)=1$. The desired result follows by setting $\beta=\sum_{a\in A} \alpha(a)m_a$, and using non-constant actions to divide through by $\sigma^2_a>0$.

\section{Proofs of the results in Section \ref{sec:choice_accuracy}}

\subsection{Proof of Proposition \ref{pro:unique_combinatorial}}

Take $l\in \RR$ as in the statement of the proposition, and define $\lambda_\pi(\theta)=l$ for all $\theta$. Note that $(\alpha,\lambda_\pi)$, with $\alpha$ uniform, is a saddle point of \eqref{eq.lagrangian.phi.information}. 

Now let $(\beta,\mu_\pi)$ be another saddle point of \eqref{eq.lagrangian.phi.information}. Since $\phi$ is essentially smooth, the Lagrange multiplier is unique (Footnote \ref{rem:unique_multi}). Therefore, $\mu_\pi=\lambda_\pi$. For every state $\theta$, let $A_\theta$ be the set of actions that pay $w$ if the realized state is $\theta$. The optimality condition for the Lagrange multiplier in state $\theta$ is:
\[
    \beta(A_\theta)\psi^\prime(w-l)+ (1-\beta(A_\theta))\psi^\prime(-l) = \psi^\prime(0).
\]
 Thus, taking any two states $\theta$ and $\tau$,
\[
    \beta(A_\theta)\left[\psi^\prime(w-l) - \psi^\prime(-l)\right]=\beta(A_\tau)\left[\psi^\prime(w-l)-\psi^\prime(-l)\right].
\]
Since $\psi^\prime$ is strictly increasing and $w>0$, we conclude that $\beta(A_\theta)=\beta(A_\tau)$. Furthermore, because each action pays $w$ in exactly $m$ states, we have:
\[
\sum_{\theta\in \Theta}\beta(A_\theta) = m.
\]
Since $\beta(A_\theta)=\beta(A_\tau)$ for all $\theta,\tau\in \Theta$, we conclude that $\beta(A_\theta)=m/n$ for all $\theta\in \Theta$. The desired result follows.

\subsection{Properties of $\ell_\gamma(w)$}

\begin{lemma}\label{lem:Lagrange_response_f}
 The Lagrange multiplier has the following properties:
\begin{enumerate}
 \item $\ell_\gamma(w)$ is strictly increasing in $w$.
 \item $\ell_\gamma(w)$ is strictly increasing in $\gamma$.
 \item $\ell_\gamma(w)$ is continuous in $w$.
 \item $\ell_\gamma(w)\rightarrow 0$ as $ w \rightarrow 0$.
 \item $\ell_\gamma(w)\rightarrow +\infty$ as $ w \rightarrow +\infty$.
 \item $\ell_\gamma(w)\rightarrow 0$ as $\gamma \rightarrow 0$.
 \item $\ell_\gamma(w)\rightarrow w$ as $\gamma \rightarrow 1$.
\end{enumerate}
\end{lemma}
\begin{proof}

 (i). Suppose $w^1>w^2$. Using the fact that $\psi^\prime$ is strictly increasing, we obtain:
\begin{align*}
    \gamma \psi^\prime\left(w^1-\ell_\gamma (w^2)\right)+(1-\gamma)\psi^\prime\left(-\ell_\gamma(w^2)\right) & > \gamma \psi^\prime\left(w^2-\ell_\gamma(w^2)\right)+(1-\gamma)\psi^\prime\left(-\ell_\gamma(w^2)\right) \\
    & = \gamma \psi^\prime\left(w^1-\ell_\gamma(w^1)\right)+(1-\gamma)\psi^\prime\left(-\ell_\gamma(w^1)\right).
\end{align*}
 We conclude that $\ell_\gamma(w^1)>\ell_\gamma(w^2)$.

 (ii). Suppose $\gamma^1>\gamma^2$. Since $\psi^\prime$ is strictly increasing,
\begin{align*}
    \gamma^1 \psi^\prime\left(w-\ell_{\gamma^2}(w)\right)+(1-\gamma^1)\psi^\prime\left(-\ell_{\gamma^2}(w)\right)& > \gamma^2 \psi^\prime\left(w-\ell_{\gamma^2}(w)\right)+(1-\gamma^2)\psi^\prime\left(-\ell_{\gamma^2}(w)\right) \\
    & = \gamma^1 \psi^\prime\left(w-\ell_{\gamma^1}(w)\right)+(1-\gamma^1)\psi^\prime\left(-\ell_{\gamma^1}(w)\right).
\end{align*}
 We conclude that $\ell_{\gamma^1}(w) > \ell_{\gamma^2}(w)$.

 (iii). Let $(w^m)$ be a sequence of rewards with limit $w$. Each $\ell_\gamma(w^m)$ satisfies $0 \leq \ell_\gamma(w^m) \leq w^m$. Thus, the sequence $(\ell_\gamma(w^m))$ is bounded. Without loss of generality, we can assume it has a limit, $\ell$. For every $m$,
\[
    \gamma \psi^\prime\left(w^m-\ell_\gamma(w^m)\right)+(1-\gamma)\psi^\prime\left(-\ell_\gamma(w^m)\right)=\psi^\prime(0).
\]
 Taking the limit as $m\rightarrow\infty$, we obtain from the continuity of $\psi^\prime$ that:
\[
    \gamma \psi^\prime(w-\ell)+(1-\gamma)\psi^\prime(-\ell)=\psi^\prime(0).
\]
 Since $\ell_\gamma(w)$ is the unique solution of this equation, we conclude that $\ell=\ell_\gamma(w)$.

 (iv). By (i), $\ell_\gamma(w)$ is increasing in $w$. Define $\ell_\gamma(0)=\inf_{w>0}\ell_\gamma(w)$. Since $\ell_\gamma(w)>0$ for all $w$, we have $\ell_\gamma(0)\geq 0$. Furthermore, for every $w>0$,
\[
    \gamma \psi^\prime(w-\ell_\gamma (w))+(1-\gamma)\psi^\prime(-\ell_\gamma (w))=\psi^\prime(0).
\]
 Taking the limit as $w\rightarrow 0$, we obtain from the continuity of $\psi^\prime$ that $\psi^\prime(-\ell_\gamma(0))=\psi^\prime(0)$. We conclude that $\ell_\gamma(0)=0$.

 (v). Define $\bar{\ell}_\gamma=\sup_{w>0}\ell_\gamma(w)$. Arguing by contradiction, suppose $\bar{\ell}_\gamma<+\infty$. Recall that $\psi^\prime(w)\rightarrow +\infty$ as $w\rightarrow+\infty$. Thus, by choosing $w$ sufficiently large, we can ensure that
\[
    \gamma \psi^\prime\left(w-\bar{\ell}_\gamma\right)+(1-\gamma)\psi^\prime\left(-\bar{\ell}_\gamma\right) > \psi^\prime(0).
\]
 This implies that $\ell_\gamma(w)>\bar{\ell}_\gamma$, contradicting  the definition of $\bar{\ell}_\gamma$. We conclude that $\bar{\ell}_\gamma=+\infty$, as desired.

 (vi). Define $\ell_0(w)=\inf_{\gamma\in (0,1)}\ell_\gamma(w)$. Since $\ell_\gamma(w)>0$ for all $\gamma$, we have $\ell_0(w)\geq 0$. Furthermore, for every $\gamma\in (0,1)$,
\[
    \gamma \psi^\prime(w-\ell_\gamma (w))+(1-\gamma)\psi^\prime(-\ell_\gamma (w))=\psi^\prime(0).
\]
 Taking the limit as $\gamma\rightarrow 0$, we obtain from the continuity of $\psi^\prime$ that $\psi^\prime(-\ell_0(w))=\psi^\prime(0)$. We conclude that $\ell_0(w)=0$.

 (vii). Define $\ell_1(w)=\sup_{\gamma\in (0,1)}\ell_\gamma(w)$. Since $\ell_\gamma(w)<w$ for all $\gamma$, we have $\ell_1(w)\leq w$. Furthermore, for every $\gamma\in (0,1)$,
\[
    \gamma \psi^\prime(w-\ell_\gamma (w))+(1-\gamma)\psi^\prime(-\ell_\gamma (w))=\psi^\prime(0).
\]
 Taking the limit as $\gamma\rightarrow 1$, we obtain from the continuity of $\psi^\prime$ that $\psi^\prime(w-\ell_1(w))=\psi^\prime(0)$. We conclude that $\ell_1(w)=w$.
\end{proof}

\subsection{Proof of Proposition \ref{pro:rep_theorem_fixed_g}}

 (i). We have:
\[
    \rho_\gamma(w) = \psi^\prime(0)-(1-\gamma)\psi^\prime(-\ell_\gamma(w)).
\]
 Since $\ell_\gamma(w)$ is strictly increasing in $w$ (Lemma \ref{lem:Lagrange_response_f}) and $\psi^\prime$ is a strictly increasing function, the right-hand side of the above equation is strictly increasing in $w$. We conclude that $\rho_\gamma(w)$ is strictly increasing in $w$.

 (ii). It follows from the continuity of $\ell_\gamma$ (Lemma \ref{lem:Lagrange_response_f}) and of $\psi^\prime$.

 (iii). It follows from the facts that $\ell_\gamma(w)\rightarrow 0$ as $w\rightarrow 0$ (Lemma \ref{lem:Lagrange_response_f}), $\psi^\prime$ is a continuous function, and $\psi^\prime(0)=1$. 

 (iv). We have 
\[
    \lim_{w \rightarrow+\infty}\rho_\gamma(w) = \psi'(0) -(1-\gamma)\lim_{w \rightarrow+\infty}\psi^\prime (-\ell_\gamma(w))=1,
\]
 where we use the fact that $\ell_\gamma(w)\rightarrow+\infty$ as $w\rightarrow+\infty$ (Lemma \ref{lem:Lagrange_response_f}), and the assumption that $\psi^\prime(t)\rightarrow 0$ as $t\rightarrow-\infty$.

 To prove the last part of the proposition, we use a guess-and-verify argument. Let $\rho_\gamma\colon (0,+\infty)\rightarrow (0,1)$ be a strictly increasing, continuous function, with $\rho_\gamma(w)\rightarrow \gamma$ as $w\rightarrow 0$ and $\rho_\gamma(w)\rightarrow 1$ as $w\rightarrow +\infty$. To ease notation, set $\rho_\gamma(0)=\gamma$. The map $\rho_\gamma$ is generated by some $\phi$ if and only if the two functions are related by the following equations:
\begin{align}\label{eq:guess_and_ver1}
    \rho_\gamma(w) & = \gamma \psi^\prime(w-\ell_\gamma(w)),\\
    1- \rho_\gamma(w) & =  (1-\gamma) \psi^\prime(-\ell_\gamma(w)),\label{eq:guess_and_ver2}
\end{align}
 where $\psi=\phi^\star$. We guess a functional form for the Lagrange multiplier:
\[
    \ell_\gamma(w)=w-\frac{w}{1+w}.
\]
 This guess allows us to define $\psi^\prime$ using \eqref{eq:guess_and_ver1} and \eqref{eq:guess_and_ver2} for $t\in (-\infty,1)$:
\[
    \psi^\prime(t)=\begin{cases}
    \frac{1}{\gamma}\rho_\gamma \left(\frac{t}{1-t}\right) &\text{if }t\in [0,1)\\
    \frac{1}{1-\gamma} \left(1-\rho_\gamma \left(\frac{\sqrt{t^2-4t}-t}{2}\right)\right) &\text{if }t< 0.
\end{cases}
\]
 To complete the construction, we define $\psi^\prime(t) = t/\gamma$ for $t\in [1,+\infty)$. 

 Using the properties of $\rho_\gamma$, one can verify that $\psi^\prime$ is strictly increasing and continuous. Moreover, the image of $\psi^\prime$ is $(0,+\infty)$. We also have that $\psi^\prime(0)=1$. Consequently, we can define $\psi$ as follows:
\[
    \psi(t)=\begin{cases}
    \int_0^t\psi^\prime(s)\dd s &\text{if }t\geq 0,\\
    -\int_t^0\psi^\prime(s)\dd s &\text{if }t< 0.
    \end{cases}
\]
 Setting $\phi=\psi^\star$, one can easily verify that $\rho_\gamma$ is the response function generated by $\phi$, with Lagrange multiplier $\ell_\gamma(w)=w-\frac{w}{1+w}$.

 \begin{remark}
     In this construction, the value of $\psi^\prime(t)$ for $t> 1$ is essentially undetermined: any other completion of $\psi^\prime$ that preserves continuity, monotonicity, and full range would work. Hence, for fixed $\gamma$, a continuum of different $\phi = \psi^\star$ can generate the same $\rho_\gamma$.
 \end{remark}

\subsection{Proof of Proposition \ref{pro:identification_resp_f}}
 Suppose $\phi_1$ and $\phi_2$ induce the same response function for all $\gamma$. Let $\ell_{i,\gamma}$, $i \in \{1,2\}$ be the corresponding multiplier. By Lemma \ref{lem:Lagrange_response_f}, $\ell_{i,\gamma}(w) \rightarrow 0$ as $\gamma\rightarrow 0$ and $\ell_{i,\gamma}(w)\rightarrow w$ as $\gamma\rightarrow 1$. Thus, for all $w>0$ and $i\in \{1,2\}$,
\begin{align*}
 \lim_{\gamma\rightarrow 0} \frac{\rho_\gamma (w)}{\gamma} & =\lim_{\gamma\rightarrow 0}\psi^\prime_i (w-\ell_{i,\gamma}(w))=\psi^\prime_i(w),\\
 \lim_{\gamma\rightarrow 1} \frac{1-\rho_\gamma (w)}{1-\gamma} & =\lim_{\gamma\rightarrow 1}\psi^\prime_i (-\ell_{i,\gamma}(w))=\psi^\prime_i(-w),
\end{align*}
where we use the fact that $\psi^\prime_i$ is continuous. Since $\psi^\prime_1(0)=1=\psi^\prime_2(0)$, we obtain that $\psi^\prime_1=\psi^\prime_2$. Given that $\psi_1(0)=0=\psi_2(0)$, $\psi^\prime_1=\psi^\prime_2$ implies $\psi_1=\psi_2$, which in turn implies $\phi_1=\phi_2$.

\subsection{Proof of Proposition \ref{pro:rep_theorem_inv_resp}}
 Starting from
\[
    \frac{\rho_\gamma(w)}{\gamma}  =\psi^\prime(w-\ell_\gamma(w))\quad\text{and}\quad
    \frac{1-\rho_\gamma(w)}{1-\gamma} =\psi^\prime(-\ell_\gamma(w))
\]
 and using the fact that $\phi^\prime = (\psi^\prime)^{-1}$, we obtain:
\begin{equation}\label{eq:response_f_invers_phi}
    w(x,y)=\phi^{\prime}(x)-\phi^\prime(y).
\end{equation}
 It is then immediate to verify that properties (i)--(vi) are satisfied. In addition, the inverse response function identifies $\phi$:
\[
    \phi^{\prime}(x) = \inf_{z\in (0,1)} w(x,z)\quad\text{and}\quad \phi^{\prime}(y) = -\inf_{z\in (1,+\infty)} w(z,y).
\]

 To prove the second part of the proposition, let $(x,y)\mapsto w(x,y)$ be a function that satisfies (i)--(vi). We define $\phi^\prime \colon(0,+\infty)\rightarrow\RR$ by:
\[
\phi^\prime (t) =  
\begin{cases}
\inf_{y\in (0,1)} w(t,y) &\text{if }t>1,\\
0                        &\text{if }t=1,\\
-\inf_{x\in (1,+\infty)} w(x,t) &\text{if }t<1.
\end{cases}
\]
\begin{claim}
The function $\phi^\prime$ satisfies the following properties:
\begin{enumerate}
    \item $\phi^\prime$ is strictly increasing.
    \item $\phi^\prime$ is continuous.
    \item The image of $\phi^\prime$ is $\RR$.
\end{enumerate}
\end{claim}
\begin{proof}
(i). First, take $t>s>1$. We have:
\begin{align*}
    \phi^\prime(t) = \inf_{y\in (0,1)} w(t,y) & = \inf_{y\in (0,1)} w(t,y) - w(s,y) + w(s,y)\\
               & = w(t,1/2) - w(s,1/2) +\inf_{y\in (0,1)} w(s,y)\\
               & > \inf_{y\in (0,1)} w(s,y) = \phi^\prime(s),
\end{align*}
 where we use the facts that $w(t,y) - w(s,y)$ is independent of $y$ and $w(t,y)> w(s,y)$. This also shows that $\phi^\prime(t)>0$ for all $t>1$. Now take $1>t>s$. We have:
\begin{align*}
    -\phi^\prime(t) = \inf_{x\in (1,+\infty)} w(x,t) & = \inf_{x\in (1,+\infty)} w(x,t) - w(x,s) + w(x,s)\\
               & = w(2,t) - w(2,s) +\inf_{x\in (1,+\infty)} w(x,s)\\
               & < \inf_{x\in (1,+\infty)} w(x,s) = - \phi^\prime(s),
\end{align*}
 where we use the facts that $w(x,t) - w(x,s)$ is independent of $x$ and $w(x,t) < w(x,s)$. This also shows that $\phi^\prime(t)<0$ for all $t<1$. We conclude that $\phi^\prime$ is strictly increasing.

(ii). First, we verify continuity at $t>1$:
\begin{align*}
    \lim_{s\rightarrow t} \phi^\prime(s) = \lim_{s\rightarrow t} \left(\inf_{y\in (0,1)} w(s,y) \right )
    &= \lim_{s\rightarrow t} \left(\inf_{y\in (0,1)} w(s,y) -w(t,y)+w(t,y)\right)\\
    & = \lim_{s\rightarrow t} \left( w(s,1/2) -w(t,1/2) + \inf_{y\in (0,1)} w(t,y)\right)\\
    & = \left(\lim_{s\rightarrow t} w(s,1/2) -w(t,1/2) \right)+ \inf_{y\in (0,1)} w(t,y) = \phi^\prime(t),
\end{align*}
    where we use again the facts that $w(t,y) - w(s,y)$ is independent of $y$ and $ w(s,y)\rightarrow w(t,y)$ as $s\rightarrow t$. Next, we verify continuity at $t<1$: 
\begin{align*}
    \lim_{s\rightarrow t} \phi^\prime(s) = - \lim_{s\rightarrow t} \left(\inf_{x\in (1,+\infty)} w(x,s) \right )
    &= -\lim_{s\rightarrow t} \left(\inf_{x\in (1,+\infty)} w(x,s) -w(x,t)+w(x,t)\right)\\
    & = -\lim_{s\rightarrow t} \left( w(2,s) -w(2,t) + \inf_{x\in (1,+\infty)} w(x,t)\right)\\
    & = \left(\lim_{s\rightarrow t} w(2,t)-w(2,s)  \right)-\inf_{x\in (1,+\infty)} w(x,t) = \phi^\prime(t).
\end{align*}
 Now, we verify right-continuity at $t=1$:
\[
    \lim_{s\downarrow 1} \phi^\prime(s) = \inf_{s \in (1,+\infty)} \phi^\prime(s)
                                     = \inf_{s \in (1,+\infty)} \inf_{y\in (0,1)} w(s,y)=0,
\]
 since $\phi'$ is decreasing and $w(s,y)\rightarrow 0$ as $s\rightarrow 1$ and $y\rightarrow 1$. Finally, we verify left-continuity at $t=1$:
\[
\lim_{s\uparrow 1} \phi^\prime(s) = \sup_{s \in (0,1)} \phi^\prime(s)
                                     = \sup_{s \in (0,1)} \left(- \inf_{x\in (1,+\infty)} w(x,s)\right)
                                     = -\inf_{s \in (0,1)}\inf_{x\in (1,+\infty)} w(x,s)=0,
\]
 since $\phi'$ is increasing and $w(s,y)\rightarrow 0$ as $s\rightarrow 1$ and $y\rightarrow 1$. We conclude that $\phi^\prime$ is continuous.

(iii). Since $\phi^\prime$ is continuous, it is enough to show that $\phi^\prime(t)\rightarrow+\infty$ as $t\rightarrow+\infty$ and $\phi^\prime(t)\rightarrow -\infty$ as $t\rightarrow 0$. We have:
\begin{align*}
\lim_{t\rightarrow+\infty}\phi^\prime(t) &=  \lim_{t\rightarrow+\infty} \left(\inf_{y\in (0,1)} w(t,y) \right)\\
& = \lim_{t\rightarrow+\infty} \left(\inf_{y\in (0,1)} w(t,y)- w(2,y) + w(2,y) \right)\\
& = \left(\lim_{t\rightarrow+\infty} w(t,1/2)- w(2,1/2)\right) + \left(\inf_{y\in (0,1)}  w(2,y) \right)=+\infty,
\end{align*}
where we use the facts that  $w(t,y)- w(s,y)$ is independent of $y$ as $w(t,y)\rightarrow+\infty$ as $t\rightarrow+\infty$. Moreover:
\begin{align*}
-\lim_{t\rightarrow 0}\phi^\prime(t) &=  \lim_{t\rightarrow 0} \left(\inf_{x\in (1,+\infty)} w(x,t) \right)\\
& = \lim_{t\rightarrow 0} \left(\inf_{x\in (1,+\infty)} w(x,t)- w(x,1/2) + w(x,1/2) \right)\\
& = \left(\lim_{t\rightarrow 0} w(2,t)- w(2,1/2)\right) + \left(\inf_{x\in (1,+\infty)}  w(x,1/2) \right)=+\infty,
\end{align*}
where we use the facts that  $w(x,t)- w(x,s)$ is independent of $x$ as $w(x,t)\rightarrow+\infty$ as $t\rightarrow 0$. We conclude that the range of $\phi^\prime$ is $\RR$.
\end{proof}

We define $\psi^\prime=(\phi^\prime)^{-1}$. Using the properties of $\phi^\prime$, one can verify that $\psi^\prime$ is strictly increasing and continuous. Moreover, the image of $\psi^\prime$ is $(0,+\infty)$. We also have that $\psi^\prime(0)=1$. Consequently, we can define $\psi$ as follows:
\[
\psi(t)=\begin{cases}
\int_0^t\psi^\prime(s)\dd s &\text{if }t\geq 0,\\
-\int_t^0\psi^\prime(s)\dd s &\text{if }t< 0.
\end{cases}
\]
Setting $\phi=\psi^\star$, we observe that $\phi^\prime$ is the derivative of $\phi$ on $(0,+\infty)$.

Let $(x,y)\mapsto \tilde{w}(x,y)$ be the inverse response function generated by $\phi$. As shown in \eqref{eq:response_f_invers_phi}, 
\[
\tilde{w}(x,y) = \phi^\prime(x)-\phi^\prime(y) = \inf_{t\in (0,1)} w(x,t)  +  \inf_{s\in (1,+\infty)} w(s,y).
\]
We obtain: for all $s^\prime$,
\begin{align*}
\tilde{w}(x,y) & = \inf_{t\in (0,1)} \left(w(x,t) -w(x,y)+w(x,y)\right)  +  \inf_{s\in (1,+\infty)} w(s,y)\\
& = \inf_{t\in (0,1)} \left(w(s^\prime,t) -w(s^\prime,y)+w(x,y)\right)  +  \inf_{s\in (1,+\infty)} w(s,y)\\
& = w(x,y) -w(s^\prime,y) + \inf_{t\in (0,1)} w(s^\prime,t)  +  \inf_{s\in (1,+\infty)} w(s,y),
\end{align*}
where we use the fact that $w(x,t) -w(x,y)$ is independent of $x$. It follows that:
\[
\tilde{w}(x,y) + \inf_{s^\prime\in (1,+\infty)} w(s^\prime,y)
 = w(x,y) + \inf_{s^\prime\in (1,+\infty)}\inf_{t\in (0,1)} w(s^\prime,t)  +  \inf_{s\in (1,+\infty)} w(s,y).
\]
Since $\inf_{s^\prime\in (1,+\infty)}\inf_{t\in (0,1)} w(s^\prime,t)=0$, we conclude that $\tilde{w}(x,y)=w(x,y)$. Thus, $(x,y)\mapsto w(x,y)$ is the inverse response function generated by $\phi$.

\subsection{Proof of Proposition \ref{pro:arrow-p-appr}}

 Throughout the proof, we allow $w$ to take negative values. Thus, fix $w\in \RR$. By symmetry arguments (see Proposition \ref{pro:unique_combinatorial}), the maxmin problem \eqref{eq.lagrangian.phi.information} for the combined decision problem $(\pi,A)$ admits a saddle point $(\alpha^w,\lambda^w_\pi)\in \Delta(A)\times \RR^\Theta$ such that
\[
\alpha^w(a_1,a_2)=\alpha^w(a_1,b_2)\quad\text{and}\quad \lambda_\pi^w(\theta_1,\theta_2)=\lambda_\pi^w(\theta_1,\tau_2)
\]
for all $a_1\in A_1$, $a_2,b_2\in A_2$, $\theta_1\in \Theta_1$, and $\theta_2,\tau_2\in \Theta_2$. Let $P^w\in \Delta(A)^\Theta$ be the corresponding stochastic choice rule. 

For short, we define $\alpha_1^w\in \Delta(A_1)$ and $\lambda_{\pi_1}^w\in \RR^{\Theta_1}$ by
\[
\alpha_1^w(a_1) = \sum_{a_2\in A_2} \alpha^w(a_1,a_2)\quad\text{and}\quad\lambda_{\pi_1}^w(\theta_1)=\frac{1}{n}\sum_{\theta_2\in \Theta_2}\lambda^w_\pi(\theta_1,\theta_2).
\]
Then, for all $(a_1,a_2)$ and $(\theta_1,\theta_2)$,
\[
\alpha^w(a_1,a_2) = \frac{1}{n}\alpha_1^w(a_1)\quad\text{and}\quad \lambda_\pi^w(\theta_1,\theta_2) = \lambda_{\pi_1}^w(\theta_1).
\]
Note that $(\alpha_1^0,\lambda_{\pi_1}^0)$ is a saddle point of the maxmin problem \eqref{eq.lagrangian.phi.information} for $(\pi_1,A_1)$. Thus, $\lambda_{\pi_1}^0$ is the Lagrange multiplier associated with $(\pi_1,A_1)$.\footnote{Since $\phi$ is essentially smooth, the Lagrange multiplier is unique (Footnote \ref{rem:unique_multi}).}

\begin{lemma}
The function $w\mapsto \lambda_{\pi_1}^w$ is directionally differentiable.
\end{lemma}
\begin{proof}
For every $r,w\in \RR$, define
\[
\begin{aligned}
\varphi(r,w) &= \gamma \psi(r+w)+(1-\gamma)\psi(r),\\
\varphi^{\prime\prime}(r,w) &= \gamma \psi^{\prime\prime}(r+w)+(1-\gamma)\psi^{\prime\prime}(r).
\end{aligned}
\]
The pair $(\alpha_1^w,\lambda_{\pi_1}^w)$ is a saddle point of the reduced maxmin problem
\[
\max_{\alpha_1\in \Delta(A_1)}\min_{\ell\in \RR^{\Theta_1}} \sum_{a_1\in A_1}\alpha_1(a_1)\sum_{\theta_1\in \Theta_1}\pi_1(\theta_1) \varphi(a_1(\theta_1)-\ell(\theta_1),w) + \sum_{\theta_1\in \Theta_1}\pi_1(\theta_1)\ell(\theta_1).
\]
Hence, $\lambda_{\pi_1}^w$ is a solution of the following minimization problem:
\begin{equation}\label{eq:unconstrained_min}
\min_{\ell\in \RR^{\Theta_1}}\left\{\max_{a_1\in A_1} \sum_{\theta_1\in \Theta_1}\pi_1(\theta_1) \varphi(a_1(\theta_1)-\ell(\theta_1),w) + \sum_{\theta_1\in \Theta_1}\pi_1(\theta_1)\ell(\theta_1)\right\}.
\end{equation}
Introducing an epigraph variable $t\in \RR$, we rewrite \eqref{eq:unconstrained_min} as a parametrized nonlinear program in the form used by \citet[Theorem 1]{ralph1995directional}. Define
\[
\begin{aligned}
F(\ell,t,w) & = t +\sum_{\theta_1\in \Theta_1}\pi_1(\theta_1)\ell (\theta_1),\\
G_{a_1}(\ell,t,w) & = \sum_{\theta_1\in \Theta_1}\pi_1(\theta_1) \varphi(a_1(\theta_1)-\ell(\theta_1),w)-t.
\end{aligned}
\]
Then \eqref{eq:unconstrained_min} is equivalent to 
\[
\begin{aligned}
\min_{\ell \in \RR^{\Theta_1},t\in \RR} \quad & F(\ell,t,w)\\
\text{s.t.} \qquad & G_{a_1}(\ell,t,w)\leq 0,\quad a_1\in A_1.
\end{aligned}
\]

We now verify assumptions (A1)--(A3) of \citet[Theorem 1]{ralph1995directional}. Assumption (A1), the required smoothness condition, follows directly from the twice continuous differentiability of $\psi$, and hence of $\varphi$.

For Assumption (A2), the Mangasarian--Fromovitz constraint qualification, it is enough to exhibit a direction along which all active constraints strictly decrease. Consider the direction that leaves $\ell$ unchanged and increases $t$ at unit speed. For every $a_1\in A_1$,
\[
\nabla_{(\ell,t)}G_{a_1} (\ell,t,w)\cdot(
\boldsymbol{0},1)
=
\frac{\partial G_{a_1}}{\partial t}(\ell,t,w)
=
-1<0.
\]
Thus the Mangasarian--Fromovitz constraint qualification holds.

It remains to verify Assumption (A3), the second-order sufficient condition. The Lagrangian is
\[
L(\ell,t,w;\eta)
=
F(\ell,t,w)
+
\sum_{a_1\in A_1}\eta(a_1)G_{a_1}(\ell,t,w),
\]
where $\eta(a_1)\geq 0$ is the multiplier on the constraint $G_{a_1}(\ell,t,w)\leq 0$. Since $L$ is affine in $t$, its Hessian with respect to $(\ell,t)$ has the block form
\[
\nabla^2_{(\ell,t)(\ell,t)}L(\ell,t,w;\eta)
=
\begin{pmatrix}
\nabla^2_{\ell\ell}L(\ell,t,w;\eta) & \boldsymbol 0\\
\boldsymbol 0^\top & 0
\end{pmatrix}.
\]
Moreover, $\nabla^2_{\ell\ell}L(\ell,t,w;\eta)$ is diagonal, with entries
\[
\frac{\partial^2 L}{\partial \ell(\theta_1)^2}(\ell,t,w;\eta)
=
\pi_1(\theta_1)
\sum_{a_1\in A_1}\eta(a_1)
\varphi^{\prime\prime}(a_1(\theta_1)-\ell(\theta_1),w).
\]

Fix a multiplier vector $\eta$ satisfying the KKT conditions. Stationarity with respect to $t$ gives
\[
0=\frac{\partial L}{\partial t}(\ell,t,w;\eta)
=
1-\sum_{a_1\in A_1}\eta(a_1),
\]
and therefore $\eta\in \Delta(A_1)$. In particular, $\eta(a_1)>0$ for at least one $a_1\in A_1$. Since $\psi''>0$, we have $\varphi''>0$. Hence, for every $\theta_1\in\Theta_1$,
\[
\frac{\partial^2 L}{\partial \ell(\theta_1)^2}(\ell,t,w;\eta)>0.
\]

Now take a nonzero direction $(x,s)\in \RR^{\Theta_1}\times \RR$ in the critical subspace associated with the constraints that have positive multipliers. Thus, for every $a_1\in A_1$ with $\eta(a_1)>0$,
\[
\nabla_{(\ell,t)}G_{a_1}(\ell,t,w)\cdot (x,s)
=
\sum_{\theta_1\in \Theta_1}
\frac{\partial G_{a_1}}{\partial \ell(\theta_1)}(\ell,t,w)x(\theta_1)
+
\frac{\partial G_{a_1}}{\partial t}(\ell,t,w)s
=
0.
\]
If $x=\boldsymbol{0}$, then the preceding equality implies $s=0$ for every constraint with positive multiplier, because $\partial G_{a_1}/\partial t=-1$. This contradicts $(x,s)\neq 0$. Therefore $x\neq \boldsymbol{0}$. It follows that
\[
(x,s)^\top
\nabla^2_{(\ell,t)(\ell,t)}L(\ell,t,w;\eta)
(x,s)
=
\sum_{\theta_1\in \Theta_1}
\frac{\partial^2 L}{\partial \ell(\theta_1)^2}(\ell,t,w;\eta)
x(\theta_1)^2
>
0.
\]
Thus Assumption (A3) holds.

By \citet[Theorem 1]{ralph1995directional}, the solution map of the parametrized program is directionally differentiable in $w$. Since the $\ell$-component of the solution is $\lambda_{\pi_1}^w$, the map $w\mapsto \lambda_{\pi_1}^w$ is directionally differentiable.
\end{proof}

We are ready to conclude the proof of the proposition. Fix $\theta=(\theta_1,\theta_2)$ and $a_1$. Define 
\[
    T(w)
    = \log\frac{\gamma}{1-\gamma} + \log \frac{\psi^\prime\left(w+a_1(\theta_1)-\lambda_{\pi_1}^w(\theta_1)\right)}{\psi^\prime\left(a_1(\theta_1)-\lambda_{\pi_1}^w(\theta_1)\right)}.
\]
Since $w\mapsto \lambda_\pi^w$ is directionally differentiable, $w\mapsto T(w)$ admits left and right derivatives. In particular, 
\[
    T(w) =  T(0) + wT^\prime_+(0) + o(w)\qquad \text{as }w\downarrow 0.
\]
Notice that
$
    T(0) = \log\frac{\gamma}{1-\gamma}.
$
Moreover,
\[
    T^\prime_+(0) =\frac{\psi^{\prime\prime}(a_1(\theta_1)-\lambda_{\pi_1}^0(\theta_1))}{\psi^\prime(a_1(\theta_1)-\lambda_{\pi_1}^0(\theta_1))} = R_\psi(a_1(\theta_1)-\lambda_{\pi_1}^0(\theta_1)).
\]
If $w$ is such that $P_\pi^{w}(a_1) > 0$, then $a_1$ is in the support of the $f$-mean of $P^{w}$. It then follows from Theorem \ref{thm:characterization} that 
\[
    \log\frac{P^{w}_{\theta}(\{a_2:a_2(\theta_2) = w\} \vert a_1)}{ P^{w}_{\theta}(\{a_2 : a_2(\theta_2) = 0\} \vert a_1)} = \log\frac{P^{w}_{\theta}(\{a_1\}\times \{a_2:a_2(\theta_2) = w\})}{ P^{w}_{\theta}(\{a_1\}\times \{a_2 : a_2(\theta_2) = 0\})} = T(w).
\]
The desired result follows.

\subsection{Proofs of the results in Section \ref{sec:second_order_p}}\label{app:proofs-second_order_p}

 We begin by deriving properties of the Lagrange multiplier. With respect to Lemma \ref{lem:Lagrange_response_f}, we use the additional hypotheses that $\psi$ is thrice continuously differentiable and $\psi''>0$.

\begin{claim}
 The Lagrange multiplier $\ell_\gamma(w)$ is twice continuously differentiable in $w$. Moreover:
\begin{enumerate}
    \item For all $w\in (0,+\infty)$, $\ell^\prime_\gamma(w)\in (0,1)$.
    \item $\ell^\prime_\gamma(w)\rightarrow 0 $ as $\gamma\rightarrow 0$.
    \item $\ell^\prime_\gamma(w)\rightarrow 1 $ as $\gamma\rightarrow 1$.
    \item For all $w\in (0,+\infty)$, 
\begin{equation}\label{eq:second_d_lambda}
    \ell^{\prime\prime}_\gamma(w)= R_{\psi^\prime}(w-\ell_\gamma(w))\ell^\prime_\gamma(w)(1-\ell^\prime_\gamma(w))^2+R_{\psi^\prime}(-\ell_\gamma(w))(\ell^\prime_\gamma(w))^2(1-\ell^\prime_\gamma(w)).
\end{equation}
\end{enumerate}
\end{claim}
\begin{proof}
 By the implicit function theorem,
\[
    \ell^\prime_\gamma(w) = \frac{\gamma \psi^{\prime\prime}(w-\ell_\gamma(w))}{\gamma \psi^{\prime\prime}(w-\ell_\gamma(w))+(1-\gamma) \psi^{\prime\prime}(-\ell_\gamma(w))}.
\]
 Since $\psi^{\prime\prime}>0$, we deduce that $\ell^\prime_\gamma(w)\in (0,1)$. In addition, because $\ell_\gamma(w)\rightarrow 0 $ as $\gamma\rightarrow 0$ and $\ell_\gamma(w)\rightarrow w $ as $\gamma\rightarrow 1$, we obtain that $\ell^\prime_\gamma(w)\rightarrow 0 $ as $\gamma\rightarrow 0$ and $l^\prime_\gamma(w)\rightarrow 1 $ as $\gamma\rightarrow 1$. Finally, differentiating $\ell^\prime_\gamma(w)$ in $w$, we obtain the desired formula for $\ell^{\prime\prime}_\gamma(w)$ after some elementary algebraic manipulation. 
\end{proof}

Next, we compute the Arrow-Pratt coefficient of the response function.  We have:
\begin{align*}
 \rho_\gamma(w) & = \gamma \psi^\prime(w-\ell_\gamma(w)),\\
 \rho^\prime_\gamma(w) & = \gamma \psi^{\prime\prime}(w-\ell_\gamma(w))(1-\ell^\prime_\gamma(w)),\\
 \rho^{\prime\prime}_\gamma(w) & = \gamma \psi^{\prime\prime\prime}(w-l_\gamma(w))(1-\ell^\prime_\gamma(w))^2-\gamma \psi^{\prime\prime}(w-\ell_\gamma(w))\ell^{\prime\prime}_\gamma(w).
\end{align*}
We obtain:
\begin{align}
    R_{\rho_\gamma}(w) =\frac{\rho^{\prime\prime}_\gamma(w)}{\rho^\prime_\gamma(w)} & = R_{\psi^\prime}(w-\ell_\gamma(w))(1-\ell^\prime_\gamma(w))-\frac{\ell^{\prime\prime}_\gamma(w)}{1-\ell^\prime_\gamma(w)}\nonumber\\
    & =  R_{\psi^\prime}(w-\ell_\gamma(w))(1-\ell^\prime_\gamma(w))^2-R_{\psi^\prime}(-\ell_\gamma(w))(\ell^\prime_\gamma(w))^2,\label{eq:prudence_index_who}
\end{align}
where the last line uses the formula for $\ell^{\prime\prime}_\gamma(w)$. In addition, we observe:
\begin{align}
    R_{\psi^\prime}(w) & = \lim_{\gamma\rightarrow 0}  R_{\rho_\gamma}(w),\label{eq:limit_g_zero}\\
    R_{\psi^\prime}(-w) & =- \lim_{\gamma\rightarrow 1}  R_{\rho_\gamma}(w),\label{eq:limit_g_one}
\end{align}
 where we apply the properties $\ell^\prime_\gamma(w)\rightarrow 0$ as $\gamma \rightarrow 0$, and $\ell^\prime_\gamma(w)\rightarrow 1$ as $\gamma \rightarrow 1$. Finally, using the formula $\phi^\prime=(\psi^\prime)^{-1}$, we obtain that for all $t\in (0,+\infty)$,
\begin{equation}\label{eq:prudence_primal_dual}
R_{\phi^\prime} (t)= \frac{\phi'''(t)}{\phi''(t)} = - \frac{R_{\psi^\prime}(\phi^\prime(t))}{\psi^{\prime\prime}(\phi^\prime(t))}.
\end{equation}

\begin{proof}[Proof of Proposition \ref{pro:response_f_concave}]
 \emph{(i) implies (ii).} Suppose $\rho_\gamma$ is concave for all $\gamma$, namely, $R_{\rho_\gamma}\leq 0$ for all $\gamma$. Then, (ii) follows from \eqref{eq:limit_g_zero} and \eqref{eq:limit_g_one}.

 \emph{(ii) implies (i).} Suppose $R_{\psi^\prime}(t)\leq 0$ for $t>0$, and  $R_{\psi^\prime}(t)\geq 0$ for $t<0$. Using \eqref{eq:prudence_index_who}, we have:
\[
    R_{\rho_\gamma}(w) =  R_{\psi^\prime}(w-\ell_\gamma(w))(1-\ell^\prime_\gamma(w))^2-R_{\psi^\prime}(-\ell_\gamma(w))(\ell^\prime_\gamma(w))^2 \leq 0,
\]
 since $\ell_\gamma(w)\in (0,w)$ and $\ell^\prime_\gamma(w)\in (0,1)$. This proves that $\rho_\gamma$ is concave.

 \emph{(ii) if and only if (iii).} We obtain from \eqref{eq:prudence_primal_dual} that:
\[
    R_{\phi^\prime} (t)\geq 0 \quad \Longleftrightarrow\quad R_{\psi^\prime}(\phi^\prime(t))\leq 0.
\]
 The equivalence of (ii) and (iii) follows from the fact that $t\geq 1$ if and only if $\phi^\prime(t) \geq 0$.
\end{proof}




 Next we prove Proposition \ref{pro:S_shape_resp}, in successive claims.

 \begin{claim}
 If $R_{\rho_\gamma}$ is decreasing for all $\gamma$, then $R_{\psi^\prime}$ is decreasing.
\end{claim}
\begin{proof}
 Suppose $w_1 \geq w_2 > 0$. Since  $R_{\rho_\gamma}$ is decreasing, 
\[
    R_{\rho_\gamma} (w_1) \leq R_{\rho_\gamma}(w_2).
\]
 Taking the limit as $\gamma\rightarrow 0$, we obtain from \eqref{eq:limit_g_zero} that:
\[
    R_{\psi^\prime} (w_1) \leq R_{\psi^\prime}(w_2).
\]
 Suppose now that $0 > w_1 \geq w_2$. Since $R_{\rho_\gamma}$ is decreasing, 
\[
    -R_{\rho_\gamma} (-w_1) \leq - R_{\rho_\gamma}(-w_2).
\]
 Taking the limit as $\gamma\rightarrow 1$, we obtain from \eqref{eq:limit_g_one} that:
\[
    R_{\psi^\prime} (w_1) \leq R_{\psi^\prime}(w_2).
\]
 Finally, suppose $w_1\geq 0\geq w_2$. For all $w_3\in (0,w_1]$,  $R_{\psi^\prime} (w_1) \leq R_{\psi^\prime}(w_3)$. By continuity, $R_{\psi^\prime} (w_1) \leq R_{\psi^\prime}(0)$. Analogously, for all $w_4\in [w_2,0)$, $R_{\psi^\prime} (w_4) \leq R_{\psi^\prime}(w_2)$. By continuity, $R_{\psi^\prime} (0) \leq R_{\psi^\prime}(w_2)$. We deduce that $R_{\psi^\prime} (w_1) \leq R_{\psi^\prime}(w_2)$. Thus $R_{\psi'}$ is decreasing.
\end{proof}

\begin{claim}\label{claim:prudence-convex}
 If $R_{\psi^\prime}$ is decreasing, then $\psi'$ is convex. 
\end{claim}
\begin{proof}
    Note that $\psi'$ is convex if and only if $\psi'''\geq 0$. Therefore, because $\psi''>0$ and $R_{\psi'} = \psi'''/\psi''$, it suffices to show that $R_{\psi'} \geq0$. We prove this by contradiction. 
    
    Suppose there exist $t_0 \in \RR$ and $a >0$ such that $R_{\psi'}(t_0) = -a <0$. Since $R_{\psi'}$ is decreasing, it follows that $R_{\psi'}(t)\leq -a$ for every $t\geq t_0$. Therefore, we have
\[
\log \psi''(t)-\log \psi''(t_0)
=
\int_{t_0}^{t}R_{\psi'}(s)\,ds
\leq -a(t-t_0) 
\qquad \forall t\geq t_0.
\]
Equivalently, exponentiating both sides delivers
\[
\psi''(t)
\leq
\psi''(t_0)e^{-a(t-t_0)}
\qquad \forall t\geq t_0.
\]
Consequently,
\[
\int_{t_0}^{+\infty}\psi''(t)\,dt
\leq
\psi''(t_0)\int_{t_0}^{+\infty}e^{-a(t-t_0)}\,dt
=
\frac{\psi''(t_0)}{a}
<+\infty.
\]
It follows that
\[
\lim_{t\to+\infty}\psi'(t)
=
\psi'(t_0)+\int_{t_0}^{+\infty}\psi''(s)\,ds
<+\infty.
\]
Meanwhile, the assumption that $\phi = \psi^\star$ is essentially smooth implies that $\psi'(\RR) = (0,+\infty)$, and hence  $\lim_{t\to+\infty}\psi'(t)=+\infty$. This delivers the desired contradiction.
\end{proof}

\begin{claim}
 If $R_{\psi^\prime}$ is decreasing, then $R_{\rho_\gamma}$ is decreasing for all $\gamma$.
\end{claim}
\begin{proof}
 To ease the exposition, we drop the subscript $\gamma$. Suppose $R_{\psi^\prime}$ is decreasing. Per Claim \ref{claim:prudence-convex}, $\psi^\prime$ is convex, which is equivalent to $R_{\psi^\prime}\geq 0$. If $w_1\geq w_2$, then 
\begin{align*}
    R_{\rho}(w_1) & = R_{\psi^\prime}(w_1-\ell(w_1))(1-\ell^\prime(w_1))^2-R_{\psi^\prime}(-\ell(w_1))(\ell^\prime(w_1))^2\\
    & \leq R_{\psi^\prime}(w_2-\ell(w_2))(1-\ell^\prime(w_2))^2-R_{\psi^\prime}(-\ell(w_2))(\ell^\prime(w_2))^2 = R_{\rho}(w_2),
\end{align*}
 where we use the facts that $w-\ell(w)$ and $\ell(w)$ are increasing in $w$, $R_{\psi^\prime}$ is decreasing and non-negative, and $\ell^\prime(w)$ is increasing in $w$---see \eqref{eq:second_d_lambda}. Thus, $R_\rho$ is decreasing.
%
\end{proof}


\begin{claim}
 If $R_{\psi^\prime}$ is decreasing, then $R_{\phi^\prime}$ is increasing.
\end{claim}
\begin{proof}
Suppose $t_1\geq t_2$. Using  \eqref{eq:prudence_primal_dual}, we have
\[
    R_{\phi^\prime} (t_1) \geq R_{\phi^\prime} (t_2)\quad\Longleftrightarrow\quad  R_{\psi^\prime}(\phi^\prime(t_1))\psi^{\prime\prime}(\phi^\prime(t_2))\leq R_{\psi^\prime}(\phi^\prime(t_2))\psi^{\prime\prime}(\phi^\prime(t_1)).
\]
 Recall that $\phi^\prime$ is increasing. Thus, since $R_{\psi^\prime}$ is decreasing, 
\[
    R_{\psi^\prime}(\phi^\prime(t_1))\leq R_{\psi^\prime}(\phi^\prime(t_2)).
\]
Since $\psi^\prime$ is convex by Claim \ref{claim:prudence-convex}, $ R_{\psi^\prime}\geq 0$ and $\psi^{\prime\prime}$ is increasing. Thus, it follows that
\[
    R_{\psi^\prime}(\phi^\prime(t_1))\psi^{\prime\prime}(\phi^\prime(t_2))\leq R_{\psi^\prime}(\phi^\prime(t_2))\psi^{\prime\prime}(\phi^\prime(t_1)),
\]
as desired.
\end{proof}

\section{Proofs of the results in Section \ref{sec:nested-entropies}}

\subsection{Proof of Proposition \ref{cor:MASE}}

Fix $p\in\Delta(\Theta)$, and let $\mu\in\Delta(N\times \Theta)$ be consistent with $p$. We show that
\begin{equation}\label{eq:finiteness}
\zeta D_{\mathrm{KL}}(\bar\mu\Vert \bar\nu)
+
\eta\sum_{i\in N}\bar\mu(i)D_{\mathrm{KL}}(\mu_i\Vert \nu_i)
<+\infty
\end{equation}
if and only if 
\[
\bar\mu(i)=p(B_i)
\qquad\text{and}\qquad
\mu_i(\theta)=p(\theta\vert B_i)
\]
for every $i\in N$ with $p(B_i)>0$ and every $\theta\in B_i$. The desired result then follows immediately. The ``if'' direction is trivial; we prove the ``only if'' direction.

Consider first the case in which $\bar\mu(i)>0$. By \eqref{eq:finiteness}, $D_{\mathrm {KL}}(\mu_i\Vert \nu_i)$ is finite. Hence, $\mu_i$ is absolutely continuous with respect to $\nu_i$: $\mu_i(B_i)=1$ and $\mu_i(B_j)=0$ for all $j\neq i$. Thus,
\[
p(\theta) = \sum_{j\in N}\mu_j(\theta)\bar{\mu}(j)=\mu_i(\theta)\bar{\mu}(i)\qquad\forall\theta\in B_i.
\]
Summing over $\theta\in B_i$, we conclude that $p(B_i)=\bar{\mu}(i)$. It follows that $p(\theta\vert B_i)=\mu_i(\theta)$.

It remains to prove that $p(B_i)>0$ implies $\bar\mu(i)>0$. Suppose $p(B_i)>0$. If $j\neq i$, then either $\bar{\mu}(j)=0$, or $\bar{\mu}(j)>0$ and, reasoning as above, $\mu_j(B_i)=0$. Thus, $p(B_i) = \mu_i(B_i)\bar{\mu}(i)$ by consistency between $p$ and $\mu$. This shows that $\bar\mu(i)>0$, as desired.

\subsection{Proof of Proposition \ref{prop:nested_logit}}
 
 With slight abuse of notation, we identify $H_\pi$ with its extension to $\mathbb R^\Theta$ defined by setting it equal to $+\infty$ outside $\Delta(\Theta)$. For each $i\in N$, define 
 $F_i \colon \RR^{\Theta} \to \RR$ as
\[	F_i(x)=\eta\log\sum_{\theta\in\Theta}\nu_i(\theta)e^{x(\theta)/\eta}.
\]
 Define, in addition, $G \colon \RR^N \to \RR$ as
\[
    G(y)=\zeta\log\sum_{i\in N}\bar\nu(i)e^{y(i)/\zeta}
\]
 The function $G$ is finite, convex, continuous, and increasing, and each $F_i$ is finite, convex, and continuous. 

 The standard log-sum-exp conjugacy formula gives
 $G^\star(\alpha) = \zeta D_{\mathrm{KL}}(\alpha\Vert\bar\nu)$ for
$\alpha\in\Delta(N)$, and $G^\star(\alpha)=+\infty$ otherwise. Similarly, $F_i^\star(m)=\eta D_{\mathrm{KL}}(m\Vert \nu_i)$ for $m\in\Delta(\Theta)$, and $F_i^\star(m)=+\infty$ otherwise.

 Theorem 2 in \cite{hiriart2006note} yields, for every $q\in\mathbb R^\Theta$,
\[
(G\circ F)^\star(q) =
\min_{\substack{\alpha_i\ge 0,\ q_i\in\mathbb R^\Theta\\
\sum_{i \in N} q_i=q}}
\left\{
G^\star(\alpha)+
\sum_{i\in N}\alpha_i F_i^\star\left(q_i/\alpha_i\right)
\right\},
\]
 with the convention that when $\alpha_i=0$, the term $\alpha_iF_i^\star(q_i/\alpha_i)$ is defined as $0$ when $q_i=0$, and $+\infty$ otherwise.

 Thus, finiteness of the objective is equivalent to the existence of a joint distribution $\mu\in\Delta(N\times\Theta)$ with marginal $\bar\mu=\alpha$, conditional beliefs $\mu_i=q_i/\alpha_i$ when $\alpha_i>0$, and marginal
 $\sum_i\mu(i,\theta)=q(\theta)$ on $\Theta$. Hence the preceding
 minimum is $+\infty$ if $q\notin\Delta(\Theta)$, while, for
 $q\in\Delta(\Theta)$, it becomes
 \[
    \min_{\substack{\mu\in\Delta(N\times\Theta)\\
    \sum_i\mu(i,\theta)=q(\theta)}}
    \left\{
    \zeta D_{\mathrm{KL}}(\bar\mu\|\bar\nu)
    +
    \eta\sum_{i\in N}\bar\mu(i)D_{\mathrm{KL}}(\mu_i\|\nu_i)
    \right\}.
\]
 By definition of nested Shannon entropy, this is precisely
$H_\pi(q)$. Therefore $(G\circ F)^\star= H_\pi$.

 Finally, $G\circ F$ is finite, convex, and continuous on $\mathbb R^\Theta$, so Fenchel-Moreau biconjugacy gives
$H_\pi^\star = ((G\circ F)^\star)^\star = G \circ F$. We obtain
\[
    H_\pi^\star(x) =
    \zeta\log\left(
    \sum_{i\in N}\bar\nu(i)
    \left(
    \sum_{\theta\in\Theta}\nu_i(\theta)e^{x(\theta)/\eta}
    \right)^{\eta/\zeta}
    \right),
\]
as claimed.

\subsection{Proof of Proposition \ref{pro:minimizers_nested_s}}

 If $\nu_i(\theta)=0$, then $\mu_i(\theta)=0$ at the optimum in \eqref{eqn:nested-shannon-entropy}; otherwise, $D_{\mathrm{KL}}(\mu_i\Vert \nu_i)=+\infty$. For the case $\nu_i(\theta)>0$, the (necessary and sufficient) first-order condition for the minimization problem \eqref{eqn:nested-shannon-entropy} is
\[
    \zeta\log \frac{\bar\mu(i)}{\bar \nu (i)} + \eta \log \frac{\mu_i(\theta)}{\nu_i(\theta)}+\zeta=\xi(\theta),
\]
 where $\xi(\theta)$ is a Lagrange multiplier on the constraint $\sum_{i\in N}\mu(\theta,i)=p(\theta)$. Substituting $\bar\mu(i)=\bar{\mu}_a(i)$ and $\mu_i(\theta)=\mu_{(a,i)}(\theta)$, we observe that the first-order condition is satisfied for 
\[
    \xi(\theta)=a(\theta)-\lambda_\pi(\theta)+c,
\]
 where $c \in \RR$ is a constant independent of $\theta$. Hence, $\mu_a$ is a solution to \eqref{eqn:nested-shannon-entropy} with $p=p_a$.

\subsection{Proof of Proposition \ref{prop:perceptual_mm}}

 Let $H_\pi$ be defined by \eqref{eq:nested_entropy_limit_case}. For every $a\in A$ and $i\in N$, define
\[
    \bar a (i) = \sum_{\theta\in \Theta} \nu_i(\theta) a(\theta).
\]
 We now prove the equivalence between the optimization problem
\begin{equation}\label{eq:problem_1}
    \max_{P\in \Delta(A)^\Theta} \sum_{a\in A,\theta\in \Theta} a(\theta) P_\theta(a)\pi(\theta) - \sum_{a\in \supp P_\pi}P_\pi(a)H_\pi(p_a)
\end{equation}
and the optimization problem
\begin{equation}\label{eq:problem_2}
    \max_{Q\in \Delta(A)^N} \sum_{a\in A,i\in N} \bar a (i) Q_i (a) \bar \nu(i) - \zeta I_S(Q,\bar\nu).
\end{equation}
 Note that \eqref{eq:problem_2} is a standard rational inattention problem, possibly with redundant actions, that is, $\bar a =\bar b$ for some $a,b\in A$ such that $a\neq b$.

 For every $Q\in \Delta(A)^N$, define $\nu\circ Q\in \Delta(A)^\Theta$ by 
\[
    (\nu\circ Q)_\theta(a) = \sum_{i\in N} \nu_\theta(i)Q_i(a).
\]
 Direct computation shows that 
\begin{equation}\label{eq:equal_EU}
    P= \nu\circ Q\quad\Rightarrow\quad \sum_{a\in A,\theta\in \Theta} a(\theta) P_\theta(a)\pi(\theta)=\sum_{a\in A,i\in N} \bar a (i) Q_i (a) \bar \nu(i).
\end{equation}
\begin{claim}
 For every $P\in \Delta(A)^\Theta$,
\begin{equation}\label{eq:equal_cost_perc}
    \sum_{a\in \supp P_\pi}P_\pi(a)H_\pi(p_a) = \zeta\inf\left\{ I_S(Q,\bar \nu):P=\nu\circ Q\right\}.
\end{equation}
 Moreover, whenever the left-hand side is finite, the infimum is attained.
\end{claim}
\begin{proof}
 Suppose first that $P= \nu\circ Q$. Note that $P_\pi = Q_{\bar \nu}$. Moreover, 
\[
    p_a = \sum_{i\in N} q_a(i) \nu_i\qquad \forall a\in \supp P_\pi = \supp Q_{\bar \nu}.
\]
Then,
\[
    H_\pi(p_a) \leq \zeta D_{\mathrm KL} ( q_a \Vert \bar \nu)\qquad \forall a\in \supp P_\pi = \supp Q_{\bar \nu}.
\]
It follows that 
\[
    \sum_{a\in \supp P_\pi}P_\pi(a)H_\pi(p_a) \leq \zeta I_S (Q,\bar \nu).
\]

Now suppose the cost of $P$ is finite. For every $a\in \supp P_\pi$, take $\bar{\mu}_a\in \Delta(N)$ such that 
\[
    \sum_{i\in N}\bar\mu_a(i)\nu_i=p_a\quad\text{and}\quad H_\pi(p_a) = \zeta D_{\mathrm KL} (\bar\mu_a\Vert \bar \nu).
\]
For notational convenience, set $p_a=\pi$ and $\mu_a = \bar\nu$ for all $a\notin \supp P_\pi$. Note that 
\[
    \sum_{i\in N} \sum_{a\in A} P_\pi(a)\bar \mu_a (i) \nu_i(\theta) = \sum_{a\in A} P_\pi(a) p_a(\theta) = \pi(\theta)\qquad \forall\theta\in \Theta.
\]
Since the set $\{\nu_i:i\in N\}$ is affinely independent, and $\sum_{i\in N}\bar \nu(i)\nu_i = \pi$, we obtain that 
\[
    \sum_{a\in A} P_\pi(a) \bar \mu_a (i) = \bar \nu (i) \qquad\forall i\in N.
\]
We can therefore define $Q\in \Delta(A)^N$ by 
\[
Q_i(a) = \frac{ P_\pi(a) \bar \mu_a (i)}{\bar \nu (i)}.
\]
We obtain that $P=\nu \circ Q$ and 
\[
\sum_{a\in \supp P_\pi}P_\pi(a)H_\pi(p_a) = \zeta I_S(Q,\bar \nu).
\]
The desired result follows.
\end{proof}

The equivalence between \eqref{eq:problem_1}  and \eqref{eq:problem_2} then follows from \eqref{eq:equal_EU} and \eqref{eq:equal_cost_perc}. If $P$ is a solution to \eqref{eq:problem_1}, then any $Q$ that attains the infimum in \eqref{eq:equal_cost_perc} is a solution to \eqref{eq:problem_2}. Conversely, if $Q$ is a solution to \eqref{eq:problem_2}, then $P=\nu\circ Q$ is a solution to \eqref{eq:problem_1}. Moreover, \eqref{eq:problem_1}  and \eqref{eq:problem_2} have the same value. Proposition \ref{prop:perceptual_mm} then is obtained by applying the results of \cite{matvejka2015rational} and \cite*{caplin2019rational} to \eqref{eq:problem_2}, and translate them back to \eqref{eq:problem_1}.

\subsection{Proof of Proposition \ref{pro:1dim_task}}

(i). Assume the nest structure satisfies the MLRP. By standard arguments (\citealp{karlin1980classes}; \citealp{milgrom1981good}), $\nu_\theta\in \Delta(N)$ and $\nu_i\in \Delta(\Theta)$ are increasing in $\theta$ and $i$, respectively, according to first-order stochastic dominance. Since $r(\theta)$ is increasing in $\theta$, 
\[
\bar r (i) = \sum_{\theta\in \Theta}\nu_i(\theta)r(\theta)
\]
is increasing in $i$. It follows that the quantity 
\[
Q_i(r) = \frac{\alpha(r)\ee^{\frac{\bar r (i)}{\zeta}}}{\alpha(r)\ee^{\frac{\bar r (i)}{\zeta}}+1-\alpha(r)}
\]
is increasing in $i$. By Proposition \ref{prop:perceptual_mm},
\[
P_\theta(r) = \sum_{i\in N}\nu_\theta(i) Q_i(r). 
\]
Thus, $P_\theta(r)$ is increasing in $\theta$, given that $\nu_\theta$ is increasing in $\theta$ with respect to first-order stochastic dominance.

(ii). As shown above, 
\[
P_\theta(r) = \sum_{i\in N} \nu_\theta(i) Q_i(r)
\]
where the quantity $Q_i(r)$ is increasing in $i$. Simple algebra shows that the psychometric function is convex at $\theta_i$ if and only if
\[
 \sum_{i\in N} \left(\frac{1}{2}\nu_{\theta_{i-1}} + \frac{1}{2} \nu_{\theta_{i+1}} \right) Q_i(r) \geq \sum_{i\in N} \nu_{\theta_i}  Q_i(r).
\]
Consequently, if $\frac{1}{2}\nu_{\theta_{i-1}} + \frac{1}{2} \nu_{\theta_{i+1}}$ first-order stochastically dominates $\nu_{\theta_{i}}$, then the psychometric function is convex at $\theta_i$.

(iii). Same argument as in (ii), with the directions reversed.

\subsection{Proof of Proposition \ref{cor:multi-tasking}}

\paragraph{Decision problem 1.} For every $\zeta>0$, let $P^\zeta\in \Delta(A_1)^\Theta$ be an optimal choice rule, and let $Q\in \Delta(A_1)^\Theta$ be defined by $Q_\theta(a)=a(\theta)$ for all $\theta\in \Theta$ and $a\in A$. Thus, $Q$ selects the optimal action in every state. Note that $Q$ generates the posteriors $p_{a_U}=\nu_U$ and $p_{a_D}=\nu_D$, with equal probability. Thus,
\[
C\left(Q,\pi\right)\leq \zeta \left(\frac{1}{2} D_{\mathrm{KL}}(\delta_U\Vert \bar{\nu}) + \frac{1}{2} D_{\mathrm{KL}}(\delta_D\Vert \bar{\nu})\right),
\]
where $\delta_U, \delta_D\in \Delta(N)$ put probability one on $U$ and $D$, respectively.

Since $P^\zeta$ is optimal and information costs are positive,
\[
\sum_{a\in A_1,\theta\in \Theta}a(\theta)P_\theta^\zeta(a)\pi(\theta)\geq \sum_{a\in A_1,\theta\in \Theta}a(\theta)Q_\theta(a)\pi(\theta) - \zeta \left(\frac{1}{2} D_{\mathrm{KL}}(\delta_U\Vert \bar{\nu}) + \frac{1}{2} D_{\mathrm{KL}}(\delta_D\Vert \bar{\nu})\right).
\]
Moreover, 
\[
\sum_{a\in A,\theta\in \Theta}a(\theta)P_\theta^\zeta(a)\pi(\theta)\leq \sum_{\theta\in \Theta}\pi(\theta)\max_{a\in A_1}a(\theta) = \sum_{a\in A_1,\theta\in \Theta}a(\theta)Q_\theta(a)\pi(\theta).
\]
We deduce that 
\[
\lim_{\zeta\rightarrow 0 }\sum_{a\in A_1,\theta\in \Theta}a(\theta)P_\theta^\zeta(a)\pi(\theta)=\sum_{\theta\in \Theta}\pi(\theta)\max_{a\in A_1}a(\theta).
\]
It follows that $P^\zeta\rightarrow Q$ as $\zeta\rightarrow 0$.

\paragraph{Decision problem 2.} The same argument applies. 

\paragraph{Decision problem 3.} Fix $\zeta>0$. By the symmetry of the decision problem, there is a saddle point $(\alpha,\lambda)$ in which $\alpha$ is uniform and $\lambda$ is constant (see Section \ref{sec:symmetric_decision_problems}).  Let $(\beta,l)$ be any other saddle point. By the product structure of the set of saddle points, $(\beta,\lambda)$ also is a saddle point. Then, for all $\theta\in \Theta$,
\begin{align*}
  \pi(\theta) &= \beta(a_{\text{diag}}) \nabla_\theta H^\star_\pi (a_{\text{diag}}-\lambda_\pi) + \left(1-\beta(a_{\text{diag}}) \right) \nabla_\theta H^\star_\pi (a_{\text{off}}-\lambda_\pi)\\
  & = \beta(a_{\text{diag}}) \nabla_\theta H^\star_\pi (a_{\text{diag}}) + \left(1-\beta(a_{\text{diag}}) \right) \nabla_\theta H^\star_\pi (a_{\text{off}}),
\end{align*}
where the second equality holds because $\lambda$ is constant and $H^\star_\pi$ is translation invariant. Analogously, for all $\theta\in \Theta$,
\[
\pi(\theta) = \alpha(a_{\text{diag}}) \nabla_\theta H^\star_\pi (a_{\text{diag}}) + \left(1-\alpha(a_{\text{diag}}) \right) \nabla_\theta H^\star_\pi (a_{\text{off}}).
\]
Since $\nabla_\theta H^\star (a_{\text{diag}})\neq \nabla_\theta H^\star (a_{\text{off}})$, we conclude that $\beta=\alpha$.

Thus, the optimal choice rule is unique and generated by a saddle point $(\alpha,\lambda)$ with $\alpha$ uniform and $\lambda$ constant (Remark \ref{rem:uniqueness}). We obtain:
\[
\frac{P_\theta(a_{\text{diag}})}{P_\theta(a_\text{off})} = \frac{ \sum_{i\in N }\nu_\theta(i)e^{\frac{a_\text{diag}(\theta)}{\eta}}\left(\sum_{\tau \in \Theta} \nu_i(\tau) e^{\frac{a_\text{diag}(\tau)}{\eta}} \right)^{\frac{\eta-\zeta}{\zeta}}}{\sum_{i\in N }\nu_\theta(i)e^{\frac{a_\text{off}(\theta)}{\eta}}\left(\sum_{\tau \in \Theta} \nu_i(\tau) e^{\frac{a_\text{off}(\tau)}{\eta}} \right)^{\frac{\eta-\zeta}{\zeta}}}= e^{\frac{a_\text{diag}(\theta)-a_\text{off}(\theta)}{\eta}}.
\]
The desired result follows.
\newpage

\begin{center}
    {\bf \Large Online Appendix}
\end{center}

\section{Symmetric decision problems}\label{sec:symmetric_decision_problems}

 Symmetry assumptions are often used to construct illustrative examples and simplify the analysis in applications. Under such assumptions, the solutions to information acquisition problems based on $f$-information inherit the symmetries of the underlying primitives, as we now explain.
 
 We formalize symmetry through invariance  with respect to a group of permutations $\Gamma$ of the state space. Specifically,  $\Gamma$ is a set of bijective functions $\gamma\colon \Theta\rightarrow \Theta$ with the following properties: the composition of any two elements of $\Gamma$ belongs to $\Gamma$, and the inverse of any element of $\Gamma$ belongs to $\Gamma$ as well. For each $x\in \mathbb{R}^\Theta$ and $\gamma\in \Gamma$, $x_\gamma\in \mathbb{R}^\Theta$ stands for the permuted vector $x_\gamma(\theta)=x(\gamma(\theta))$. 

 A decision problem $\D=(\pi,A)$ is said to be \textit{invariant} with respect to a group $\Gamma$ if $\pi_\gamma=\pi$ and $\{a_\gamma:a\in A\} = A$ for all $\gamma\in \Gamma$. A simple example arises when $\pi$ is uniform and $A$ is the set of bets that pay 1 in one state and 0 otherwise. This decision problem is invariant under all bijections $\gamma$. A more general form of symmetry arises in decision problems with a priori homogeneous actions (Definition \ref{def:homo_actions}).

 A function $f_\pi\colon \RR^\Theta_+\rightarrow\RRcvx$ is said to be \emph{invariant} with respect to a group $\Gamma$ if  $f_\pi(x_\gamma)=f_\pi(x)$ for all $x\in \RR^\Theta_+$ and $\gamma\in \Gamma$. An example is provided by Csisz\'ar information, whose associated transformation $f_\pi$ is invariant under any permutation $\gamma$ for which the prior is invariant.

\begin{proposition}\label{pro:symmetry}
 Consider a decision problem $\D$ and a transformation $f_\pi$ that are invariant with respect to a group $\Gamma$ of permutations of the state space. Then, the maxmin problem (\ref{eq:maxmin_problem}) has an invariant saddle point $(\alpha,\lambda)$, meaning that:
 \begin{enumerate}
    \item $\alpha(a_\gamma) = \alpha(a)$ for all $a\in A$ and $\gamma\in \Gamma$.
    \item $\lambda_\gamma=\lambda$ for all $\gamma\in \Gamma$.
 \end{enumerate}
 The resulting choice rule has the following symmetry property: 
 $
    P_{\gamma(\theta)} (a) = P_{\theta}(a_\gamma)
 $
 for all $\theta\in \Theta$, $a\in A$, and $\gamma\in \Gamma$.
\end{proposition}

 Pivotal to Proposition \ref{pro:symmetry} is the following lemma, which shows that the conjugate $f^\star_\pi$ inherits the symmetry properties of $f_\pi$:

\begin{lemma}\label{lem:f_invariant}
 A transformation $f_\pi$ is invariant under a permutation $\gamma$ if and only if its conjugate $f^\star_\pi$ also is invariant, meaning that $f^\star_\pi(x_\gamma)=f^\star_\pi(x)$ for all $x\in \RR^\Theta$. Moreover,
 $
    \nabla_{\gamma(\theta)} f^\star_\pi (x) = \nabla_\theta f^\star_\pi (x_\gamma)
 $
 for all $\theta\in \Theta$.
\end{lemma} 

\begin{proof}
 If $f_\pi$ is invariant, then for all $x\in\mathbb{R}^\Theta$
\begin{align*}
    f^\star_\pi(x_\gamma)  = \sup_{y\in \mathbb{R}^\Theta_+} \sum_{\theta\in \Theta} x_\gamma(\theta)y(\theta) - f_\pi(y_\gamma) 
    = \sup_{y\in \mathbb{R}^\Theta_+} \sum_{\theta\in \Theta} x(\theta)y_{\gamma^{-1}}(\theta) - f_\pi(y) = f^\star_\pi(x).
\end{align*}
 Thus, $f_\pi$ invariant implies $f^\star_\pi$ invariant. An analogous argument shows that $f^\star_\pi$ invariant implies $f_\pi$ invariant. To prove the last part of the claim, set $x^\star = \nabla f^\star_\pi (x)$. By the subdifferential inequality, for all $y\in \mathbb{R}^\Theta$,
 \[
    f^\star_\pi (y) - f^\star_\pi (x) \geq \sum_{\theta\in\Theta} x^\star(\theta)(y(\theta)-x(\theta)).
 \]
 Since $f^\star_\pi$ is invariant, $f^\star_\pi (y_\gamma) = f^\star_\pi (y)$ and $ f^\star_\pi (x_\gamma) = f^\star_\pi (x) $. Moreover, simple algebra shows that 
 \[
    \sum_{\theta\in\Theta} x^\star(\theta)(y(\theta)-x(\theta)) =\sum_{\theta\in\Theta} x^\star_\gamma(\theta)(y_\gamma(\theta)-x_\gamma(\theta)).
 \]
 We obtain that for all $y\in \mathbb{R}^\Theta$,
 \[
    f^\star_\pi (y_\gamma) - f^\star_\pi (x_\gamma) \geq \sum_{\theta\in\Theta} x^\star_\gamma(\theta)(y_\gamma(\theta)-x_\gamma(\theta)). 
 \]
 Since $ \mathbb{R}^\Theta = \{y_\gamma: y \in \mathbb{R}^\Theta\}$, we deduce that $x^\star_\gamma  = \nabla f^\star_\pi (x_\gamma)$, as desired.
 \end{proof}

 \begin{proof}[Proof of Proposition \ref{pro:symmetry}]

 Let $(\alpha,\lambda)$ be a saddle point of the maxmin problem (\ref{eq:maxmin_problem}). For every $\gamma\in\Gamma$, we define $\alpha_\gamma$ so that $\alpha_\gamma(a)=\alpha(a_{\gamma^{-1}})$ for all $a\in A$. We claim that $(\alpha_\gamma,\lambda_\gamma)$ is also a saddle point of (\ref{eq:maxmin_problem}).

 First we show that $\alpha_\gamma$ is a best response to $\lambda_\gamma$, that is, 
 \[
    \sum_{a\in A}\alpha_\gamma(a)f^\star_\pi(a\pi-\lambda_\gamma)=\max_{a\in A} f^\star_\pi(a\pi-\lambda_\gamma).
 \]
 We begin by observing that
 \begin{align*}
    \sum_{a\in A}\alpha_\gamma(a)f^\star_\pi(a\pi-\lambda_\gamma) 
    & = \sum_{a\in A}\alpha_\gamma(a)f^\star_\pi(a\pi_\gamma-\lambda_\gamma) \\
    & = \sum_{a\in A}\alpha_\gamma(a)f^\star_\pi((a_{\gamma^{-1}}\pi-\lambda)_\gamma)\\
    & = \sum_{a\in A}\alpha (a_{\gamma^{-1}})f^\star_\pi(a_{\gamma^{-1}}\pi-\lambda) = \sum_{a\in A}\alpha (a)f^\star_\pi(a\pi-\lambda),
 \end{align*}
 where the first equality uses the invariance of $\pi$, the third equality the invariance of $f_\pi$ (which implies the invariance of $f^\star_\pi$), and the last equality the invariance of $A$. An analogous argument demonstrates that 
 \[
    \max_{a\in A} f^\star_\pi(a\pi-\lambda_\gamma) = \max_{a\in A} f^\star_\pi(a\pi-\lambda).
 \]
 Since $\alpha$ is a best response to $\lambda$,
 \[
    \sum_{a\in A}\alpha (a)f^\star_\pi(a\pi-\lambda) = \max_{a\in A} f^\star_\pi(a\pi-\lambda).
 \]
 It follows that $\alpha_\gamma$ is a best response to $\lambda_\gamma$.

 Next we show that $\lambda_\gamma$ is a best response to $\alpha_\gamma$, that is, 
 \[
    \lambda_\gamma \in \arg\min_{l\in \mathbb{R}^\Theta} \sum_{a\in A}\alpha_\gamma(a)f^\star_\pi(a\pi-l)+\sum_{\theta\in \Theta}l(\theta).
 \]
 Reasoning as above, 
 \[
    \sum_{a\in A}\alpha_\gamma(a)f^\star_\pi(a\pi-\lambda_\gamma) + \sum_{\theta\in\Theta}\lambda_\gamma(\theta) = \sum_{a\in A}\alpha (a)f^\star_\pi(a\pi-\lambda) + \sum_{\theta\in\Theta}\lambda(\theta).
 \]
 In addition, for all $l\in \mathbb{R}^\Theta$,
 \[
    \sum_{a\in A}\alpha_\gamma(a)f^\star_\pi(a\pi-l)+\sum_{\theta\in \Theta}l(\theta) = \sum_{a\in A}\alpha(a)f^\star_\pi(a\pi-l_{\gamma^{-1}})+\sum_{\theta\in \Theta}l_{\gamma^{-1}}(\theta).
 \]
 Hence, since $\lambda$ is a best response to $\alpha$,
 \[
    \lambda \in \arg\min_{l\in \mathbb{R}^\Theta} \sum_{a\in A}\alpha_\gamma(a)f^\star_\pi(a\pi-l_{\gamma^{-1}})+\sum_{\theta\in \Theta}l_{\gamma^{-1}}(\theta).
 \]
 Since $\mathbb{R}^\Theta = \{l_{\gamma^{-1}}:l\in \mathbb{R}^\Theta \}$, it follows that $\lambda_\gamma$ is a best response to $\alpha_\gamma$.

 Since the choice $\gamma\in \Gamma$ was arbitrary, any pair $(\alpha_\gamma,\lambda_\gamma)$ is a saddle point of (\ref{eq:maxmin_problem}). We define $\bar{\alpha}\in \Delta(A)$ and $\bar{\lambda}\in\mathbb{R}^\Theta$ as follows:
 \[
    \bar{\alpha}=\frac{1}{\vert\Gamma\vert}\sum_{\gamma\in \Gamma} \alpha_\gamma \quad\text{and}\quad \bar{\lambda} = \frac{1}{\vert\Gamma\vert}\sum_{\gamma\in \Gamma} \lambda_\gamma
 \]
 where $\vert\Gamma\vert$ is the cardinality of $\Gamma$. Since the saddle points of (\ref{eq:maxmin_problem}) form a convex product set in $\Delta(A)\times \mathbb{R}^\Theta$, we deduce that $(\bar{\alpha},\bar{\lambda})$ is a saddle-point as well. Since $\Gamma$ is a group, $\bar{\alpha}(a)=\bar{\alpha}(a_\gamma)$ for all $a\in A$ and $\gamma\in \Gamma$, and $\bar{\lambda}_\gamma=\bar{\lambda}$ for all $\gamma\in \Gamma$. We conclude that $(\bar{\alpha},\bar{\lambda})$ is an invariant saddle point of (\ref{eq:maxmin_problem}).

 The resulting optimal choice rule is given by 
 \[
    P_\theta(a) = \bar{\alpha}(a) \nabla_\theta f^\star (a\pi-\bar{\lambda}).
 \]
 For every $\gamma\in \Gamma$, we have that
 \begin{align*}
    P_{\gamma(\theta)}(a) = \bar{\alpha}(a) \nabla_{\gamma(\theta)} f^\star (a\pi-\bar{\lambda}) & = \bar{\alpha}(a) \nabla_{\theta} f^\star (a_\gamma \pi_\gamma-\bar{\lambda}_\gamma) \\
    & = \bar{\alpha}(a_\gamma) \nabla_{\theta} f^\star (a_\gamma \pi-\bar{\lambda})=P_\theta(a_\gamma)
 \end{align*}
 where the first line uses the relation $\nabla_{\gamma(\theta)} f^\star (x) = \nabla_\theta f^\star (x_\gamma)$ (Lemma \ref{lem:f_invariant}), and the second line the invariance of $\bar{\alpha}$, $\pi$, and $\bar{\lambda}$. 
\end{proof}

\section{IIA properties}\label{sec:IIA}
 
 \citet{matvejka2015rational} introduce two IIA properties for costly information acquisition.
 \begin{definition}\label{def:IIA-actions}
    A cost function $C$ satisfies \emph{IIA with respect to actions} if, for every decision problem $\D=(\pi, A)$ and optimal choice rule $P = \left(A, (P_\theta)_{\theta \in \Theta} \right)$, 
     \begin{equation}
    a(\theta)=b(\theta) \text{ and } a(\tau)=b(\tau)\ \  \implies \ \ \frac{P_\theta(a)}{P_\theta(b)} = \frac{P_\tau(a)}{P_\tau(b)}
    \end{equation}
for all actions $a,b\in \supp (P_\theta) \cap \supp (P_\tau)$  and every pair of states $\theta, \tau \in \Theta$.
\end{definition}

This property says that, if two actions $a$ and $b$ coincide on the event $\{\theta,\tau\}$, then the relative likelihood of choosing $a$ rather than $b$ is the same in states $\theta$ and $\tau$, independently of how the two actions differ on the complementary event $\Theta\setminus\{\theta,\tau\}$. In this sense, the property is reminiscent of Savage's sure-thing principle, but applied to stochastic choice rules.

IIA with respect to actions is satisfied by mutual information and, more broadly, by Csiszár information:
\[
a(\theta)=b(\theta) \ \  \implies \ \ 
\frac{P_\theta(a)}{P_\theta(b)}= \frac{\alpha(a)(\phi^\star)^\prime (a(\theta)-\lambda_\pi(\theta))}{\alpha(b)(\phi^\star)^\prime (b(\theta)-\lambda_\pi(\theta))}= \frac{\alpha(a)}{\alpha(b)}.
\]

 \begin{definition}\label{def:IIA-consequences}
    A cost function $C$ satisfies \emph{IIA with respect to consequences} if, for every decision problem $\D=(\pi, A)$ and optimal choice rule $P = \left(A, (P_\theta)_{\theta \in \Theta} \right)$, 
     \begin{equation}\label{eq.IIA.1}
    a(\theta)=a(\tau) \text{ and } b(\theta)=b(\tau)\ \  \implies \ \ \frac{P_\theta(a)}{P_\theta(b)} = \frac{P_\tau(a)}{P_\tau(b)}
    \end{equation}
 for all actions $a,b\in \supp (P_\theta) \cap \supp (P_\tau)$  and every pair of states $\theta, \tau \in \Theta$.
\end{definition}

 To interpret this condition, observe that by Bayes' rule
 \[
    \frac{P_\theta(a)}{P_\theta(b)} = \frac{P_\tau(a)}{P_\tau(b)} \ \ \iff \ \ \frac{p_a(\theta)}{p_a(\tau)} = \frac{p_b(\theta)}{p_b(\tau)},
 \]
 where $p_a,p_b\in \Delta(\Theta)$ denote the decision maker's posterior beliefs conditional on choosing actions $a$ and $b$, respectively. Thus, \eqref{eq.IIA.1} says that, if actions $a$ and $b$ are both constant on the event $\{\theta,\tau\}$, then, conditional on this event, they induce the same posterior odds. In this sense, the two actions are informationally equivalent signals about whether the state is $\theta$ or $\tau$.
 
 Mutual information satisfies IIA with respect to consequences:
 \[
 \log\frac{P_\theta(a)}{P_\theta(b)}= \log\frac{P_\pi(a)}{P_\pi(b)}+
 \frac{a(\theta)-b(\theta)}{\kappa}.
 \]
 The next proposition shows that mutual information is the only smooth Csiszár information cost with this property.

\begin{samepage}
\begin{proposition}\label{prop:csiszar-IIA}
Suppose that $|\Theta| \geq 5$. For any Csiszár information cost $C$ such that $\phi$ is essentially smooth and is thrice continuously differentiable with $\phi''>0$ on $(0,+\infty)$, the following statements are equivalent:
\begin{enumerate}
\item $C$ satisfies IIA with respect to consequences.
\item $C$ is proportional to mutual information: there exists $\kappa>0$ such that $C=\kappa I_S$.
\end{enumerate}
\end{proposition}
\end{samepage}
This finding is related in spirit to \citet*[Proposition 2]{matvejka2015rational}. Their result shows that, in a given decision problem $\D=(\pi,A)$, a stochastic choice rule $P$ satisfies IIA with respect to actions and consequences if and only if it admits the representation 
\[
P_\theta(a) = \frac{\alpha(a)\ee^{v(a(\theta))}}{\sum_{b\in A}\alpha(b) \ee^{v(b(\theta))}}
\]
for some $\alpha\in \Delta(A)$ and $v:\RR\rightarrow\RR$. Importantly, however, their representation does not require $\alpha$ to be optimal. Hence, one cannot conclude that $P$ arises from mutual information, even after transforming payoffs by $v$. By restricting attention to Csiszár information, we obtain an exact link between an IIA property and mutual information.

\begin{proof}
That (ii) implies (i) follows from \citet{matvejka2015rational}. We prove the converse by contradiction. Suppose that $C$ is not proportional to mutual information. As argued in the proof of Proposition \ref{prop:phi-PS-disjoint}, this implies that the Arrow--Pratt coefficient of $\psi=\phi^\star$ is strictly monotone on some open interval $X$. Since $X$ is open, it intersects either $(0,+\infty)$ or $(-\infty,0)$. We consider the first case; the second is analogous. 

We now construct a decision problem $(\pi,A)$ and an optimal choice rule $P=(A,(P_\theta)_{\theta\in \Theta})$ with  saddle point $(\alpha,\lambda)$ such that IIA with respect to consequences is violated. 

Enumerate the state space as $\Theta=\{\theta_1,\ldots,\theta_n\}$. By hypothesis, $n\geq 5$. Let $A=\{a,b,c\}$. We specify the prior, payoffs, $f$-mean, and Lagrange multiplier state by state as follows.
 
 \emph{State $\theta_1$.} Fix $\bar x,\underline x\in X$ such that
$\bar x>\underline x>0.$ Action $a$ pays $\bar{x}$, action $b$ pays $\underline{x}$, and action $c$ pays $-1$. The prior-adjusted Lagrange multiplier takes value $0$. Since $\bar{x}>\underline{x}>0$, we can find $y\in (0,1)$ such that 
 \begin{equation*}\label{selectivity_state1}
 y\left(\frac{1}{2}\psi^\prime(\bar{x})+\frac{1}{2}\psi^\prime(\underline{x})\right)+(1-y)\psi^\prime(-1)=1.
 \end{equation*}
 Set $\alpha(a)=\alpha(b)=y/2$ and $\alpha(c)=1-y$. Then \eqref{eq:lambda_separable_case-1} holds.

 \emph{State $\theta_2$.} Pick $\epsilon>0$ sufficiently small so that $\bar{x}-\epsilon\in X$, $\underline{x}-\epsilon\in X$, and $\underline{x}-\epsilon>0$. Since $\psi^\prime(t)\rightarrow0$ as $t\rightarrow-\infty$, there exists $z>0$ such that  
 \begin{equation*}\label{selectivity_state2}
 y\left(\frac{1}{2}\psi^\prime(\bar{x}-\epsilon)+\frac{1}{2}\psi^\prime(\underline{x}-\epsilon)\right)+(1-y)\psi^\prime(-z)=1.
 \end{equation*}
 Then, in state $\theta_2$, action $a$ pays $\bar{x}$, action $b$ pays $\underline{x}$, and action $c$ pays $-z+\epsilon$. The prior-adjusted Lagrange multiplier takes value $\epsilon$. Then \eqref{eq:lambda_separable_case-1} holds.
 
 \emph{State $\theta_3$.} Same as $\theta_1$, but swap the payoffs of $a$ and $b$.

 \emph{State $\theta_4$.} Same as $\theta_2$, but swap the payoffs of $a$ and $b$.

 \emph{State $\theta_5$.} Here, $a$ and $b$ pay $-1$, and action $c$ pays $1$. By the intermediate value theorem, there exists $w\in [-1,1]$ such that  
\begin{equation}\label{eq:selectivity_state5}
 y \psi^\prime (-1-w) + (1-y)\psi^\prime(1-w)=\psi^\prime(0).
 \end{equation}
 Set $w$ as the prior-adjusted Lagrange multiplier in state $\theta_5$. Then \eqref{eq:lambda_separable_case-1} holds.

 \emph{States $\theta_6,\ldots,\theta_n$.} Here, all actions pay zero, and the prior-adjusted Lagrange multiplier is zero. Then \eqref{eq:lambda_separable_case-1} trivially holds.

We now complete our construction by selecting the prior. Since $\underline{x}-\epsilon>0$ and $z>0$,
 \[
\frac{1}{4}\left[\psi(\bar{x})+\psi(\underline{x})+\psi(\bar{x}-\epsilon)+\psi(\underline{x}-\epsilon)\right]>\frac{1}{2}\left[\psi(-1)+\psi(-z)\right].
 \]
 Moreover, $\psi (-1-w)<\psi (1-w)$. Thus, there exists $\zeta\in (0,1)$ such that 
 \begin{align}\label{selectivity_end}
     &\notag\frac{\zeta}{4}[\psi(\bar{x})+\psi(\underline{x})+\psi(\bar{x}-\epsilon)+\psi(\underline{x}-\epsilon)]+(1-\zeta)\psi(-1-w)\\ 
   =& \frac{\zeta}{2}[\psi(-1)+\psi(-z)]+(1-\zeta) \psi (1-w).
 \end{align}
 We set the prior as follows:
 \[
\pi(\theta_1)=\ldots=\pi(\theta_4)=\frac{5\zeta}{4n},\quad\pi(\theta_5)=\frac{5(1-\zeta)}{n},\quad\pi(\theta_6)=\ldots=\pi(\theta_n)=\frac{1}{n}.
 \]
It follows from \eqref{selectivity_end} that $\alpha$ is a best response to $\lambda$ in the maxmin problem \eqref{eq:maxmin_problem}. This concludes our construction. 

We are ready to show that IIA with respect to consequences is violated. Observe that 
\[
a(\theta_1)=a(\theta_2)=\bar{x}\quad\text{and}\quad b(\theta_1)=b(\theta_2)=\underline{x}.
\]
Moreover, denoting by $R_\psi$ the Arrow--Pratt coefficient of $\psi$,
\begin{align*}
\log \frac{P_{\theta_1}(a)}{P_{\theta_1}(b)}  = \log \psi^\prime (\bar{x})-\log \psi^\prime (\underline{x}) 
  & = \int_{\underline{x}}^{\bar{x}}R_\psi(x)\dd x \\
 &  \neq  \int_{\underline{x}}^{\bar{x}} R_\psi(x-\varepsilon)\dd x\\
& = \log \psi^\prime (\bar{x}-\epsilon)-\log \psi^\prime (\underline{x}-\epsilon)
=
\log \frac{P_{\theta_2}(a)}{P_{\theta_2}(b)},
\end{align*}
where we use the fact that $R_\psi = (\log \psi')^\prime$ is strictly monotone on $X$. We deduce that IIA with respect to consequences is violated.
\end{proof}

\section{Comparison of nested entropy and neighborhood-based costs}\label{sec:online.appendix.nests}

 In this section we show that the neighborhood-based model \eqref{eqn:neighborhood} cannot produce the behavioral pattern described in Proposition~\ref{cor:multi-tasking}, regardless of the neighborhood structure. In particular, we show that, under any such cost function, if the choice accuracy in both problems 1 and 2 is nearly perfect, then choice accuracy in problem 3 must also be nearly perfect and the cost function itself must be nearly identically zero. 

 Formally, we call a neighborhood structure $\mathcal{B}$ \emph{nonredundant} if it contains no singleton neighborhoods, i.e., $B \in \mathcal{B}$ implies $|B|\geq 2$. Since singleton neighborhoods do not contribute to the entropy \eqref{eqn:neighborhood}, nonredundancy is an innocuous assumption that merely simplifies notation. 

\begin{proposition}\label{cor:neighborhood}
 Fix any finite index set $I$, nonredundant neighborhood structure $\mathcal{B}$, and convergent sequence of coefficients $(\kappa_i^n)_{i \in I} \to (\kappa^*_i)_{i \in I} \in \RRcvx^I_+$. For each decision problem $j \in\{1,2,3\}$ and $n \in \mathbb{N}$, let $P^{j,n} \in \Delta(A_j)^\Theta$ 
be an optimal stochastic choice rule under the neighborhood-based cost defined via \eqref{eqn:neighborhood} 
with this neighborhood structure and coefficients $(\kappa_i^n)_{i \in I}$. If it holds that for every $a_1 \in A_1$ and $a_2 \in A_2$,
 \[
    \lim_{n \to \infty} P^{1,n}_\theta(a_1) = a_1(\theta) \quad  \text{ and } \quad  \lim_{n \to \infty} P^{2,n}_\theta(a_2) = a_2(\theta),  
    \]
    then it also holds that 
 \[
   \kappa^*_i = 0 \text{ for all $i \in I$} \quad \text{ and for every $a_3 \in A_3$, } \quad \lim_{n \to \infty} P^{3,n}_\theta(a_3) = a_3(\theta).
 \]
\end{proposition}

The contrast between Propositions \ref{cor:multi-tasking} and \ref{cor:neighborhood} reflects the difference between the ways nested Shannon and neighborhood-based cost functions aggregate costs across nests/neighborhoods. Namely, the nested Shannon model allows us to decouple the operations of learning \emph{about} nests and learning \emph{within} nests, while the neighborhood-based model generally does not.\footnote{The ``deterministic categorization'' special case discussed in Section \ref{ssec:HW} is an exception.}

To illustrate, consider a neighborhood-based cost with neighborhood structure $\mathcal{B}' = \{U,D\}$ and strictly positive coefficients. Under this cost function, it is free to learn about the first dimension of the state, as doing so does not require distinguishing the states within $U$ and $D$. However, it is costly to learn about the second dimension, which does require distinguishing the states within $U$ and $D$. This implies that choice accuracy is perfect in problem 1, and imperfect in problems 2 and 3. By symmetric reasoning, the neighborhood-based cost with neighborhood structure $\mathcal{B}'' = \{L,R\}$ makes it costly to learn about the first dimension and free to learn about the second dimension; this implies perfect choice accuracy in problem 2, and imperfect choice accuracy in problems 1 and 3. In either case, learning \emph{about} one dimension of the state requires distinguishing between states \emph{within} the neighborhoods that hold the other dimension fixed, rendering both uni- and multi-dimensional learning costly.

 By contrast, under the nested Shannon cost with nests $\{U,D,L,R\}$, the decision maker can learn exclusively about one dimension of the state while learning nothing about the other. In problems 1 and 2, this flexibility effectively allows the decision maker to \emph{choose} between facing the neighborhood structure $\mathcal{B}'$ or $\mathcal{B}''$, resulting in perfect choice accuracy in both problems. Choice accuracy is only imperfect in problem 3, where learning about both dimensions is necessary. We conclude that this feature of the nested Shannon cost is crucial for modeling multi-dimensional learning. 

 \subsection{Proof of Proposition~\ref{cor:neighborhood}}

 Define $P^1 \in \Delta(A_1)^\Theta$ and $P^2 \in \Delta(A_2)^\Theta$ as $P^1_\theta(a_1) = a_1(\theta)$ and $P^2_\theta(a_2) = a_2(\theta)$ for all $a_1\in A_1,a_2 \in A_2$, and $\theta \in \Theta$. The associated unconditional action probabilities and posteriors are given by 
    \begin{align*}
        &P^1_\pi(a_1) = 1/2 \quad \text{ and } \quad p^1_{a_1}(\theta) = \frac{a_1(\theta)}{2} \quad \text{ for all } a_1 \in A_1, \, \theta \in \Theta, \\
        & P^2_\pi(a_2) = 1/2 \quad \text{ and } \quad p^2_{a_2}(\theta) = \frac{a_2(\theta)}{2}  \quad \text{ for all } a_2 \in A_2, \, \theta \in \Theta.
    \end{align*}
    Suppose that $\lim_{n \to \infty} P^{1,n} = P^1$ and $\lim_{n \to \infty} P^{2,n}_\theta = P^2_\theta$, which implies that 
     \begin{align*}
        &\lim_{n \to \infty} P^{1,n}_\pi(a_1) = P^1_\pi(a_1) \quad \text{ and } \quad \lim_{n \to \infty}p^{1,n}_{a_1} = p^1_{a_1} \quad \text{ for all } a_1 \in A_1, \\
        & \lim_{n \to \infty} P^{2,n}_\pi(a_2) = P^2_\pi(a_2) \quad \text{ and } \quad \lim_{n \to \infty}p^{2,n}_{a_2} = p^2_{a_2} \quad \text{ for all } a_2 \in A_2.
    \end{align*}
    
    Throughout the proof, we adopt the following notational conventions. First, for each vector of coefficients $\kappa \in \RR^I_+$, we denote by $C (\cdot; \kappa) \colon \E \to \RR_+$ the associated neighborhood-based cost defined via \eqref{eqn:neighborhood}. Second, for each $i \in I$, we extend the KL divergence $D_\text{KL}(\cdot \Vert \pi_i)$ on $\Delta(B_i)$ to the orthant $\RR^{B_i}_+$ by defining (with minor abuse of notation) the map $D_\text{KL}(\cdot \Vert \pi_i) \colon \RR^{B_i}_+ \to \RR$ as 
    \[
    D_\text{KL} (x \Vert \pi_i) = \sum_{\theta \in B_i} x(\theta) \log \left(\frac{x(\theta)}{\pi_i(\theta)} \right).
    \]
    This extension, which is without loss of generality, allows us to take derivatives of $D_\text{KL}(\cdot \Vert \pi_i)$ in the usual way on $\RR^{B_i}_+$. Finally, we define the maps $H^n \colon \RR^\Theta_+ \to \RR$ and $H \colon \RR^\Theta_+ \to \RR$ as
    \[
    H^n(x) = \sum_{i \in I} \kappa^n_i \, \overline{x}(i)  D_\text{KL} (x_i \Vert \pi_i) \quad \text{ and } \quad H(x) = \sum_{i \in I} \kappa^*_i \, \overline{x}(i)  D_\text{KL} (x_i \Vert \pi_i),
    \]
    where $\overline{x}(i) = \sum_{\theta \in B_i} x(\theta)$ and $x_i(\theta) = x(\theta) / \overline{x}(i)$ for all $i \in I$ and $\theta \in B_i$ (with the standard continuous extension at points where $\overline{x}(i) = 0$ for some $i \in I$). This is shorthand notation for the entropy  \eqref{eqn:neighborhood} with coefficients $\kappa^n$ and $\kappa^*$, respectively, extended to the orthant. 

    We first show, via three claims, that the coefficients converge to $\kappa^*_i = 0$ for all $i \in I$. 

    \begin{claim}\label{claim:HW-claim1}
         $\kappa^*_i < +\infty$ for every $i \in I$.
    \end{claim}
    \begin{proof}
        Suppose, towards a contradiction, that there exists $i \in I$ with $\kappa^*_i = +\infty$. Since $|B_i|\geq 2$ by the nonredundancy assumption, there is some $E \in \{U,D,L,R\}$ such that $B_i \cap E \neq \emptyset$ and $B_i \backslash E \neq \emptyset$. We suppose here that  $E = U$; the other cases are analogous and hence omitted. 

        Consider the decision maker's cost in problem 1. For each $n \in \mathbb{N}$, it holds that
        \begin{align*}
        C(P^{1,n}; \kappa^n) & = P^{1,n}_\pi(a_U) H^n (p^{1,n}_{a_U}) + (1- P^{1,n}_\pi(a_U)) H^n (p^{1,n}_{a_D}) \\
        & \geq \kappa^n_i \, \left[ P^{1,n}_\pi(a_U)\, \overline{p}^{1,n}_{a_U}(i) \,  D_\text{KL} \left( p^{1,n}_{a_U,i} \Vert \pi_i \right)\right]. 
        \end{align*}
        Note that $\lim_{n \to \infty} \overline{p}^{1,n}_{a_U}(i) = \frac{1}{2} |B_i \cap U| \geq 1/2$, where the inequality is by $B_i \cap U \neq \emptyset$. Moreover, $\lim_{n \to \infty} p^{1,n}_{a_U,i} =  p^{1,}_{a_U,i}$ and $\supp(p^{1}_{a_U,i}) \subsetneq \supp(\pi_i) = B_i$, where the strict inclusion is by $B_i \backslash U \neq \emptyset$. Thus, by continuity of KL divergence, given any $\epsilon \in (0,1)$ and sufficiently large $n$, 
        \begin{align*}
        C(P^{1,n}; \kappa^n)  \geq  \kappa_i^n \, \left[ (1-\epsilon ) \, \frac{1}{4} \, D_\text{KL}\left(p^{1}_{a_U,i} \Vert \pi_i \right) \right].
        \end{align*}
        Since the term in brackets is strictly positive and $\kappa_i^n \to \kappa^*_i = +\infty$, we obtain $C(P^{1,n}; \kappa^n) \to +\infty$. This contradicts the optimality of $P^{1,n}$ in decision problem 1 for large $n$, as desired.
    \end{proof}

     \begin{claim}\label{claim:HW-claim2}
         It holds that
         \begin{align}
         \lim_{n \to \infty} C(P^{1,n}; \kappa^n) &=  C(P^1; \kappa^*) = \frac{1}{2} H(p^{1}_{a_U}) +\frac{1}{2} H (p^{1}_{a_D}), \label{eqn:HW-cost-conv-1}\\
         \lim_{n \to \infty} C(P^{2,n}; \kappa^n) &=  C(P^2; \kappa^*) =\frac{1}{2} H(p^{2}_{a_L}) +\frac{1}{2} H(p^{2}_{a_R}). \label{eqn:HW-cost-conv-2}
          \end{align}
        Let $\D_j = \{Q\in \Delta(A_j)^\Theta: Q_\pi(a)\ge 1/4 \text{ for every } a\in A_j\} $. In each decision problem $j \in \{1,2\}$, it holds that
        \begin{equation}\label{eqn:HW-cost-conv-3}
        P^j \in \argmax_{Q \in \D_j} \sum_{\theta \in \Theta} \pi(\theta) \sum_{a \in A_j} Q_\theta (a) a(\theta)  - C(Q; \kappa^*),
         \end{equation}

    \end{claim}
    \begin{proof}
    For each $j \in \{1,2\}$, define the map $C^j \colon \D_j \times \RR_{+}^I \to \RR_+$ as 
    \[
    C^j(Q; \kappa) = \sum_{a \in A^j} Q_\pi(a) \left[ \sum_{i \in I} \kappa_i \, \overline{q}_a(i) \, D_\text{KL} \left( q_{a,i} \Vert \pi_i\right) \right], 
    \]
    where $\{q_a\}_{a \in A^j} \subseteq \Delta(\Theta)$ are the posteriors induced by $Q$. In words, $C^j(\cdot ; \kappa)$ is the neighborhood-based cost with coefficients $\kappa$, restricted to the subdomain of stochastic choice rules on $A_j$ where each action is taken with at least $1/4$ ex-ante probability. 
    
    Fix $j \in \{1,2\}$. We assert that $C^j$ is jointly continuous. To this end, take any convergent sequence $(Q^k,\kappa^k)$ in $\D_j \times \RR^I_+$ with limit point $(Q,\kappa)$. By the triangle inequality, 
    \begin{equation}\label{eqn:HW-triangle-1}
        \left|C^j (Q^k,\kappa^k) - C^j(Q,\kappa) \right|  \leq \left| C^j (Q^k,\kappa^k) - C^j(Q^k,\kappa)\right| + \left| C^j (Q^k,\kappa) - C^j(Q,\kappa)\right|.
    \end{equation}
    We consider each term on the RHS of \eqref{eqn:HW-triangle-1} in turn. For the first term, we have
    \begin{align*}
    \left| C^j (Q^k,\kappa^k) - C^j(Q^k,\kappa)\right| & \leq \sum_{a \in A_j} Q_\pi^k(a) \sum_{i \in I} \left| \kappa^k_i - \kappa_i\right| \, \overline{q}^k_a(i) \, D_\text{KL} \left( q^k_{a,i} \Vert \pi_i\right) \\
    & \leq \sum_{i \in N} \left| \kappa^k_i - \kappa_i\right| \times \sup_{p_i \in \Delta(B_i)} D_\text{KL} \left( p_i \Vert \pi_i\right) \\
    & \to 0 \quad \text{ as } k \to +\infty,
    \end{align*}
    where the first line is by the triangle inequality and the final line uses the fact that $D_\text{KL} \left( \cdot \Vert \pi_i\right)$ is bounded on $\Delta(B_i)$. For the second term, note that $Q^k \to Q$ implies $Q^k_\pi \to Q_\pi$ and $(q^k_a)_{a \in A_j} \to (q_a)_{a \in A_j}$ (being that $A_j$ is finite and actions are taken with at least 1/4 ex-ante probability). It follows that
    \begin{align*}
    \lim_{k \to \infty} C^j(Q^k; \kappa) & =  \sum_{a \in A^j} \lim_{k \to \infty}Q^k_\pi(a)\sum_{i \in I} \kappa_i \, \lim_{k \to \infty}  \left[  \overline{q}^k_a(i) \, D_\text{KL} \left( q^k_{a,i} \Vert \pi_i\right) \right] \\
    & = \sum_{a \in A^j} Q_\pi(a)\sum_{i \in I} \kappa_i \,  \overline{q}_a(i) \, D_\text{KL} \left( q_{a,i} \Vert \pi_i\right) = C^j(Q;\kappa).
    \end{align*}
    Since both terms on the RHS of \eqref{eqn:HW-triangle-1} converge to $0$ as $k \to \infty$, we obtain $\lim_{k \to \infty} C^j(Q^k; \kappa^k) = C^j(Q; \kappa)$. We conclude that $C^j$ is jointly continuous, as asserted. 

    Since Claim \ref{claim:HW-claim1} establishes that $\kappa^* \in \RR^I_+$, continuity of the $C^j$ directly implies \eqref{eqn:HW-cost-conv-1} and \eqref{eqn:HW-cost-conv-2}. Moreover, note that continuity of the $C^j$ also implies, via Berge's Theorem of the Maximum, that the correspondences $\mathcal{Q}^j : \RR^I_+ \rightrightarrows  \D_j$ defined as
    \[
    \mathcal{Q}^j (\kappa) = \argmax_{Q \in \D_j} \sum_{\theta \in \Theta} \pi(\theta) \sum_{a \in A^j} Q_\theta (a) a(\theta)  - C^j(Q; \kappa)
    \]
    are upper hemi-continuous. Since \(P^j_\pi(a)=1/2\) for every $a\in A_j$, the convergence $P^{j,n}\to P^j$ implies that $P^{j,n}\in \D_j$ for all sufficiently large $n$. As $P^{j,n}$ is optimal in the unrestricted problem, it is therefore also optimal on $\D_j$, so $P^{j,n}\in \mathcal{Q}^j(\kappa^n)$ for all sufficiently large $n$. This implies that $P^j \in \mathcal{Q}^j(\kappa^*)$ for each $j \in \{1,2\}$. This establishes \eqref{eqn:HW-cost-conv-3}, completing the proof of the claim. 
    \end{proof}

     \begin{claim}\label{claim:HW-claim3}
         $\kappa^*_i = 0$ for every $i \in I$. 
    \end{claim}
    \begin{proof}
    Suppose, towards a contradiction, that there exists $k \in I$ with $\kappa^*_k >0$. Since $|B_k|\geq 2$ by the nonredundancy hypothesis, there is some $E \in \{U,D,L,R\}$ such that $B_k \cap E \neq \emptyset$ and $B_k \backslash E \neq \emptyset$. We suppose here that $E = U$; the other cases are specular and hence omitted. 

    Consider decision problem $1$. Since $\supp(p^1_{a_U}) = U$ and $B_k \backslash U \neq \emptyset$, it follows that $\supp(p_{a_U,k}) \subsetneq B_k = \supp(\pi_k)$. We show that this yields a contradiction to \eqref{eqn:HW-cost-conv-3} in Claim \ref{claim:HW-claim2}. 
    
    To this end, for each $\epsilon \in (0,1)$, define $Q^\epsilon \in \Delta(A_1)^\Theta$ as $Q^\epsilon_\theta(\cdot) = \epsilon/2 + (1-\epsilon)P^1_\theta (\cdot)$ for all $\theta \in \Theta$, so that $Q^\epsilon_\pi(\cdot) = 1/2$ and the associated posteriors are $q^\epsilon_{a} = \epsilon \pi + (1-\epsilon) p^1_{a} \in \Delta(\Theta)$ for each $a \in A_1$. Note that $Q^\epsilon \in \D_1$. For the limit coefficients $\kappa^*$, the value of decision problem 1 under $Q^\epsilon$ is 
    \[
    V(\epsilon) := \sum_{\theta \in \Theta} \pi(\theta) \sum_{a \in A_1} Q_\theta^\epsilon (a) a(\theta) - C(Q^\epsilon;\kappa^*) = 1 - \epsilon/2 - C(Q^\epsilon; \kappa^*).
    \]
    Note that $V(0) = 1 - C(P^1)$ is the value under $P^1$. Moreover, $V(\epsilon) > V(0)$ if and only if
    \begin{equation}\label{eqn:HW-value-ineq}
    \frac{2}{\epsilon} \cdot \left[ C(P^1; \kappa^*) - C(Q^\epsilon; \kappa^*) \right] > 1.
    \end{equation}
    Thus, to obtain a contradiction to \eqref{eqn:HW-cost-conv-3}, it suffices to show \eqref{eqn:HW-value-ineq} for some $\epsilon>0$. Note that
    \begin{align*}
        C(P^1; \kappa^*) - C(Q^\epsilon; \kappa^*) & = \frac{1}{2} \left(H(p^1_{a_U}) - H(q^\epsilon_{a_U}) \right) + \frac{1}{2} \left(H(p^1_{a_D}) - H(q^\epsilon_{a_D}) \right) \\
        & \geq \frac{1}{2} \nabla H(q^\epsilon_{a_U}) \cdot \left(p^1_{a_U} - q^\epsilon_{a_U}\right) + \frac{1}{2} \nabla H(q^\epsilon_{a_D}) \cdot \left(p^1_{a_D} - q^\epsilon_{a_D}\right) \\
        & = \frac{\epsilon}{2} \left(\nabla H(q^\epsilon_{a_U}) \cdot \left(p^1_{a_U} - \pi \right) + \nabla H(q^\epsilon_{a_D}) \cdot \left(p^1_{a_D} - \pi \right)\right),
    \end{align*}
    where the inequality holds because $H$ is convex and differentiable at full-support beliefs. Therefore, to show that \eqref{eqn:HW-value-ineq} holds for some $\epsilon>0$, it suffices to show that
    \begin{equation}\label{eqn:HW-inada}
    \nabla H(q^\epsilon_{a_U}) \cdot \left(p^1_{a_U} - \pi \right) + \nabla H(q^\epsilon_{a_D}) \cdot \left(p^1_{a_D} - \pi \right) \to +\infty \quad \text{ as } \epsilon \to 0.
    \end{equation}
    We establish \eqref{eqn:HW-inada} in what follows.
    
    First, note that for any $\theta \in \Theta$ and full-support $p \in \Delta(\Theta)$, it holds that
    \begin{align*}
    \nabla_\theta H(p) &= \frac{\partial}{\partial p(\theta)} \sum_{i \in I} \kappa_i \overline{p}(i) D_\text{KL}(p_i \Vert \pi_i) \\
    &= \sum_{i \in I \, : \,  \theta \in B_i} \kappa_i \left( \frac{\partial \overline{p}(i)}{\partial p(\theta)} D_\text{KL}(p_i \Vert \pi_i) + \overline{p}(i) \sum_{\tau \in B_i} \nabla_\tau D_\text{KL}(p_i \Vert \pi_i)\frac{\partial p_i(\tau)}{\partial p(\theta)} \right) \\
    & = \sum_{i \in I \, : \,  \theta \in B_i} \kappa_i \big(  D_\text{KL}(p_i \Vert \pi_i) + \nabla_\theta D_\text{KL}(p_i \Vert \pi_i) -  \nabla D_\text{KL}(p_i \Vert \pi_i) \cdot p_i \big) \\
    & = \sum_{i \in I \, : \,  \theta \in B_i} \kappa_i \log \left( \frac{p_i(\theta)}{\pi_i(\theta)} \right),
    \end{align*}
    where the second line is by the chain rule and the third and fourth lines are by direct calculation. Now, take any $a \in A_1$. The above display implies that
    \begin{align}
        \nabla H(q^\epsilon_a) \cdot \left(p^1_a - \pi \right) &= \sum_{\theta \in \Theta} \nabla_\theta H(q^\epsilon_a)  \left(p^1_a(\theta) - \pi(\theta)\right)  \notag \\
        & = \sum_{i \in I} \kappa_i \, \sum_{\theta \in B_i} \log \left( \frac{q^\epsilon_{a,i}(\theta)}{\pi_i(\theta)} \right) \left(p^1_{a}(\theta) - \pi(\theta)\right), \label{eqn:HW-inada-2}
    \end{align}
    where the second line is by the preceding display and interchanging the order of summation. Moreover, note that
    \begin{equation}\label{eqn:HW-post-perturb}
    q^\epsilon_{a,i}(\theta) = \frac{\epsilon \pi(\theta) + (1-\epsilon) p^1_a(\theta)}{\epsilon \overline{\pi}(i) + (1-\epsilon) \overline{p}^1_a(i)} \quad \text{ for all $i \in I, \, \theta \in \Theta$.}
    \end{equation}
    We assert that each term in the sum in \eqref{eqn:HW-inada-2} is non-negative. Fix any $i \in I$ and $\theta \in B_i$. We prove the assertion for $a = a_U$; the case where $a = a_D$ is analogous. Note that $\overline{p}^1_{a_U}(i) = \frac{1}{2} |B_i \cap U|$. 

    \emph{Case 1: $p^1_{a_U}(\theta) > \pi(\theta)$.} This implies $p^1_{a_U}(\theta) = 1/2$. Plugging this, $\overline{\pi}(i) = |B_i|/4$, and $\pi_i(\theta) = 1 / |B_i|$ into \eqref{eqn:HW-post-perturb}, we see that $q^\epsilon_{a,i}(\theta) \geq \pi_i(\theta)$ if and only if $|B_i| \geq |B_i \cap U|$. Since the latter inequality trivially holds, we conclude that $\log \left( q^\epsilon_{a_U,i}(\theta) / \pi_i(\theta) \right) \left(p^1_{a_U}(\theta) - \pi(\theta)\right) \geq 0$.

    \emph{Case 2: $p^1_{a_U}(\theta) < \pi(\theta)$.} This implies $p^1_{a_U}(\theta) = 0$. If $B_i \cap U = \emptyset$, then  \eqref{eqn:HW-post-perturb} implies $q^\epsilon_{a_U,i}(\theta) = \pi_i(\theta)$. If $B_i \cap U \neq \emptyset$, then \eqref{eqn:HW-post-perturb} implies $q^\epsilon_{a_U,i}(\theta) < \pi_i(\theta)$. In either case, we obtain $\log \left( q^\epsilon_{a_U,i}(\theta)/ \pi_i(\theta) \right) \leq 0$. We conclude that $\log \left( q^\epsilon_{a_U,i}(\theta) / \pi_i(\theta) \right) \left(p^1_{a_U}(\theta) - \pi(\theta)\right) \geq 0$.

    This proves the assertion. It follows that, for every $\epsilon \in (0,1)$, 
    \begin{align}
    \begin{split}\label{eqn:HW-inada-3}
    \nabla H(q^\epsilon_{a_U}) \cdot \left(p^1_{a_U} - \pi \right) + \nabla H(q^\epsilon_{a_D}) \cdot &\left(p^1_{a_D} - \pi \right) \\
    & \geq \kappa_k \, \sum_{\theta \in B_k}  \log \left( \frac{q^\epsilon_{a_U,k}(\theta)}{\pi_k(\theta)} \right) \left(p^1_{a_U}(\theta) - \pi(\theta)\right) \\
    & \geq \kappa_k \, \sum_{\theta \in B_k \backslash U}  \log \left( \frac{q^\epsilon_{a_U,k}(\theta)}{\pi_k(\theta)} \right) \left(p^1_{a_U}(\theta) - \pi(\theta)\right),
    \end{split}
    \end{align}
    where $k \in I$ is the index that, by supposition, satisfies $\kappa_k >0$, $B_k \cap U \neq \emptyset$, and $B_k \backslash U \neq \emptyset$. Plugging the definition of $p^1_{a_U}$ and \eqref{eqn:HW-post-perturb} into the final expression, we obtain
    \begin{align*}
    \kappa_k \, \sum_{\theta \in B_k \backslash U}  \log \left( \frac{q^\epsilon_{a_U,k}(\theta)}{\pi_k(\theta)} \right) \left(p^1_{a_U}(\theta) - \pi(\theta)\right) & = \frac{\kappa_k}{4} \, \sum_{\theta \in B_k \backslash U} \log\left( \frac{\epsilon \, |B_k| + 2(1-\epsilon) \, |B_k\cap U|}{\epsilon \, |B_k|}\right) \\
    & \to +\infty \quad \text{ as } \epsilon \to 0,
    \end{align*}
    where the limit is infinite because $\kappa_k >0$, $B_k \backslash U \neq \emptyset$, and $|B_k\cap U| \geq 1$. Plugging this into \eqref{eqn:HW-inada-3} then establishes \eqref{eqn:HW-inada}, and hence the desired contradiction. We conclude that $\kappa_k = 0$. 
    \end{proof}

Claim \ref{claim:HW-claim3} delivers the first conclusion of the proposition. It remains to show that $\kappa^* = \mathbf{0}$ implies that $\lim_{n \to \infty} P^{3,n}_\theta(a) = a(\theta)$ for all $\theta \in \Theta$ and $a \in A_3$. Suppose, towards a contradiction, that there exist some $\tau \in \Theta$ and $a \in A_3$ such that $a(\tau) = 1$ and yet $\liminf_{n \to \infty} P^{3,n}_\tau(a) <1$. Then there is a subsequence $(P^{3,n_k})_{k \in \mathbb{N}}$ such that
\[
\limsup_{k \to \infty} \sum_{\theta \in \Theta} \pi(\theta) \sum_{a \in A_3} P^{3,{n_k}}_\theta(a) a(\theta) - C(P^{3,n_k}; \kappa^{n_k}) \leq \limsup_{k \to \infty} \sum_{\theta \in \Theta} \pi(\theta) \sum_{a \in A_3} P^{3,{n_k}}_\theta(a) a(\theta) <1.
\]
Define $P^3 \in \Delta(A_3)^\Theta$ as $P^3_\theta (a) = \mathbf{1}(a(\theta) = 1)$ for all $\theta \in \Theta$ and $a \in A_3$. Since $\kappa^* = 0$, 
\[
\lim_{k \to \infty} \sum_{\theta \in \Theta} \pi(\theta) \sum_{a \in A_3} P^{3}_\theta(a) a(\theta) - C(P^{3}; \kappa^{n_k}) = 1 - C(P^3; \kappa^*) = 1.
\]
This implies that, for sufficiently large $k$, $P^{3,n_k}$ is not optimal in decision problem 3. This delivers the desired contradiction, and thereby completes the proof.

\section{Consideration sets}\label{section:OA:consideration-sets}

The next proposition delivers bounds for the consideration sets under posterior-separable costs and $f$-information.

\begin{proposition}\label{pro:consideration_set}
Let $n$ be the cardinality of $\Theta$.
\begin{enumerate}
\item Under posterior-separable costs, each decision problem $(\pi,A)$ admits an optimal choice rule $P\in \Delta(A)^\Theta$ such that $\vert \supp P_\pi\vert \leq n$.
\item Under $f$-information, each decision problem $(\pi,A)$ admits an optimal choice rule $P\in \Delta(A)^\Theta$ such that $\vert \supp P_\pi\vert \leq n+1$.
\end{enumerate}
\end{proposition}

Part (i) is known in the literature. In what follows, we provide a self-contained proof using Theorem \ref{thm:characterization}.

\begin{proof}[Proof of (i)]

Let $P$ be an optimal choice rule, and denote by $m$ and $l$ the cardinalities of $A$ and $\supp P_\pi$, respectively. If $l \leq n$, then the desired result holds. Suppose therefore that $l>n$. Next we construct another optimal choice rule $Q$ such that $\vert \supp Q_\pi \vert < l $. By induction on $l$, this implies that there exists an optimal choice rule whose consideration set has no more than $n$ actions.

Let $(\alpha,\lambda)$ be a saddle point of \eqref{eq:maxmin_problem} that generates $P$. Notice that $\alpha$, which is equal to $P_\pi$, is a solution of the following system of linear equations (label the system's independent variable by $\beta\in \mathbb{R}^A$):
\begin{align}\label{eq:ps_lin_system1}
\sum_{a\in A}\beta(a) \nabla_\theta H^\star_\pi(a-\lambda_\pi) & = \pi(\theta), &\theta\in \Theta,\\
\beta(a) & = 0, & a \notin \supp P_\pi.\label{eq:ps_lin_system111}
\end{align}
This linear system has $n+m-l$ equations and $m$ unknowns. Since $l>n$, there must be a non-zero vector $\beta$ such that 
\begin{align}\label{eq:ps_lin_system2}
\sum_{a\in A}\beta(a) \nabla_\theta H^\star_\pi(a-\lambda_\pi) & = 0, &\theta\in \Theta,\\
\beta(a) & = 0, & a\notin \supp P_\pi \label{eq:ps_lin_system222}.
\end{align}
Note that both $\beta$ and $-\beta$ are non-zero solutions of (\ref{eq:ps_lin_system2}) and (\ref{eq:ps_lin_system222}). Hence, we can assume without loss of generality that $\beta(a)>0$ for some $a\in A$. 

We define $\gamma\in \RR^A$ as follows: for all $a\in A$,
\[
\gamma(a) = \alpha(a) - \beta(a) \min_{b: \beta(b)>0} \frac{\alpha(b)}{\beta(b)}.
\]
\begin{claim}\label{claim:ps_suppor_bounds}
The vector $\gamma$ has the following properties:
\begin{enumerate}
\item $\gamma(a)\geq 0$ for all $a\in A$.
\item $\gamma(a)=0$ for some $a\in \supp P_\pi$.
\item $\gamma$ is a solution of (\ref{eq:ps_lin_system1}) and (\ref{eq:ps_lin_system111}).
\item $\gamma\in \Delta(A)$.
\item $(\gamma,\lambda)$ is a saddle point of \eqref{eq:maxmin_problem}.
\end{enumerate}
\end{claim}
\begin{proof}
(i). If $\beta(a) \leq 0$, then $\gamma(a) \geq \alpha(a)\geq 0$. If $\beta(a)>0$, then
\[
\gamma(a)\geq 0\quad\Longleftrightarrow\quad \frac{\alpha(a)}{\beta(a)}\geq \min_{b:\beta(b)>0}\frac{\alpha(b)}{\beta(b)}.
\]
Thus, $\gamma(a)\geq 0$ also when $\beta(a)>0$. This proves (i).

(ii). Take any $a$, with $\beta(a)>0$, such that 
\[
\frac{\alpha(a)}{\beta(a)} = \min_{b:\beta(b)>0}\frac{\alpha(b)}{\beta(b)}.
\]
Then, $\gamma(a)=0$. Moreover, (\ref{eq:ps_lin_system222}) ensures that $a\in\supp P_\pi $. This proves (ii).

(iii). This follows from $\alpha$ being a solution of (\ref{eq:ps_lin_system1})--(\ref{eq:ps_lin_system111}) and $\beta$ being a solution of (\ref{eq:ps_lin_system2})--(\ref{eq:ps_lin_system222}). 

(iv). By (i), $\gamma(a)\geq 0$ for all $a\in A$. Since $H^\star_\pi$ is translation invariant, 
\[
\sum_{\theta\in \Theta}\nabla_\theta H^\star_\pi(a-\lambda_\pi)=1.
\]
It follows from (\ref{eq:ps_lin_system1}) that
\begin{align*}
\sum_{a\in A}\gamma(a) & = \sum_{a\in A}\gamma(a)\left(\sum_{\theta\in\Theta}\nabla_\theta H^\star_\pi(a-\lambda_\pi)\right) \\
& = \sum_{\theta\in\Theta}\left(\sum_{a\in A}\gamma(a) \nabla_\theta H^\star_\pi(a-\lambda_\pi)\right) = \sum_{\theta\in\Theta} \pi(\theta)=1.
\end{align*}
We conclude that $\gamma\in \Delta(A)$.

(v). Since $\supp \gamma\subseteq \supp P_\pi$ and $(P_\pi,\lambda)$ is a saddle point, we have:
\[
\min_{a\in \supp\gamma} H^\star_\pi (a-\lambda_\pi)\geq \min_{a\in \supp P_\pi}H^\star_\pi(a-\lambda_\pi)=\max_{a\in A}H^\star_\pi(a-\lambda_\pi).
\]
Hence, it follows from (\ref{eq:ps_lin_system1}) that $(\gamma,\lambda)$ is a saddle point.
\end{proof}

Let $Q$ be the optimal choice rule generated by $(\gamma,\lambda)$. Under posterior separability, $Q_\pi=\gamma$. Thus, $\supp Q_\pi\subseteq \supp P_\pi$ by (\ref{eq:ps_lin_system111}). Moreover, $\supp Q_\pi\neq \supp P_\pi$. Indeed, by (ii) of Claim \ref{claim:ps_suppor_bounds}, there exists $a\in \supp P_\pi$ such that $\gamma(a)=0$. Given that $\gamma(a)=0$, we must have $Q_\pi(a)=0$. It follows that $\supp Q_\pi \neq \supp P_\pi$. Overall, we conclude that  $\supp Q_\pi$ is a proper subset of $\supp P_\pi$. This shows that $\vert \supp Q_\pi\vert < \vert \supp P_\pi\vert=l$, as desired.
\end{proof}

The structure of the proof of part (ii) parallels that of part (i). We repeat several steps to emphasize both the analogies and the differences.

\begin{proof}[Proof of (ii)]

Let $P$ be an optimal choice rule for $(\pi,A)$, and denote by $m$ and $l$ the cardinalities of $A$ and $\supp P_\pi$, respectively. If $l \leq n+1$, then the desired result holds. Suppose therefore that $l>n+1$. Next we construct another optimal choice rule $Q$ such that $\vert \supp Q_\pi \vert < l $. By induction on $l$, this implies that there exists an optimal choice rule whose consideration set has no more than $n+1$ actions.

Let $(\alpha,\lambda)$ be a saddle point of \eqref{eq:maxmin_problem} that generates $P$. Notice that $\alpha$ is a solution of the following system of linear equations (label the system's independent variable by $\beta\in \mathbb{R}^A$):
\begin{align}\label{eq:lin_system1}
\sum_{a\in A}\beta(a) \nabla_\theta f^\star_\pi(a\pi-\lambda) & = 1, &\theta\in \Theta,\\
\sum_{a\in \supp P_\pi}\beta(a) & = \sum_{a\in \supp P_\pi}\alpha(a),\label{eq:lin_system11}\\
\beta(a) & = \alpha(a), & a \notin \supp P_\pi.\label{eq:lin_system111}
\end{align}
This linear system has $n+1+m-l$ equations and $m$ unknowns. Since $l>n+1$, there must be a non-zero vector $\beta$ such that 
\begin{align}\label{eq:lin_system2}
\sum_{a\in A}\beta(a) \nabla_\theta f^\star_\pi(a\pi-\lambda) & = 0, &\theta\in \Theta,\\
\sum_{a\in \supp P_\pi }\beta(a) & = 0,\label{eq:lin_system22}\\
\beta(a) & = 0, & a\notin \supp P_\pi \label{eq:lin_system222}.
\end{align}
Note that both $\beta$ and $-\beta$ are non-zero solutions of (\ref{eq:lin_system2})--(\ref{eq:lin_system222}). Hence, we can assume without loss of generality that $\beta(a)>0$ for some $a\in A$. 

We define $\gamma\in \RR^A$ as follows: for all $a\in A$,
\[
\gamma(a) = \alpha(a) - \beta(a) \min_{b: \beta(b)>0} \frac{\alpha(b)}{\beta(b)}.
\]
\begin{claim}\label{claim:suppor_bounds}
The vector $\gamma$ has the following properties:
\begin{enumerate}
\item $\gamma(a)\geq 0$ for all $a\in A$.
\item $\gamma(a)=0$ for some $a\in \supp P_\pi$.
\item $\gamma$ is a solution of (\ref{eq:lin_system1})--(\ref{eq:lin_system111}).
\item $\gamma\in \Delta(A)$.
\item If $\gamma(a)>0$, then $\alpha(a)>0$.
\item $(\gamma,\lambda)$ is a saddle point of \eqref{eq:maxmin_problem}.
\end{enumerate}
\end{claim}
\begin{proof}
(i). If $\beta(a) \leq 0$, then $\gamma(a) \geq \alpha(a)\geq 0$. If $\beta(a)>0$, then
\[
\gamma(a)\geq 0\quad\Longleftrightarrow\quad \frac{\alpha(a)}{\beta(a)}\geq \min_{b:\beta(b)>0}\frac{\alpha(b)}{\beta(b)}.
\]
Thus, $\gamma(a)\geq 0$ also when $\beta(a)>0$. This proves (i).

(ii). Take any $a$, with $\beta(a)>0$, such that 
\[
\frac{\alpha(a)}{\beta(a)} = \min_{b:\beta(b)>0}\frac{\alpha(b)}{\beta(b)}.
\]
Then, $\gamma(a)=0$. Moreover, (\ref{eq:lin_system222}) ensures that $a\in\supp P_\pi $. This proves (ii).

(iii). This follows from $\alpha$ being a solution of (\ref{eq:lin_system1})--(\ref{eq:lin_system111}) and $\beta$ being a solution of (\ref{eq:lin_system2})--(\ref{eq:lin_system222}). 

(iv). By (i), $\gamma(a)\geq 0$ for all $a\in A$. By (\ref{eq:lin_system11}) and (\ref{eq:lin_system111}), 
\begin{align*}
\sum_{a\in A}\gamma(a) & = \sum_{a\in \supp P_\pi} \gamma(a) + \sum_{a\notin \supp P_\pi} \gamma(a) \\
&= \sum_{a\in \supp P_\pi} \alpha(a) + \sum_{a\notin \supp P_\pi} \alpha(a) = \sum_{a\in A}\alpha(a)=1.
\end{align*}
It follows that $\gamma\in \Delta(A)$.

(v). For $a\notin \supp P_\pi$, we have $\gamma(a)=\alpha(a)$. For $a\in \supp P_\pi$, we have: 
\[
P_\pi(a) = \alpha(a) \sum_{\theta\in \Theta}\pi(\theta) \nabla_\theta f^\star_\pi(a\pi-\lambda).
\]
Thus, $\supp P_\pi\subseteq\supp\alpha$. Thus, in any case, $\gamma(a)>0$ implies $\alpha(a)>0$, as desired.

(vi). Since $\supp \gamma\subseteq \supp \alpha$ and $(\alpha,\lambda)$ is a saddle point, we have:
\[
\min_{a\in \supp\gamma} f^\star_\pi (a\pi-\lambda)\geq \min_{a\in \supp\alpha}f^\star_\pi(a\pi-\lambda)=\max_{a\in A}f^\star_\pi(a\pi-\lambda).
\]
Hence, it follows from (\ref{eq:lin_system1}) that $(\gamma,\lambda)$ is a saddle point.
\end{proof}

Let $Q$ be the optimal choice rule generated by $(\gamma,\lambda)$. We claim that $\vert \supp Q_\pi\vert < l$. To verify this claim, first we observe that $\supp Q_\pi\subseteq \supp P_\pi$. Indeed, $Q_\pi(a)>0$ implies 
\[
\gamma(a)>0\quad\text{and}\quad\sum_{\theta\in \Theta}\pi(\theta)\nabla_\theta f^\star_\pi(a\pi-\lambda)>0.
\]
Since $\gamma(a)>0$ implies $\alpha(a)>0$ (see (v) of Claim \ref{claim:suppor_bounds}), we obtain that $Q_\pi(a)>0$  implies 
\[
P_\pi(a) = \alpha(a) \sum_{\theta\in \Theta}\pi(\theta)\nabla_\theta f^\star_\pi(a\pi-\lambda)>0.
\]
This proves that $\supp Q_\pi\subseteq \supp P_\pi$. We also note that $\supp Q_\pi\neq \supp P_\pi$. Indeed, by (ii) of Claim \ref{claim:suppor_bounds}, there exists $a\in \supp P_\pi$ such that $\gamma(a)=0$. Given that $\gamma(a)=0$, we must have $Q_\pi(a)=0$. It follows that $\supp Q_\pi \neq \supp P_\pi$. Overall, we conclude that  $\supp Q_\pi$ is a proper subset of $\supp P_\pi$. This shows that $\vert \supp Q_\pi\vert < \vert \supp P_\pi\vert=l$, as desired.
\end{proof}

\section{Nested $f$-information}\label{sec:nested_f}

 The construction underlying nested Shannon entropy extends to general $f$-information costs. The key is to work with likelihood vectors, which are the arguments of $f$-information generators, rather than posterior beliefs, which are the arguments of entropy functions.

\begin{definition}\label{def.personal_state_space_general}
 A  \emph{generalized nest structure} consists of a finite set $N$ of nests, a vector $y^*\in \RR_+^N$, and a profile $z^*=(z_i^*)_{i\in N}$ of vectors $z_i^*\in \RR_+^\Theta$ such that:
\begin{enumerate}
    \item $y^*(i)>0$ for every $i\in N$;
    \item $\sum_{i\in N} y^*(i)z_i^*(\theta)=1$ for every $\theta\in\Theta$.
\end{enumerate}
\end{definition}

 The vector $y^*$ assigns weights to nests, while each $z_i^*$ assigns weights to states conditional on nest $i$. The nest structure studied in Section \ref{sec:nested-entropies} is the special case in which
 \begin{equation}\label{eqn:gen-nest-special}
    y^*=\bar{\nu} \qquad\text{and}\qquad z_i^*(\theta)=\frac{\nu_i(\theta)}{\pi(\theta)}.
 \end{equation}

 Given a generalized nest structure $(N,y^*,z^*)$, for every $\pi \in \Delta^\circ(\Theta)$, let the functions
\[
    g_\pi\colon \RR_+^N\to\overline{\RR}_+
    \qquad\text{and}\qquad
    h_{(\pi,i)}\colon \RR_+^\Theta\to\overline{\RR}_+,
    \quad i\in N,
\]
 be convex and lower semicontinuous, with $g_\pi(y^*)=0$ and $h_{(\pi,i)}(z_i^*)=0$ for every $i\in N$.

\begin{definition}\label{def.nested_f_information}
 \emph{Nested $f$-information} is generated by $f  = (f_\pi)_{\pi \in \Delta^\circ(\Theta)}$, defined as
 \[
    f_\pi(x) = \inf \left\{ g_\pi(y) + \sum_{i\in N}y(i)h_{(\pi,i)}(z_i) \right\} \qquad \forall x\in\RR_+^\Theta,
\]
 where the infimum is taken over all $y\in \RR^N_+$ and $z_i\in \RR^\Theta_+$, $i\in N$, such that $\sum_{i\in N}y(i)z_i=x$.
\end{definition}

 The interpretation parallels that of nested Shannon entropy, with likelihood vectors replacing posterior beliefs. In particular, $g$ governs the cost of acquiring information across nests, whereas $h_i$ governs the cost of acquiring information within nest $i$.

 The convex conjugate of $f$ admits the following representation: for every $x\in \RR^\Theta$,
 \[
    f^\star_\pi(x) = g^\star_\pi\left(h^\star_\pi\left(x\right) \right)
    \qquad\text{with}\qquad h^\star_\pi(x)
    =
    \left(h_{(\pi,i)}^\star(x)\right)_{i\in N}.
 \]
 Thus, $f^\star_\pi$ has a two-layer structure. The outer function $g^\star$ captures the across-nest dimension, while the inner functions $h_{(\pi,i)}^\star$ capture the within-nest dimension. Finally, $f$ satisfies Assumption~1 whenever $g$ and each $h_{(\pi,i)}$, $i\in N$, satisfy the corresponding conditions in Assumption~1.\footnote{In the analogues of Assumption~1(iii) for $g$ and $h_i$, the reference vector $\boldsymbol{1}$ should be replaced by $y^*$ and $z_i^*$, respectively.}

 Applying Theorem~\ref{thm:characterization} and the chain rule, optimal stochastic choice rules take the form
\[
    P_\theta(a) = \alpha(a)\sum_{i\in N}\nabla_i g^\star_\pi (h^\star_\pi(a\pi-\lambda))\nabla_\theta h^\star_{(\pi,i)}(a\pi-\lambda)).
\]
 This expression resembles the attribute-based stochastic choice rules commonly studied in discrete-choice theory (e.g., \citealp*{strzalecki2025stochastic}). The distinction is that, in our setting, the attributes are defined over states rather than actions.

We conclude by highlighting a leading special case of nested $f$-information:

\begin{example}[Nested Entropies]
    Consider the extension of nested Shannon entropy (Definition \ref{def:nested_s_entropy}) in which the KL divergences are replaced by arbitrary entropy functions. 
    
    Given any prior $\pi$ and corresponding nest structure $(N,\nu)$ (Definition \ref{def.personal_state_space}), let $G_\pi : \Delta(N) \to \RRcvx_+$ be an entropy with $G_\pi(\bar{\nu}) = 0$, and let $H_{\pi} = (H_{(\pi,i)})_{i \in N}$ be a profile of entropies $H_{(\pi,i)}: \Delta(\Theta) \to \RRcvx_+$ with $H_{(\pi,i)}(\nu_i) = 0$. The associated \emph{nested entropy} $F_\pi$ is 
    \[
    F_\pi(p)= \inf \left\{G_\pi(\bar{\mu}) + \sum_{i \in N } \bar{\mu}(i) H_{(\pi,i)}(\mu_i) \right\} \qquad \forall p \in \Delta(\Theta),
    \]
    where the infimum is taken over all $\mu \in \Delta(N \times \Theta)$ such that $\sum_{i \in N} \bar{\mu}(i) \mu_i = p$ (i.e., the marginal of $\mu$ on $\Theta$ is $p$). The induced posterior-separable cost can be represented as nested $f$-information as follows: consider the generalized nest structure in \eqref{eqn:gen-nest-special}, define 
    \[
    g_\pi(x) = \begin{cases}
        G_\pi(x), & \text{if $x \in \Delta(N)$},
        \\
        +\infty, & \text{otherwise},
    \end{cases}
    ~~~ \text{and} ~~~
     h_{(\pi,i)}(x) = \begin{cases}
          H_{(\pi,i)}(x\pi), &\text{if $\sum_{\theta \in \Theta} x(\theta)\pi(\theta)=1$}, \\
          +\infty, & \text{otherwise,}
     \end{cases}
    \]
    and let $f_\pi$ be the generator induced by the entropy $F_\pi$ through \eqref{eq:f_pi_entropy} (as in Example \ref{exa:CDL}). It can then be directly verified that $f_\pi$ is induced by $g_\pi$ and $h_{\pi}$ as in Definition \ref{def.nested_f_information}. 

    Using Theorem~\ref{thm:characterization} and the chain rule, the optimal posteriors in \eqref{eq:PS-FOC-dual} decompose as
    \[
    p_a = \nabla F^\star_\pi(a-\lambda_\pi) = \sum_{i \in N} \nabla_i G^\star_\pi\left( H^\star_{\pi} (a - \lambda_\pi) \right) \nabla H^\star_{(\pi,i)} (a - \lambda_\pi).
    \]
    This generalizes the nested logit decomposition of posterior beliefs from Proposition \ref{prop:perceptual_mm}.
\end{example}

\end{document}